\documentclass[journal]{IEEEtran}
\IEEEoverridecommandlockouts
\usepackage{amsmath,amssymb,amsfonts,bm}
\usepackage{algorithmic}
\usepackage{graphicx}
\usepackage[nospace]{cite}
\usepackage{textcomp}
\usepackage{booktabs}
\usepackage{xcolor}
\usepackage{hhline}
\usepackage{float}

\def\BibTeX{{\rm B\kern-.05em{\sc i\kern-.025em b}\kern-.08em
    T\kern-.1667em\lower.7ex\hbox{E}\kern-.125emX}}
\usepackage[nospace]{cite}
\graphicspath{ {./images/} }
\usepackage{url}
\usepackage{color}
\usepackage{float}
\usepackage{amssymb}
\usepackage{multirow}
\usepackage{caption}
\usepackage{subcaption}

\author{Yang Zhao,~\IEEEmembership{Graduate Student Member,~IEEE,}
Wenchao Zhai,~\IEEEmembership{Member,~IEEE,} 
Jun Zhao,~\IEEEmembership{Member,~IEEE,} \\
Tinghao Zhang,~\IEEEmembership{Graduate Student Member,~IEEE,}
Sumei Sun,~\IEEEmembership{Fellow,~IEEE,} 
Dusit Niyato,~\IEEEmembership{Fellow,~IEEE,}
and Kwok-Yan Lam,~\IEEEmembership{Senior Member,~IEEE}
\thanks{Yang Zhao, Jun Zhao, Tinghao Zhang,  Dusit Niyato and Kwok-Yan~Lam are with School of Computer Science and Engineering, Nanyang Technological University, Singapore. (Email: s180049@e., junzhao, tinghao001@e., dniyato, kwokyan.lam\}@ntu.edu.sg).
}
\thanks{Wenchao Zhai is with College of Information Engineering, Jiliang University, Hangzhou, China. (Email: zhaiwenchao@cjlu.edu.cn).
}
\thanks{Sumei Sun is with Institute for Infocomm Research (I2R), Agency for Science, Technology and Research (A*STAR), Singapore, and
Infocomm Technology Cluster, Singapore Institute of Technology, Singapore.
(Email: sunsm@i2r.a-star.edu.sg).}
\thanks{*Part of this paper is published in the FICC conference.}
}

\begin{document}

\title{A Comprehensive Survey of 6G Wireless Communications}

\maketitle

\begin{abstract}
While fifth-generation (5G) communications are being rolled out worldwide, sixth-generation (6G) communications have attracted much attention from both the industry and the academia. Compared with 5G, 6G will have a wider frequency band, higher transmission rate, spectrum efficiency, greater connection capacity, shorter delay, broader coverage, and more robust anti-interference capability to satisfy various network requirements. This survey presents an insightful understanding of  6G wireless communications by introducing requirements, features, critical technologies, challenges, and applications. First, we give an overview of 6G from perspectives of technologies, security and privacy, and applications. Subsequently, we introduce various 6G technologies and their existing challenges in detail, e.g., artificial intelligence (AI), intelligent surfaces, THz, space-air-ground-sea integrated network, cell-free massive MIMO, etc. Because of these technologies, 6G is expected to outperform existing wireless communication systems regarding the transmission rate, latency, global coverage, etc. Next,  we discuss security and privacy techniques that can be applied to protect data in 6G. Since edge devices are expected to gain popularity soon, the vast amount of generated data and frequent data exchange make the leakage of data easily. Finally, we predict real-world applications built on the technologies and features of 6G; for example, smart healthcare, smart city, and smart manufacturing will be implemented by taking advantage of AI.
\end{abstract}


\section{Introduction}
As 5G communication networks are being deployed commercially~\cite{5Gcountries}, the academic and industry start developing 6G wireless communication systems. Currently, the rapid growth of data-centric intelligent systems has brought significant challenges to 5G wireless systems. For example, the haptic Internet-based telemedicine requires that the delay of air interface is less than $0.1$ millisecond (ms)~\cite{matti2019key}. But the existing delay is only $1$ms, which is not satisfying. 5G's ubiquitous mobile ultra-broadband, ultrahigh data density, and ultrahigh-speed-with-low-latency communications cannot fully satisfy the high computation, low latency, etc., which are required by the 6G's applications~\cite{chowdhury20196g, saad2019vision, ho2019next, mollah2019emerging}. Compared with 5G, 6G has stricter requirements in power consumption, lower latency, higher reliability, privacy and security, etc. Also, it provides the better service of quality (QoS)  and broader coverage than wireless communications in the past.  6G is considered to be a revolutionary generation of wireless communication because of the growing roles of intelligence, autonomy will bring new vigor and vitality into the communications.

\begin{figure}[!h]
    \centering
     \includegraphics[scale=0.56]{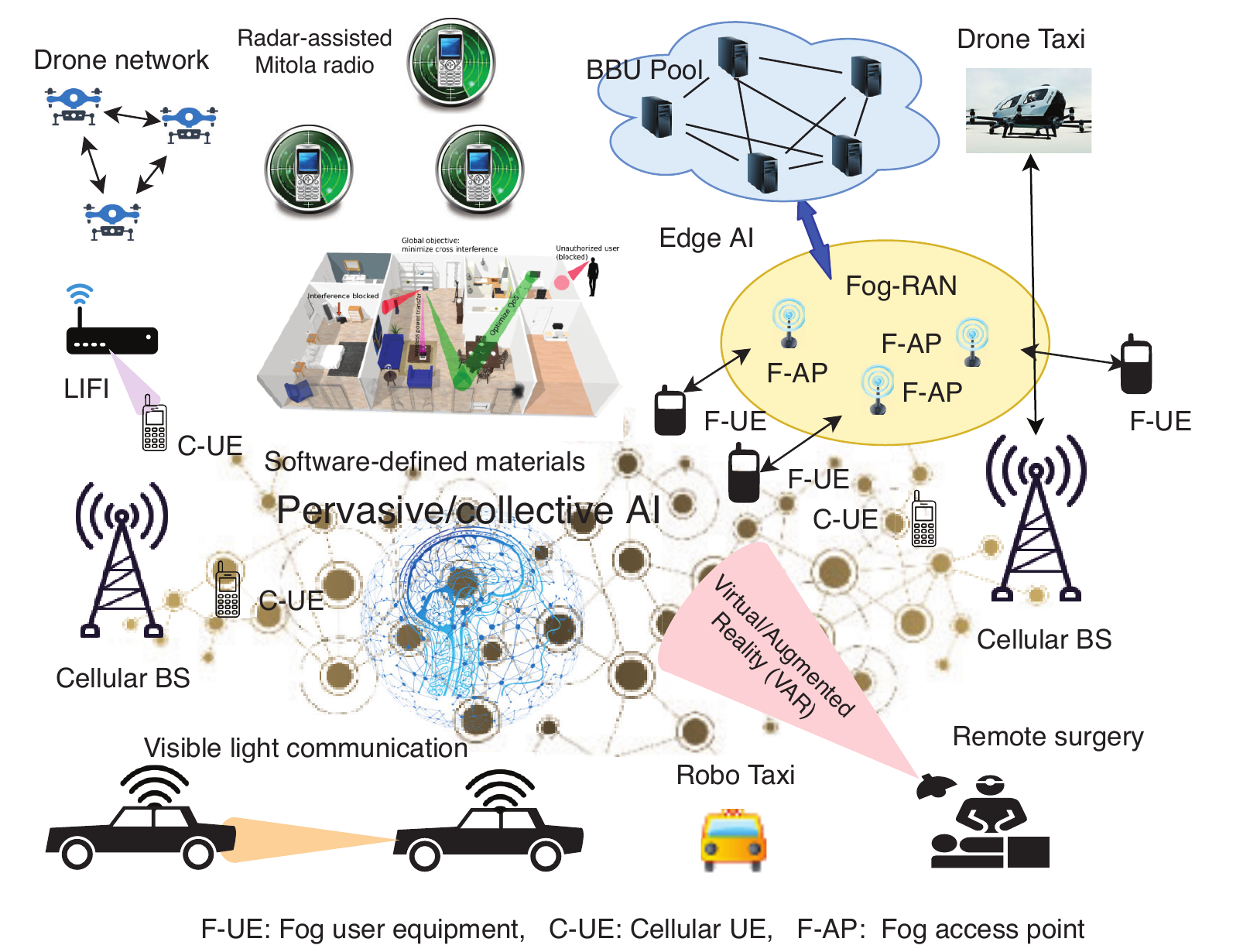}
    \caption{The vision of 6G.}
    \label{fig:vision}
\end{figure}

To be more specific, Table~\ref{table} lists details of key performance indicators (KPI) of 6G. Various candidate technologies have been proposed to overcome the bottleneck of existing wireless communication systems to meet 6G's requirements.  For example, artificial intelligence (AI) is expected to enable a significant paradigm shift in 6G wireless networks, including machine learning, deep learning, etc.~\cite{loven2019edgeai}. The rapid consumption of the spectrum resources makes the efficient utilization of the spectrum highly important. In order to improve the spectral efficiency, various techniques will be introduced in 6G, including existing technologies like index modulation (IM) and free duplex. Moreover, new technologies such as intelligent reflecting surface (IRS) may enhance the spectrum efficiency by leveraging the passive reflecting elements without using the external power source. Furthermore, because the consumption of the spectrum frequency is fast in the existing wireless communication systems, 6G also explores the sub-terahertz (sub-THz), visible light communication (VLC), and terahertz (THz), which use frequencies ranging from 100 GHz to 3 THz~\cite{david2019defining, tariq2019speculative, saad2019vision, giordani2019towards, xiao2017millimeter, andrews2016modeling, zhu2019millimeter, nawaz2019quantum, corre2019sub}. Besides, to obtain the target of the global coverage, 6G wireless communications use the space-air-ground-sea integrated network (SAGSIN) to connect the communication systems from the sky to the deep sea~\cite{giordani2020satellite}.  Also, the blockchain is used to manage and share the spectrum resources, which may eliminate the central authority because blockchains are immutable, untampered decentralized databases~\cite{underwood2016blockchain}. In addition to the spectral efficiency, the improvement of the efficiency of the energy consumption is quite necessary as well. Therefore, the simultaneous wireless information and power transfer (SWIPT) technology is leveraged to prolong the battery's span to reduce the power consumption.

Additionally, security and privacy technologies are being eye-catching due to the explosive growth of data generated by edge devices. As the 6G generation is coming, the number of edge devices will increase rapidly. Therefore, more security and privacy related techniques will be explored to protect data's confidentiality. Finally, when above technologies are getting mature, emerging applications illustrated by Fig.~(\ref{fig:vision}), such as the smart healthcare and the holography radio, will be launched to enable 6G to be a part of our life. More new services will also be supported, for instance, smart wearable devices, computing reality devices, autonomous vehicles, implants, sensing, and 3D mapping~\cite{chowdhury20196g}.

\begin{table}[!h]
\centering
 \caption{Requirements and Features of 6G~\cite{ho2019next, piran2019learning,dang2020should,nayak20206g}.}
\renewcommand{\arraystretch}{1.5}
\label{table}
\begin{tabular}{l|l}
        \toprule
\textbf{Requirements}                       & \textbf{6G}                                                                                                                                                                     \\ \hhline{==}
Service types                               & MBRLLC/mURLLC/HCS/MPS                                                                           \\ \hline
Service level                               & Tactile         \\ \hline
Device types                                & Smart implants/ CRAS/ \\&XR and BCI equipment/ \\&Sensors and DLT devices \\ \hline
Jitters                                     & 1 $\mu$sec        \\ \hline
Individual data rate                        & 100 Gbps  \\ \hline
Peak DL data rate                           & $\geq$ 1 Tbps   \\ \hline
Latency                                    & $0.1$ msec                                                                                                                                                                        \\ \hline
Mobility                                    & up to 1000 km/h                                                                                                                                                                 \\ \hline
Reliability                                 & up to 99.99999\%                                                                                                                                                                \\ \hline
Frequency bands                             & \begin{tabular}[c]{@{}l@{}}- sub-THz band\\ - Non-RF, e.g, optical, VLC, laser $\cdots$\end{tabular}                     \\ \hline
Multiplexing                                & Smart OFDMA plus IM \\  \hline
Power consumption                           & Ultra low                                                                                                                                                                       \\ \hline
Processing delay                            & $\leq$ 10ns                                                                                                                                                        \\ \hline
Maximum rate                &100Gb s$^{-1}$ \\ \hline
Security and privacy                        & Very high                                                                                                                                                                       \\ \hline
Network orientation                         & Service-centric                                                                                                                                                                 \\ \hline
Wireless power transfer/ \\Wireless charging & Support (BS to devices power transfer)                                                                                                                                         \\ \hline
Smart city components                       & Integrated                                                                                                                                                                      \\ \hline
Autonomous V2X                              & Fully                                                                                                                                                                           \\ \hline
Localization precision                      & 1 cm on 3D     
                                               \\ \hline
\end{tabular}
\end{table}


\textbf{Contributions.} The contributions of this survey are summarized below.
\begin{itemize}
    \item Existing papers on 6G pay more attention to predicting technologies that may be used in the future, and few of them give a summary. Our paper surveys almost all of the existing visions of 6G and summarizes them in detail, including both physical and network layer technologies. Furthermore, we give a deep insight into current technologies potentially utilized in 6G according to how these technologies satisfy the requirements of future 6G network, thus facilitating the understanding of their respective advantages comprehensively.
    \item We highlight some technologies that may be essential for 6G, including index modulation, artificial intelligence, intelligent surfaces, simultaneous wireless information, power transfer, etc. These technologies' advantages and challenges are discussed in our manuscript, which gives future work directions for 6G applications. 
    \item Not only the principles of the technologies themselves are presented, but their relevant variations are also provided. Therefore, readers can have a comprehensive understanding of these technologies and their evolution in solving sophisticated problems.
    \item Applications of  potential 6G technologies are also investigated in our manuscript. By summarizing the applications in various scenarios, we help recognize the 6G network's differences from its 5G counterpart. Therefore, it is easy to have an exact blueprint for the development of a future network.
\end{itemize}

\textbf{Organization.} The rest of paper is organized as follows. Section~\ref{sec:overview} gives an overview of emerging technologies that enable the paradigm shift in 6G wireless networks. From Section~\ref{sec:IM} to Section~\ref{sec:other-technologies}, we present each technology of 6G in detail. Section~\ref{sec:security} discusses potential security and privacy problems and their  corresponding solutions  existing in the 6G wireless communications. Section~\ref{sec:applications} lists some applications that will be enabled by 6G technologies. Section~\ref{sec:conclusion} concludes this paper.

\textbf{Notations.} Boldfaced letters denote vector and boldfaced capital letters represent matrix. $||\cdot||_F$ represents the Frobenious norm of a matrix. $|\cdot|$ is the magnitude of a complex number. The subscripts $(\cdot)^T$, $(\cdot)^*$, and $(\cdot)^H$ define the operation of transpose, conjugate, and Hermitian transpose, respectively. $\textbf{I}_N$ represents an $N \times N$ identity matrix. Table~\ref{table:notions} summarizes abbreviations used in this paper.


\begin{table}[htp]
\centering
\caption{Table of Abbreviations.}
\label{table:notions}
\begin{tabular}{|c|c|}
\hline
5G                     & Fifth generation                                                                               \\ \hline
6G                     & Sixth generation                                                                               \\ \hline
IM                     & Index Modulation                                                                               \\ \hline
IoE                    & Internet of Everything                                                                         \\ \hline
SWIPT                  & \begin{tabular}[c]{@{}c@{}}Simultaneous Wireless Information\\ and Power Transfer\end{tabular} \\ \hline
AI                     & Artificial Intelligence                                                                        \\ \hline
IRS                    & Intelligent Reflecting Surfaces                                                                \\ \hline
LIS                    & Large Intelligent Surfaces                                                                     \\ \hline
THz                    & Terahertz Communications                                                                       \\ \hline
SR                     & Symbiotic  Radio               
                                                \\ \hline
VLC                    & Visible Light Communication                                                                    \\ \hline
D2D                    & Device-to-Device Communication                                                                 \\ \hline
CFmMM                  & Cell-Free massive MIMO                                                                         \\ \hline
SAGSIN                  & Space-Air-Ground-Sea Integrated Network                                                   \\ \hline
NIB                    & Network in Box                                                                                 \\ \hline
AIoT                   & Artificial Internet of Things                                                                  \\ \hline
mMIMO                  & massive multiple  input  multiple  output                                                      \\ \hline
FD                     & Full-duplex                                                                                    \\ \hline
APM                    & Amplitude Phase Modulation                                                                     \\ \hline
PSK                    & Phase-Shift Keying                                                                             \\ \hline
QAM                    & Quadrature Amplitude Modulation                                                                \\ \hline
ML                     & Maximum Likelihood                                                                             \\ \hline
SD-IM                  & Spatial-Domain IM                                                                              \\ \hline
SSK                    & Spatial Shift Keying                                                                           \\ \hline
FD-IM                  & Frequency-Domain IM                                                                            \\ \hline
TD-IM               & Time-Domain IM                                                                     \\ \hline
CD-IM                  & Channel-Domain IM                                                                              \\ \hline
SM                  & Spatial modulation                                                                 \\ \hline
OFDM                   & Orthogonal Frequency Division Multplexing                                                      \\ \hline
TDD                    & Time Division Duplex                                                                           \\ \hline
RF                     & Radio Frequency                                                                                \\ \hline
MMSE                   & Minimum Mean Square Error                                                                      \\ \hline
CSI                    & Channel State Information                                                                      \\ \hline
QIMMA                 & Quadrature IM Multiple Access                                                                                  \\ \hline
NOMA                   & Non-Orthogonal Multiple Access                                                                 \\ \hline
BICM                   & Bit-Interleaved Coded Modulation                                                               \\ \hline
UM-MIMO                & Ultra-Massive MIMO                                                                             \\ \hline
LIM                    & Large Intelligent Metasurface                                                                  \\ \hline
SDS                    & Software-Defined Surface                                                                       \\ \hline
SDM                    & Software-Defined Metasurface                                                                   \\ \hline
EM                     & Electromagnetic                                                                                \\ \hline
QoS                    & Quality of Service                                                                             \\ \hline
BS                     & Base Station                                                                                   \\ \hline
UE                     & User Equipment                                                                                 \\ \hline
SINR                   & Signal-to-Interference-Plus-Noise Ratio                                                        \\ \hline
BER                    & Bit-Error Rate                                                                                \\ \hline
HD                     & Half-Duplex                                                                                    \\ \hline
RFID                   & Radio Frequency Identification                                                                 \\ \hline
TS                     & Time Switching                                                                                 \\ \hline
PS                     & Power Splitting                                                                                \\ \hline
SIC                    & Successive Interference Cancellation                                                           \\ \hline
SER                    & Symbol Error Rate                                                                              \\ \hline
PEP                    & Pairwise Error Probability                                                                     \\ \hline
SNR                    & Signal-to-Noise Ratio                                                                          \\ \hline
\end{tabular}
\end{table}

\section{An Overview of 6G Technologies}~\label{sec:overview}

6G is expected to outperform 5G in multiple specifications. We highlight six of them in Fig.~(\ref{fig:vision}), including the frequency, individual data rate, peak data rate, spectral efficiency, mobility, and latency. From Fig.~(\ref{fig:vision}), 6G requires the lower latency,  higher frequency and data transmission rate, faster mobility, and better spectral efficiency than 5G. In the following, we present an overview of promising technologies that contribute to the development of 6G wireless communications;  Subsequently, we discuss security and privacy problems and introduce potential applications.

\begin{figure}[!h]
    \centering
     \includegraphics[scale=0.55]{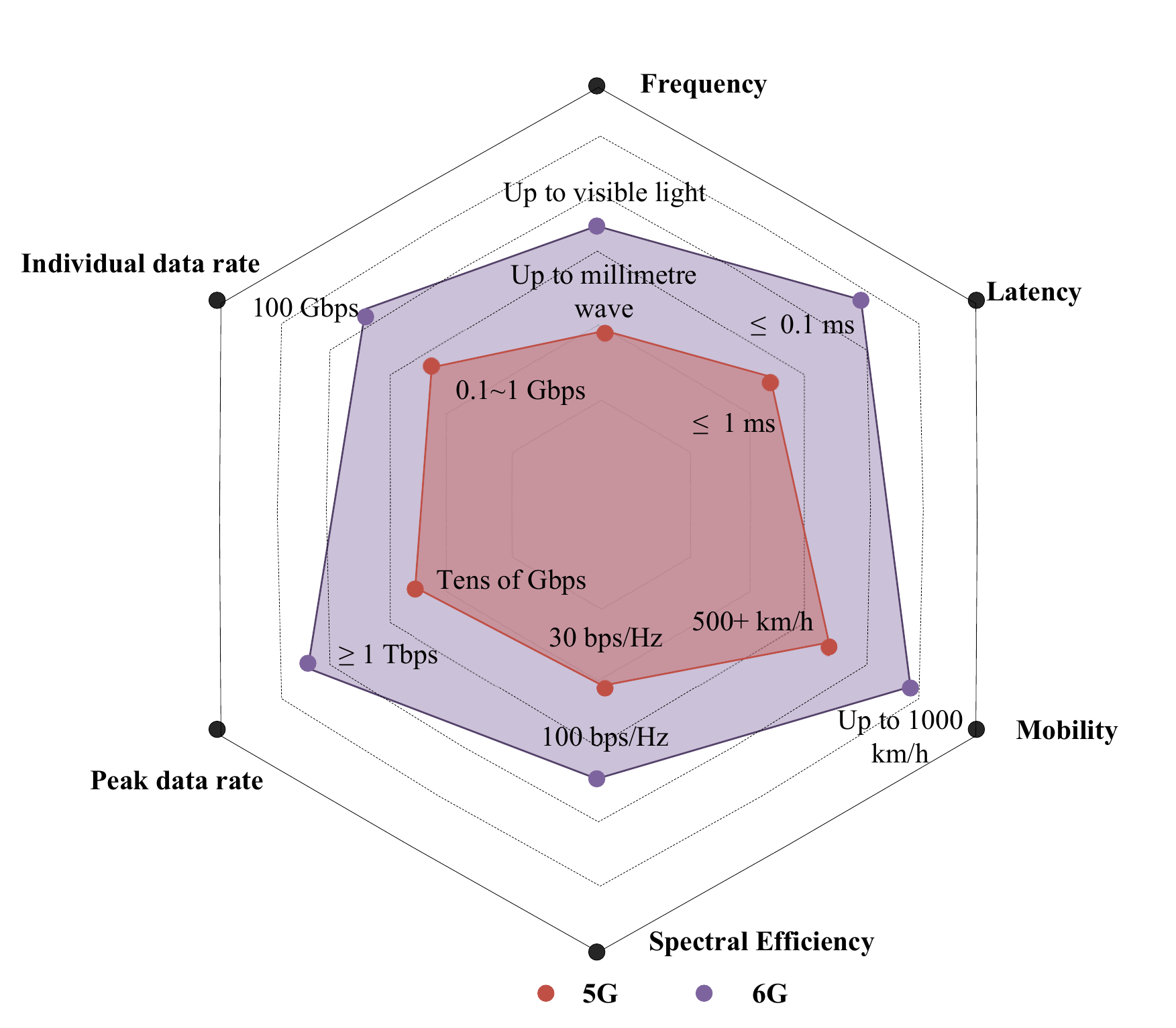}
    \caption{Comparison between 5G and 6G communications.}
    \label{fig:compare_5g_6g}
\end{figure}

\subsection{Technologies}

In this section, we give a brief introduction to the most eye-catching technologies pertaining to 6G, including  artificial intelligence (AI),  intelligent surfaces, terahertz (THz),  symbiotic radio (SR), free duplex, cell-free massive MIMO (CFmMM), space–air–ground-sea integrated network (SAGSIN), blockchain-based network, index modulation (IM), simultaneous wireless information and power transfer (SWIPT),  visible light communication (VLC), device-to-device (D2D),  and network in box~\cite{dang2020should, kato2020ten,shafin2020artificial,tan2020thz,elmeadawy2020enabling,gui20206g,chen2020vision,xiaohutowards}. These technologies have potentials to lay the foundation for future's communications and applications.

\textbf{Artificial Intelligence.} Artificial intelligence (AI) provides intelligence and automation to wireless networks by imitating human thought processes and intelligent behaviors. As 6G is envisioned to offer full automation, AI becomes one of the critical and highly valued technologies in the future's 6G wireless communication systems.  AI is also considered to be a modern tool for data analysis. By learning from big data, machine learning models are built to make correct decisions automatically. Thus, AI technologies, for example, machine learning, are expected to play an essential role in networks when they are too dynamic and complicated to be analyzed by humans.  In particular,  scenarios involve complex processes such as the joint optimization in network design, resource management, and resource allocation.

\textbf{Intelligent Surfaces.} There are two kinds of intelligent surfaces, including the large intelligent surfaces (LIS) and intelligent reflecting surfaces (IRS). Both of them are considered as promising candidate technologies for 6G. The concept of deploying antenna arrays as the LIS in massive MIMO systems is initially proposed by Hu~\emph{et al.}~\cite{Hu2018LIS,Hu2017LIS}. Compared with the beamforming technology, where many antennas are required to make the signals focused,  the LIS is electromagnetically active in the physical environment and impose little restrictions on how antennas spread. Therefore, the LIS can avoid the adverse effects of antenna correlations. However, due to the surfaces' active property, the LIS consumes many power resources and are not power efficient.

To overcome the shortness of power consumption, the IRS is proposed by Wu~\emph{et al.}~\cite{Wu2019IRS} to replace the active antennas used by the LIS. The IRS uses passive reflecting elements to reflect signals intelligently such that the data rate and the network coverage can be improved. Thus, the IRS is considered to tune the wireless environments to increase the spectrum and energy efficiency while consuming little power. Therefore, by controlling the incident waves' reflection characteristics, the signal quality can be improved significantly.

\textbf{Terahertz.} Terahertz (THz) is considered as one of the key technologies for the 6G wireless communications~\cite{tariq2019speculative, saad2019vision, giordani2019towards, xiao2017millimeter, andrews2016modeling, zhu2019millimeter}. 5G defines below 100 GHz as the millimeter-wave bands, whereas 100 GHz - 3 THz is categorized as THz band in 6G~\cite{sarieddeen2019next}. The above 90 GHz band is merely used for scientific service, which has not been fully explored. Therefore, it is envisioned to support the increased wireless network capacity~\cite{corre2019sub}. The THz also enables the ultra-high bandwidth and ultra-low latency communication paradigms, which caters to the needs of many emerging applications such as autonomous driving, Internet of Things.  However, as the frequency increases, the path loss increases as well, which makes it challenging for long-range communications. Thus, the THz is particularly suitable for high bit-rate short-range communications~\cite{nawaz2019quantum}.

\textbf{Symbiotic Radio.} Symbiotic radio (SR)~\cite{liang2020symbiotic} is a new technique that maintains the advantages and make up the disadvantages of cognitive radio (CR) and ambient backscattering communications (AmBC). SR is considered as one of the promising solutions to build 6G as a spectrum and energy-efficient communication system. In addition, IRS technology can further contribute to improving the performance of the transmission by enhancing the backscattering link signal.

\textbf{Free Duplex.} 6G will eliminate the difference between FDD and TDD, and the frequency sharing is based on the requirements, which we call free duplex~\cite{lu20206g, zhao20196g, akhtar2020shift}. Hence, the spectrum resource allocation in 6G will be more efficient and effective. Free duplex technology will help 6G increase its transmission rate, throughput, and reduce its transmission latency.

\textbf{Cell-Free Massive MIMO.} Cellular wireless networks are based on cellular topology. An area is then divided into multiple cells according to the topology, and each cell is served by one base station. The drawback of the cell-based wireless network is that if the device is at the edge the cell, its signal is pretty weak. By tackling the low communication ability at the edge of the cell, the cell-free  massive multiple-input-multiple-output  (CFmMM) concept is proposed, which removes the cells of the network. The idea of CFmMM means that one device is no longer attached to a single base station; instead, all base stations coherently serve the device in an area.

\textbf{Space-Air-Ground-Sea Integrated Network.} 6G wireless communication networks will integrate space-air-ground-sea networks to achieve the global coverage~\cite{giordani2020satellite}, i.e., 6G will construct a space-air-ground-sea integrated network (SAGSIN). As expected in the white paper released in January 2020, the 6G network should cover environments, including sky (10,000km) and sea (20 nautical miles)~\cite{White2020}. By integrating the three networks, SAGSIN can cope with various users and services' growing traffic demands.

\textbf{Blockchain-based Network.} Blockchains are distributed databases which are constructed based on the theory of the hash tree, and they are tamper-proof and hard to reverse~\cite{underwood2016blockchain}. Blockchains have attributes like auditability, data integrity, and transparency~\cite{xu2020blockchain}. Thus, blockchains can be used to manage spectrum resources without using a centralized authority. Besides, blockchains are also suitable to protect data's security and privacy or control access.  Fan~\emph{et al.}~\cite{fan2017blockchain} propose an efficient and secure blockchain-based privacy-preserving scheme, combining access policy, and encryption technology to guarantee data privacy. Kotobi~\emph{et al.}~\cite{kotobi2018secure} use blockchain as a decentralized database to improve the access protocols and secure spectrum sharing in mobile cognitive radio networks. Yang~\emph{et al.}~\cite{yang2017blockchain} present a trusted authentication architecture based on blockchain for could radio over a fiber network.

\textbf{Index Modulation.} Index modulation (IM) conveys the source information bits through the classical APM signals and the index selection of resource entities. Therefore, IM can improve the transmission rate and is potentially to be used in 6G. Chau~\emph{et al.}~\cite{Chau2001SSK} suggest that information bits can be transmitted through the index of the antennas in MIMO systems. They name such a technique as space shift keying (SSK) and combine the SSK with classical linear modulation, amplitude-phase modulation (APM); for example, space modulation (SM) is proposed based on the same idea in SSK. In SM technology, the source information bits are divided into two parts: the index of the transmit antennas and the other parts for the APM. Therefore, SM can significantly increase the transmission rate by sending the spare information bits through the traditional APM transceiver's antenna index. Apart from the antennas, other resource entities can also be indexed to transmit the additional information bits. These resource entities include time slots, sub-carriers, and channel state. Also, this modulation class, where the spare information bits are transmitted through the index of resources, is called index modulation. Orthogonal frequency division multiplexing access (OFDMA) plus IM will be essential technologies to significantly increase the throughput to support more users to access the 6G network.

\textbf{Simultaneous Wireless Information and Power Transfer.} 6G is supposed to be a complex network where a large variety of smart devices are accessed to the system and are required to communicate with others at anytime, and the lifetime of the battery-charging modules is also required to fulfill the constraints of ultra-low power consumption as listed in Table~\ref{table}. To prolong the life span of various devices in the network, simultaneous wireless information and power transfer (SWIPT) technology is proposed. SWIPT enables sensors to be charged exploiting wireless power transfer; thereafter, battery-free devices can be supported in 6G, reducing the network's power consumption substantially~\cite{Wang2014SWIPT}. Zhang~\emph{et al.}~\cite{Zhang2015SWIPT} describe SWIPT technology from the perspective of a scientific hypothesis and engineering practice in detail. Subsequently, in~\cite{Liu2016SWIPT,Yang2017SWIPT,Gong2017SWIPT,Alsaba2018SWIPT}, performance on the outage probability, throughput, and sum rate for non-orthogonal multiple access (NOMA) networks with SWIPT are derived. Bariah~\emph{et al.}~\cite{Bariah2019SWIPT,Li2019SWIPT} first conduct error probability analysis of the NOMA-based networks with SWIPT. The error probability performance of SWIPT is severely affected by the power splitting factor, which provides a principle for designing the SWIPT system in practical engineering.

\textbf{Visible Light Communication.} Visible light communication (VLC) uses visible light between 400 and 800 THz (780–375 nm) to communicate information. It provides ultra-high bandwidth (THz), zero electromagnetic interference, free abundant unlicensed spectrum, and very-high-frequency reuse~\cite{haas2018lifi}. Therefore, VLC helps to develop the short range communications of 6G networks~\cite{strinati20196g}. Besides, 1G-5G wireless networks have utilized micro-wave communications over the sub-6 GHz band, whose resources are almost used up~\cite{zhu2019millimeter}.  Correspondingly,  since more smart devices are integrated into the network, and an explosive growth appears for the area traffic capacity, the high capacity becomes an essential requirement of 6G. To fulfill the requirement of capacity or data rate, VLC technologies are potential to be used in 6G.

\textbf{Device-to-Device.} Device-to-Device (D2D) communication represents the direct communication between devices without going through base stations, and the communication can be done under licensed (i.e., cellular network) or unlicensed spectrum (i.e., WiFi)~\cite{kar2020critical, malik2020survey}.  D2D  improves the throughput, energy efficiency, delay, and fairness of the communication~\cite{asadi2014survey, zhang2014social}. As the number of edge devices is explosively increasing, D2D is gaining more attention and getting more widely implemented~\cite{zhang2020beyond}.

\subsection{Security and Privacy} In 6G, every edge device is envisioned to connect to the Internet, and AI applications will be widely used by edge devices. Most AI applications are data-driven, which raises the concern of security and privacy of the collected data~\cite{sun2020machine}. For example, wearable devices may collect clients' health status everyday, but these data are private and sensitive. Clients' privacy may be compromised if the data are leaked. Moreover, Akhtar~\emph{et al.}~\cite{akhtar2018threat} investigate the adversarial attacks against deep learning models and show how the attacks happen in practice, whereas some other security and privacy issues are identified in~\cite{mcmahan2017communication}. Lovén~\emph{et al.}~\cite{loven2019edgeai} identify the challenges of the edge AI and aim to improve edge computing and AI-based approaches security via security systems. Zhou~\emph{et al.}~\cite{zhou2018robust} combine deep learning and edge computing, developing robust mobile crowdsensing that can conduct data validation and local processing. Sattiraju~\emph{et al.}~\cite{sattiraju2019ai} analyze the feasibility of utilizing machine learning techniques such as recurrent neural networks in the physical layer and propose an unsupervised learning algorithm to improve the physical layer security.

\subsection{Applications} Empowered by 6G, a wide range of AI applications will evolve into ``connected intelligence''\cite{Letaief2019IEEE}, hence facilitating every aspect of our daily life. For example, advanced AI approaches can be employed in network management or autonomy to save manpower~\cite{piran2019learning,loven2019edgeai}. In addition, 6G revitalizes smart healthcare by providing real-time health monitoring~\cite{Nayak2020Healthcare}, high-precision medical treatment~\cite{tariq2019speculative}, and reliable privacy protection~\cite{nayak20206g}. With the advent of 6G, Industry 4.0 will be fully realized as smart manufacturing will achieve high-precision manufacturing~\cite{rajatheva2020white}. Intelligent robots connected by ubiquitous 6G network enable manufacturing systems to carry out complex and dangerous tasks without risking people's life~\cite{Nayak2020Healthcare}. Moreover, the smart home that equips with intelligent IoT devices will provide a comfortable living environment to people~\cite{Nayak20Communications}, and 6G allows the smart home to ensure the residents' security. In terms of traffic and transportation, the sophisticated sensing and planning algorithms can be deployed for traffic optimization~\cite{Nayak20Communications}. Other applications such as smart grid~\cite{Fang12} and unmanned aerial vehicle~\cite{Wang20Security} will also be enhanced with the aid of 6G.

\begin{figure}[!htb]
    \centering
    \includegraphics[scale=0.7]{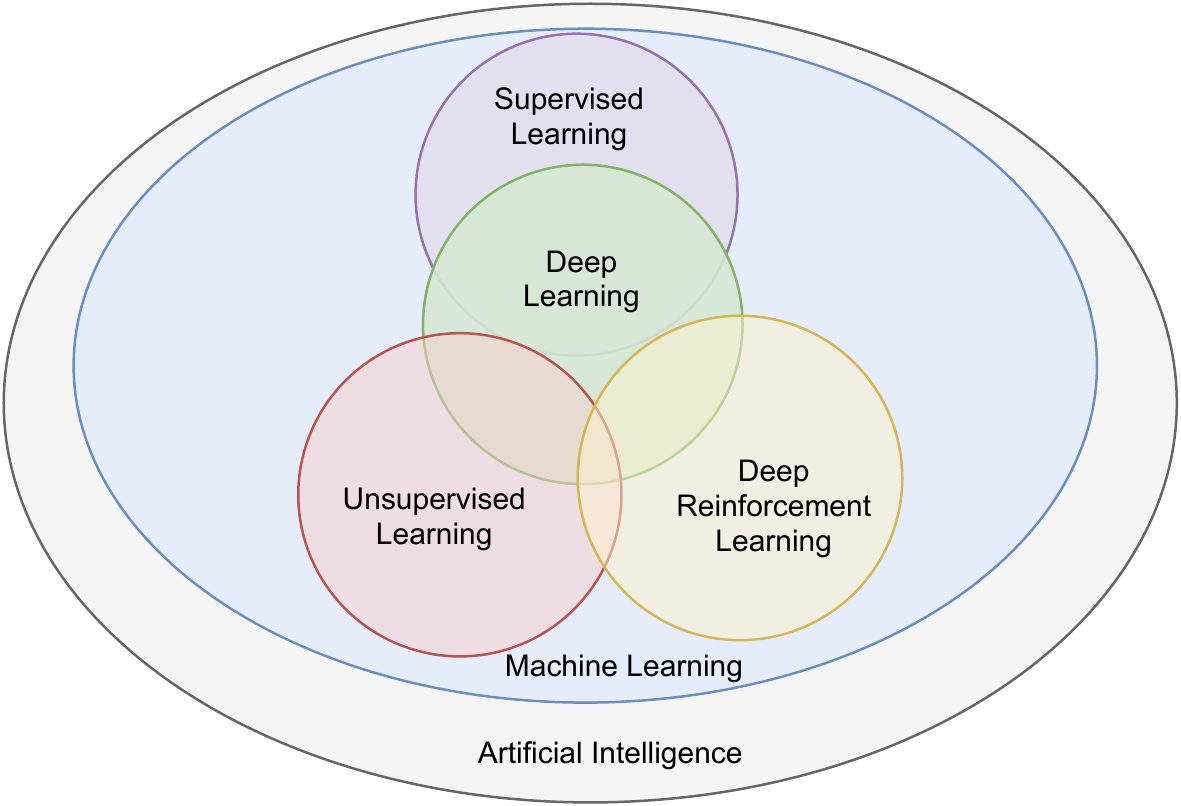}
    \caption{The relationship of AI, ML and DL~\cite{zhang2020towards}.}
    \label{fig:AI_relation}
\end{figure}

\begin{figure*}[!htb]
    \centering
    \includegraphics[scale=0.75]{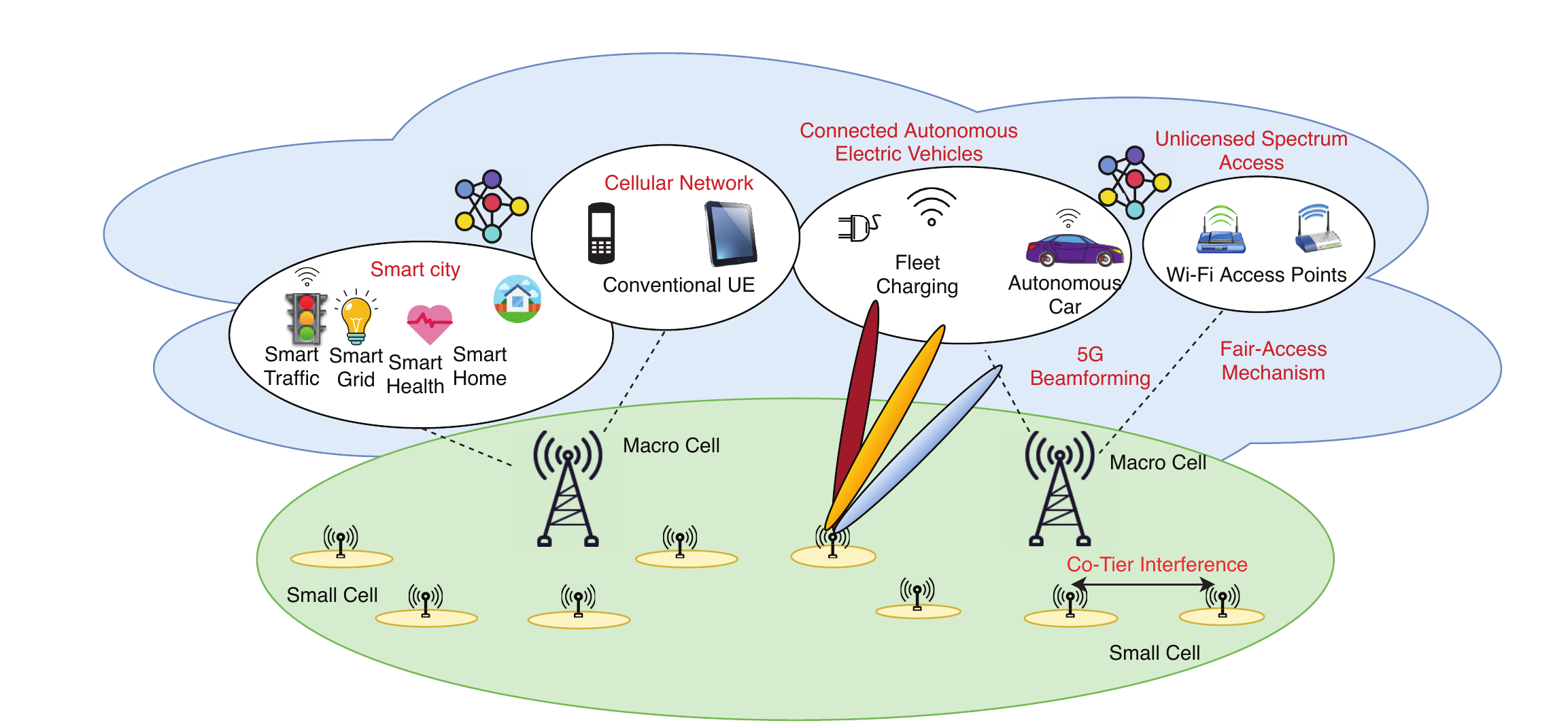}
    \caption{An AI-Enabled 6G wireless network and related applications.}
    \label{fig:AI}
\end{figure*}

\section{Artificial Intelligence} \label{sec:AI}

Artificial Intelligence (AI) provides intelligence to wireless networks by simulating human thought processes and intelligent behaviors~\cite{Letaief2019IEEE, elsayed2019ai}. Machine learning is the widely recognized the start-of-the-art technology to achieve AI, which learns from datasets to find out their pattern. Commonly used machine learning methods include deep learning, supervised learning, unsupervised learning, and deep reinforcement learning~\cite{mao2018deep}. Depending on the labels of the training datasets and types of training tasks, different machine learning approaches are applied. Deep learning is a branch of ML, which uses the principle of backpropagation for building models. Besides, the supervised learning, unsupervised learning, and deep reinforcement learning also can leverage deep learning methods for training. The relation among aforementioned machine learning methods are illustrated in Fig.~(\ref{fig:AI_relation}). In 6G, more advanced training approaches are introduced, including model-driven deep learning, federated learning, and explainable artificial intelligence. Specifically, model-driven deep learning uses the professional experience to build a model instead of pure data-driven training~\cite{zappone2019model,he2019model}. Federated learning is a new machine learning technique that enables users to train models with local data without sending raw data to the central server. Since existing deep learning training is still a black-box, scientists are unable to explain its working principles. However, to apply deep learning training methods in real-world applications, the explanation is required to gain trust. Hence, the explainable artificial intelligence is motivated. 

With the increasing popularity of AI technologies, 6G enables various applications in communications such as network planning, network optimization, network operation and maintenance, and autonomous vehicles~\cite{david2019defining, zong20196g, letaief2019roadmap, elsayed2019ai, zappone2019model, gacanin2019autonomous, strinati20196g, stoica20196g, zhang20196g, zhangLin20196g, piran2019learning} as shown in Fig.~(\ref{fig:AI}). In particular, AI, residing in new local ``clouds'' and ``fog'' environments, helps create many novel applications using sensors that can be embedded into every corner of our life~\cite{giordani2019towards}. 
For example, Table~\ref{ml_techniques} lists applications and their corresponding deep learning methods and tools in the physical layer, MAC layer, and network layer. Deep learning, which is considered as the vital ingredient of AI technologies, has been widely used in wireless networks~\cite{mao2018deep}. It will play an essential role in various areas, including semantic communications, holistic management of communication, computation, caching, control resources areas, etc., which may contribute to the paradigm-shift of 6G. In the following, we introduce popular AI methods and their applications in detail.


\begin{table*}[!htbp]
\centering
\caption{Machine Learning Techniques Overview.}
\renewcommand\arraystretch{1.5}
\begin{tabular}{l|l|l|l}
\toprule
\textbf{ML Types} & \textbf{Description} & \textbf{ML Structures} & \textbf{References} \\ \hhline{====}
Supervised Learning & \begin{tabular}[c]{@{}l@{}}Supervised learning refers to algorithms that \\ determine a predictive model using a \\ set of labeled data.\end{tabular} & \begin{tabular}[c]{@{}l@{}}Support vector machine (SVM), \\ K-nearest neighbors (KNN), \\ Multilayer Perceptron (MLP),\\ Bayesian Learning (BL),\\ Deep Neural Network (DNN),\\ Long Short Term Memory (LSTM), \\ Linear regression (LR), \\ Support Vector Regression (SVR), \\ Gaussian Process (GP), \\ Regression, \\ Self-supervised learning\end{tabular} & \begin{tabular}[c]{@{}l@{}}~\cite{tang2019future},~\cite{piran2019learning},~\cite{xiao2020towards},~\cite{kato2020ten}, \\~\cite{jameel2020machine},~\cite{sun2020machine},~\cite{guo2020explainable},~\cite{du2020machine}, \\~\cite{han2020artificial}, ~\cite{elsayed2019ai},~\cite{liu2019machine},~\cite{liu2020data}  \end{tabular} \\ \hline
Unsupervised Learning & \begin{tabular}[c]{@{}l@{}}Unsupervised learning refers to algorithms that\\ can automatically find structure in the data with no \\ labels and with minimal human supervision.\end{tabular} & \begin{tabular}[c]{@{}l@{}}K-means, EM,\\  Clustering,\\ PCA, \\ Generative Adversarial  Network  (GAN)\end{tabular} & \begin{tabular}[c]{@{}l@{}}~\cite{piran2019learning},~\cite{kato2020ten},~\cite{tang2019future},~\cite{jameel2020machine}, \\~\cite{yang2020artificial}, ~\cite{shafin2020artificial},~\cite{huang2020general}, \\~\cite{giordani2020toward},~\cite{sattiraju2019ai}   \end{tabular}  \\ \hline
Reinforcement Learning & \begin{tabular}[c]{@{}l@{}}Reinforcement learning refers to algorithms that\\ seek to intelligent agents to learn \\ correct decisions by trial and error and  \\ pursue a long term reward.\end{tabular}  & \begin{tabular}[c]{@{}l@{}} Q-learning,\\Deep Q-learning, \\ Advantage Actor Critic (A2C), \\
Asynchronous Actor-Critic Agents (A3C), \\ Dueling DQN, \\ Double DQN, \\Proximal Policy Optimization (PPO) \end{tabular} & \begin{tabular}[c]{@{}l@{}}~\cite{dai2019blockchain},~\cite{tang2020deep},~\cite{xiong2019deep},~\cite{zhao2020deep}, \\~\cite{al2020multiple}, ~\cite{he2020beamspace},~\cite{huang2020reconfigurable},~\cite{zhang2020uav}, \\~\cite{rout20206g},~\cite{du2020mec}, ~\cite{wang2020multi}   \end{tabular}  \\ \hline
Federated Learning & \begin{tabular}[c]{@{}l@{}}Federated learning is a privacy-preserving machine \\ learning technique, which allows users to hold \\ data locally and send trained models to the \\ central server for obtaining the global model. \end{tabular} & \begin{tabular}[c]{@{}l@{}} Decentralized Learning \end{tabular}  & \begin{tabular}[c]{@{}l@{}}~\cite{xiao2020towards},~\cite{lu2020low},~\cite{aledhari2020federated},~\cite{yang2020federated}, \\~\cite{khan2020federated},~\cite{chen2020joint},~\cite{khan20206g},~\cite{qu2020empowering}, \\~\cite{fadlullah2020hcp},~\cite{zhao2020federated},~\cite{yang2020energy}\end{tabular}  \\ \bottomrule
\end{tabular}
\label{ml_techniques}
\end{table*}

\textbf{Supervised Learning.} The supervised learning trains the machine learning model using labelled training data~\cite{piran2019learning}. Thus, supervised learning is particularly suitable for tasks which can be trained using the historical data. There are some well developed algorithms that are used in 6G for classification or regression, for example, support vector machines, linear regression, logistic regression, linear discriminant analysis, naive Bayes, k-nearest neighbors, and decision tree, etc.  Supervised learning techniques can be used in both physical layer and network layer in 6G. In physical layer, we can utilize supervised learning for channel states estimation where transmitted signals known by the receiver can be used for training~\cite{bai2019deep, qin2019deep}. Besides, supervised learning techniques can be deployed for  data-driven tasks such as channel decoding~\cite{wang2017deep}, caching~\cite{doan2018content}, and delay mitigation~\cite{piran2019learning} and so on in the network layer. 

\begin{figure}[!h]
    \centering
    \includegraphics[scale=0.8]{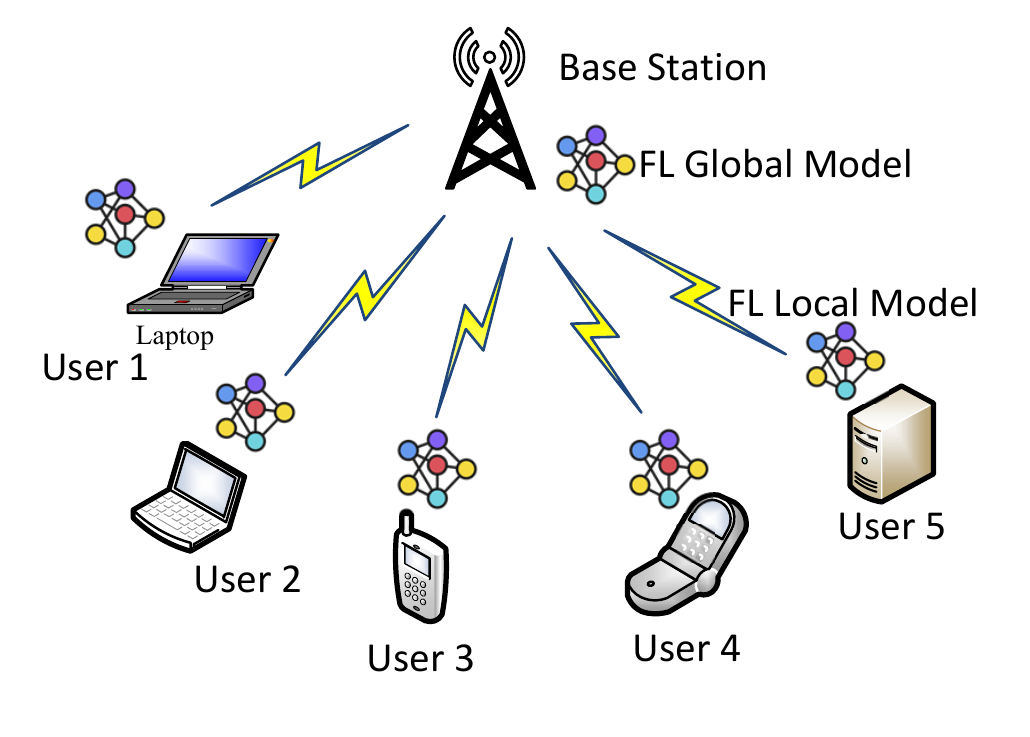}
    \caption{Federated Learning.}
    \label{fig:FL}
\end{figure}

\textbf{Unsupervised Learning.}
The unsupervised learning is leveraged to find undefined patterns in the dataset without labels. Commonly used unsupervised learning techniques include clustering,  autoencoders, deep belief nets, generative adversarial networks, and the expectation–maximization algorithm. In the physical layer, unsupervised learning techniques are applicable to optimal modulation, channel-aware feature-extraction,  etc. In addition, unsupervised learning technologies can be used for anomaly detection~\cite{rajendran2019crowdsourced},  localization~\cite{yang2020machine}, traffic control~\cite{huang2018unsupervised}, and parameter optimization~\cite{nikbakht2020unsupervised}, etc. in network layer.

\textbf{Deep Reinforcement Learning.}
Deep reinforcement learning (DRL) leverages Markov decision models to select the next ``action'' based on the state transition models~\cite{mao2018deep}. DRL technique is considered as one of the promising solutions to maximize the long term cumulative reward by sequential decision-making~\cite{piran2019learning}. The agent observes the state of the environment, chooses an action from available actions, obtains rewards, and then the state changes to the next state during the learning process~\cite{ali20206g}. In DRL, a neural network is used to estimate states instead of mapping every solution. DRL is used to solve resource allocation problems in 6G~\cite{elsayed2019ai, zhangLin20196g}. As 6G wireless networks serve a wider variety of users in the future, the radio-resource will become extremely scarce. Hence, efficient solutions for radio-resource allocation are urgent and challenging~\cite{elsayed2019ai}.

\textbf{Model-Driven Deep Learning.} The model-driven approach is to train an artificial neural network (ANN) with prior information based on domain knowledge~\cite{zappone2019model, gacanin2019autonomous, he2019model}. The model-driven approach is more suitable for most communication devices than the pure data-driven deep learning approach, because it does not require  tremendous computing resources and considerable time to train what the data-driven method needs~\cite{he2019model}. The approach to apply model-driven deep learning proposed by Zappone~\emph{et~al.}~\cite{he2019model} includes two steps:  first, we can use theoretical models derived from wireless communication problems as prior expert information; second, we can subsequently tune ANN with small sets of live data even though initial theoretical models are inaccurate.

\textbf{Federated Learning.}
Federated Learning (FL) aims to  train a machine  learning model with training data remaining distributed at clients in order to protect data owners' privacy~\cite{konevcny2016federated}. The idea of FL is illustrated in Fig.~(\ref{fig:FL}), where users' devices train local models and then send trained local models to the base station for aggregation. Since users' data are still maintained in the devices, the privacy of their data can be well preserved. As 6G heads towards a distributed architecture, FL technologies can contribute to enabling the shift of AI moving from a centralized cloud-based model to the decentralized devices based~\cite{letaief2019roadmap, shafin2019artificial,tariq2019speculative, yang2021federated}. In addition, since the edge computing and edge devices are gaining popularity, AI computing tasks can be distributed from a central node to multiple decentralized edge nodes. Thus, FL is one of the essential machine learning methods to enable the deployment of accurately generalized models across multiple devices~\cite{cousik2019cogrf}.

\textbf{Explainable Artificial Intelligence.}
There will be a large scale of applications such as autonomous driving and remote surgery in 6G era. Since these applications are closely related to humans' life, a mistake may incur miserable disasters. Therefore, it is very necessary to make AI explainable for building trust between humans and machines.  Currently, most AI approaches in PYH and MAC layers of 5G wireless networks are inexplicable~\cite{guo2019explainable}. Some AI applications like autonomous driving and remote surgery, are considered to be widely used in 6G, which requires explainability to enable trust. AI decisions should be explainable and understood by human experts to be considered as reliable. Existing methods, including visualization with case studies, hypothesis testing, and didactic statements, can improve the explainability of deep learning.

\section{Intelligent Surfaces}~\label{sec:irs}

\begin{figure}[!t]
     \centering
     \begin{subfigure}[b]{0.4\textwidth}
         \centering
         \includegraphics[width=\textwidth]{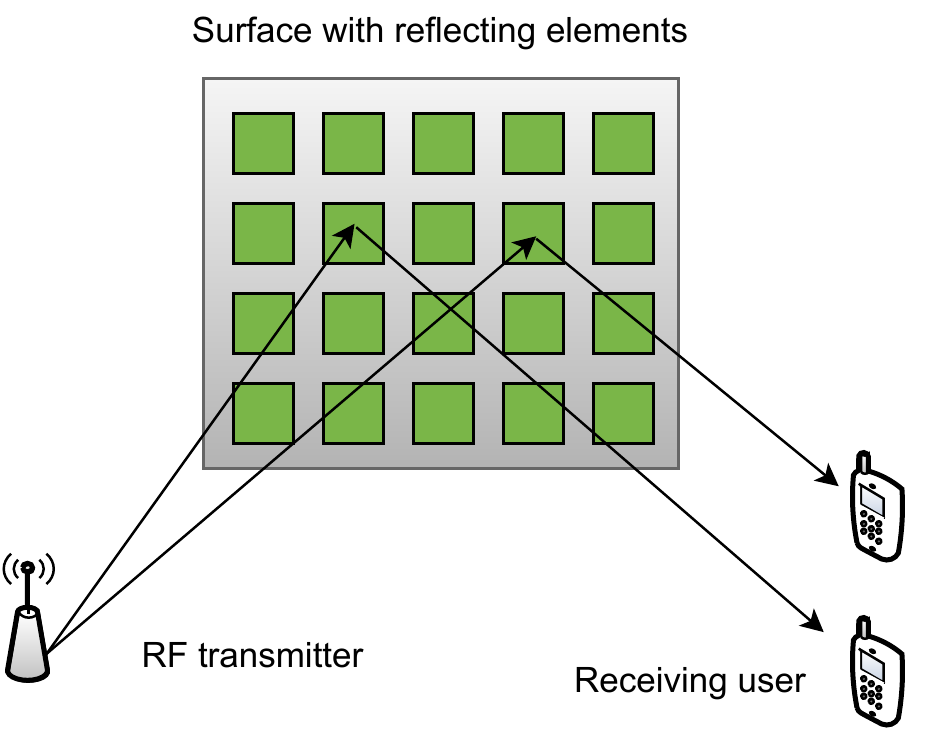}
         \caption{Intelligent Reflecting Surface.}
         \label{fig:irs}
     \end{subfigure}
     \hfill
     \hfill
     \begin{subfigure}[b]{0.35\textwidth}
         \centering
         \includegraphics[width=\textwidth]{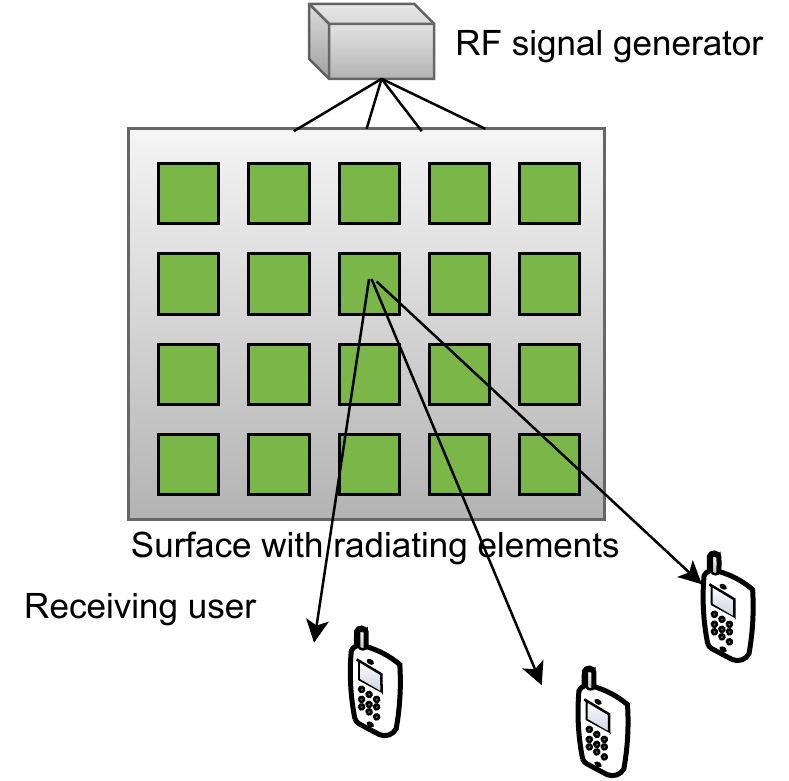}
         \caption{Large Intelligent Surface.}
         \label{fig:lis}
     \end{subfigure}
        \caption{Compare IRS with LIS.}
        \label{fig:intelligent_surfaces}
\end{figure}

Intelligent surfaces are new and revolutionary technologies for significantly improving the performance of the wireless communication networks.  Currently, two types of intelligent surfaces  attract researchers' attention, including the large intelligent surface (LIS) and intelligent reflecting surface (IRS). As shown in Fig.~(\ref{fig:intelligent_surfaces}), the LIS is useful for constructing  an intelligent and active environment with integrated electronics and an external signal generator~\cite{faisal2019ultra, hu2018beyond}; the IRS utilizes many low-cost passive reflecting elements to reflect beamforming. In the following, we present an overview of the LIS, IRS, and other similar technologies.



\textbf{LIS.} The concept of deploying antenna arrays as the LIS in massive MIMO systems is originally proposed by Hu~\emph{et~al.}~\cite{hu2017potential}. The LIS is electromagnetically active in the physical environment, where each part of an LIS can send and receive electromagnetic fields. Buildings, streets, and walls are expected to be electronically active after decorating with LISs~\cite{faisal2019ultra}. As radio frequency circuits and signal processing units are embedded in the surface, the entire surface of the LIS can be used to transmit and receive communication signals. The LIS has the following main favorable features~\cite{faisal2019ultra}:
(i) Generate perfect LoS indoor and outdoor propagation environments.
(ii) They put little restriction on the  spread of antenna elements. Hence, mutual coupling effects and antenna correlations can be easily avoided, such that sub-arrays are large and the channel is well-conditioned for propagation. Thus, the LIS can be realized via THz Ultra-Massive MIMO (UM-MIMO). The LIS is very useful for applications with low-latency such as wireless virtual/augmented reality and vehicular communications.

\textbf{IRS.} The IRS is considered as a promising candidate to improve the signal quality at the receiver by modifying the phase of incident waves~\cite{hu2018beyond, nadeem2019large, jung2019performance, de2019non, hu2017potential, hu2018user, hu2017cramer,gong2020toward}. IRSs are made of electromagnetic (EM) material that are electronically controlled with integrated low-cost passive reflecting elements, so  that they can contribute to forming the smart radio environment~\cite{basar2019wireless}. The highly probabilistic wireless channel is tuned into a deterministic space by using the software-controlled propagation of the EM waves in the smart radio environment realized by the IRS. The IRS helps to enhance the communication between a source and a destination by reflecting the incident wave~\cite{basar2019wireless, ozdogan2019intelligent, liang2019large}. By adjusting the reflection coefficients, the IRS enables the reflected signals being coherently added to the receiver without adding additional noise~\cite{liang2019large}. Besides, the IRS can modify the signal phase and increase signal power~\cite{faisal2019ultra}. In particular, by utilizing local tuning, graphene-based plasmonic reconfigurable metasurfaces can obtain  some benefits, including the beam focusing, beam steering, and  control on wave vorticity~\cite{liaskos2018new}. Unlike the LIS, the IRS uses a passive array architecture for the reflecting purpose~\cite{qingqing2019towards}. In the following, we list some features of the IRS~\cite{basar2019wireless, qingqing2019towards} :
\begin{itemize}
    \item They comprise low-cost passive elements which are controlled by the software programming.
    \item They do  not  require a specific energy source to support the transmission.
    \item They do not need any backhaul connections to exchange traffics.
    \item The IRS is a configurable surface, so that points on its surface can shape the wave impinging upon it.
    \item They are fabricated with the low profile, lightweight, and conformal geometry such that they can be easily deployed.
    \item They work in the FD mode.
    \item No self-interference.
    \item The noise level does not increase.
\end{itemize}

\textbf{Other Similar Technologies.} Besides LIS and IRS, some other similar technologies have appeared in recent years. 
\begin{itemize}
\item Large intelligent metasurface (LIM): LIM uses a special metallic material called meta-atom to form its surface, which is more flexible in manipulating electromagnetic waves~\cite{He2019LIM}.
\item Smart reflect arrays: smart reflect arrays put more emphasis upon the reflection function other than the transmission, reception and waveguiding functions~\cite{Tan2016SRA,Tan2018SRA,Nie2019SRA}.
\item Software-defined surface (SDS) / software-defined meta-surface (SDM): the SDS/SDM introduces the thought of software-defined radio into smart surfaces and by controlling the smart surface units through programmable fashion~\cite{Basar2019IRS,Liaskos2019IRS}.
\end{itemize}

As the above mentioned technologies have almost similar functions with LIS or IRS, and in our survey, we put stress on the investigation of these intelligent surfaces' quality of service (QoS) enhancement, therefore, we use the name IRS in our following manuscript for general purpose. 

\begin{figure}[htbp]
    \centering
    \includegraphics[scale=0.5]{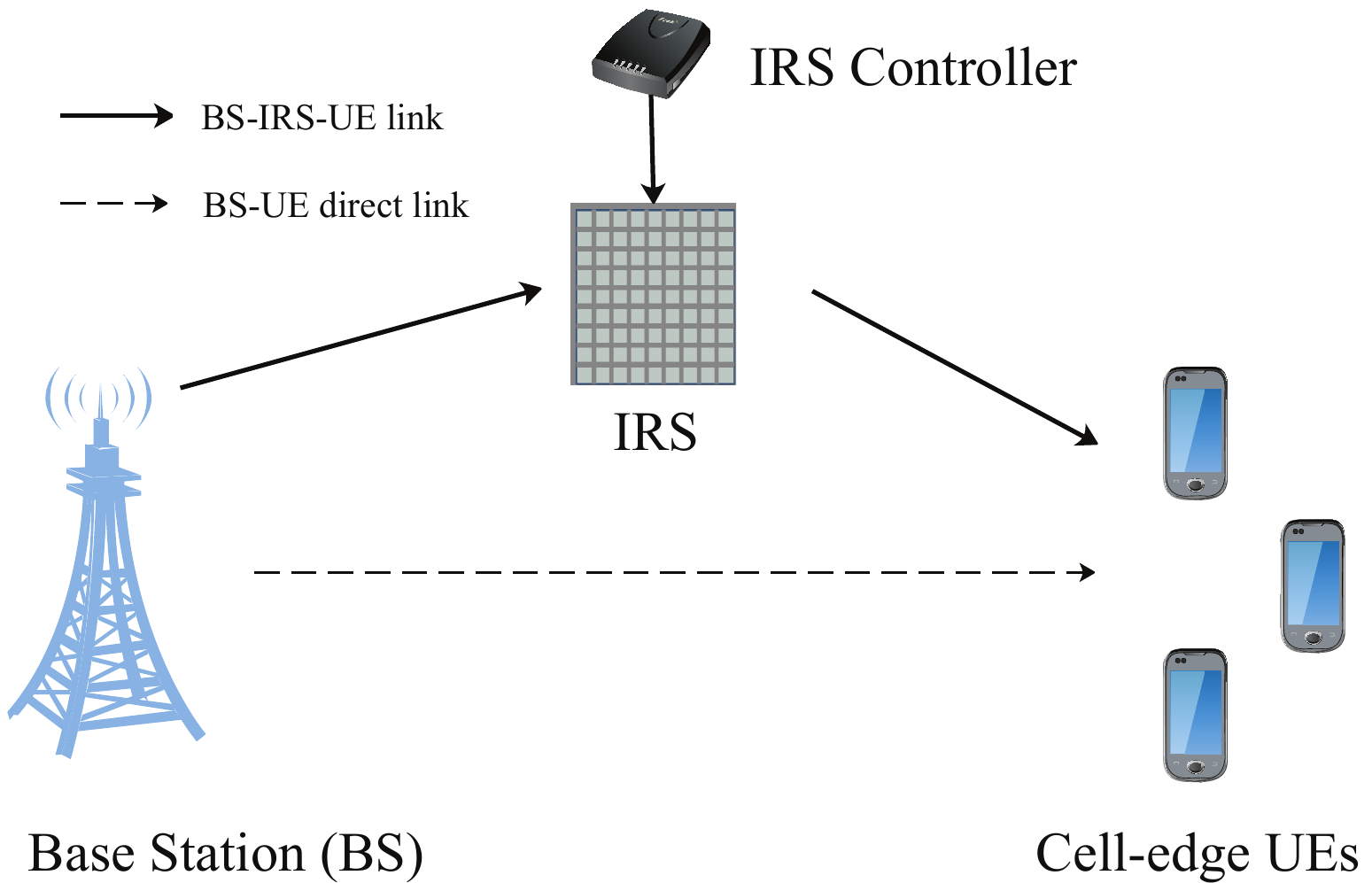}
    \caption{Structure of IRS system.}
    \label{fig:IRS-arc}
\end{figure}

The structure of IRS is depicted in Fig.~(\ref{fig:IRS-arc}). The base station (BS) equipped with $N_T$ antennas is communicating with one or multiple single-antenna cell-edge user equipment (UE). Since the channel fading of the BS-UE direct link is too large to guarantee the QoS, an IRS of $N_I$ units is utilized to facilitate the communication.

Suppose there are (is) $N_\text{UE}$ UE(s) in the IRS-aided wireless communication systems ($N_\text{UE} \ge 1$). The received signal for the $n$th UE is given by \begin{equation}\label{ISRx}
y_{\text{I},n}=(\textbf{h}_{\text{I},n}\mathbf{\Phi}\textbf{H}_\text{I,B}+\textbf{h}^T_{\text{B},n})\textbf{x}+w_n, {\kern 5pt}  n=1,2,...,N_\text{UE},
\end{equation}
where $\textbf{h}_{\text{I},n}\in\mathbb C^{1\times N_I}$ and $\textbf{h}_{\text{B},n}\in\mathbb C^{1\times N_T}$ are the channel fading vectors between the IRS and the $n$th UE and the BS and the $n$th UE, respectively, $\textbf{H}_\text{I,B}\in\mathbb C^{N_I\times N_T}$ is the channel fading matrix between the BS and the IRS, $w_n$ is the additive white Gaussian noise at the $n$th UE, $\textbf{x}\in\mathbb C^{N_T\times 1}$ is the transmitting signal vector of the BS, and $\mathbf{\Phi}\in\mathbb C^{N_I\times N_I}$ is a diagonal matrix whose diagonal elements are the corresponding reflect coefficients of the IRS, i.e.,
\begin{equation}\label{Phi}
\mathbf{\Phi}=\text{diag}(\phi_1,\phi_2,...,\phi_{N_I})
\end{equation}
with $\phi_i (i=1,...,N+I)$ is the reflect coefficient of the IRS's $i$th element. By defining $\boldsymbol{\varphi}=[\phi_1,...,\phi_{N_I}]^T$, the received signal can be rewritten as
\begin{equation}\label{ISRx1}
y_{\text{I},n}=(\boldsymbol{\varphi}\text{diag}(\textbf{h}_{\text{I},n})\textbf{H}_\text{I,B}+\textbf{h}^T_{\text{B},n})\textbf{x}+w_n, {\kern 5pt} n=1,2,...,N_\text{UE}.
\end{equation}
Compared with Eq.~(\ref{ISRx}), Eq.~(\ref{ISRx1}) is more convenient to be processed.

However, $\textbf{h}^T_{\text{B},n}$ is often omitted since the distance between the BS and UE is long and the channel fading degrades considerably fast. The reflect coefficient has two parts: phase shift $\theta_i$ and amplitude attenuation $\alpha_i$, i.e.,
\begin{equation}\label{AP}
\phi_i=\alpha_ie^{1j\theta_i}{\kern 8pt}(i=1,...,N_I)
\end{equation}
with $1j$ is the imaginary unit.

According to whether the models for the phase shift and amplitude attenuation are continuous or not, the IRS can be classified into continuous mode and discrete mode. For the phase shift, $\phi_i$ can take any value in the range of $[0, 2\pi)$ in the continuous mode, however, $\phi_i$ can only take a finite number of values in $[0, 2\pi)$ in the discrete mode. Similarly for the amplitude attenuation, when there is no attenuation for the coefficients, $\alpha_i=1$, and when the amplitude attenuation is in the continuous mode, $\alpha_i$ can take any value in $(0, 1]$ or a finite number of values between 0 and 1 in the discrete mode. However, when the phase shift or the amplitude attenuation are in the discrete mode, the optimization problem of maximizing the capacity or minimizing the transmitting power as discussed later are non-convex. The generally-used method is to solve the problem considering the phase shift and amplitude attenuation in continuous mode, and choose the discrete values that are close to the solved continuous solutions. The problem with no amplitude attenuation also makes the constraint non-convex, and in such a case the optimization can be done first by neglecting this non-convex constraint and then by  normalizing the phase shifts to fulfill the constraint of unit modulus~\cite{Yuille2003Convex,Lipp2016Convex,Boyd2004Convex}.

\begin{table*}[!htp]
\centering
\caption{Developments of IRS Technology.}
\renewcommand\arraystretch{2}
\label{table:IRS}
\begin{tabular}{l|l}
\toprule
\textbf{Developments}                             & \textbf{Key IRS Technologies}        \\                                                                                                                                                             \hhline{==}
IRS-aided multi-user systems~\cite{Tan2016SRA,Tan2018SRA,Nie2019SRA}      &\begin{tabular}[c]{@{}l@{}} - Maximization of transmission rate with power constraint\\  - Minimization of transmitting power with received SINR constraint\end{tabular}\\
\hline
IRS-aided OFDM systems~\cite{Yang2019IRS}                      & Maximization of capacity with power constraint \\
\hline
IRS systems with hardware impairment~\cite{Hu2018IRS}          & Capacity performance is investigated   \\
\hline
IRS-aided SM systems~\cite{Basar2019IRS}                       & BER performance of IRS-SM/IRS-SSK is investigated\\
\hline
Channel estimation                       & \begin{tabular}[c]{@{}l@{}} - MMSE principle~\cite{Nadeem2020IRS} \\ - Matrix factorization, ambiguity elimination, matrix completion~\cite{He2020IRS} \\ - Joint optimization of channel estimation and IRS reflection~\cite{Zheng2019IRS} \end{tabular} \\
\hline
HMIMOS systems~\cite{huang2020holographic}  & Constructive interference is utilized to promote performance \\
\hline 
Phase-dependent amplitude model for IRS schemes                & \begin{tabular}[c]{@{}l@{}} - Lorentzian resonance response $\phi_i=(1j+e^{1j\eta_i})/2$ ~\cite{Di2020IRS}\\ - Lumped circuit model $\alpha_i=(1-\alpha_{\min})[(1+\sin(\theta_i-\eta))/2]^k+\alpha_{\min}$~\cite{Pozar2012IS,Koziel2013IS}\end{tabular}\\
\bottomrule
\end{tabular}
\end{table*}


Nadeem~\emph{et al.}~\cite{Nadeem2020IRS} propose a channel estimation protocol based on the MMSE principle. The protocol first estimates the channel coefficients between BS and UEs by turning off the IRS, and then estimates the BS-IRS-UE link by turning on the IRS unit in turn. Finally, MMSE approach is utilized to obtain a comprehensive estimation of all channels. This protocol is complicated and time consuming due to the fact that each IRS should be turned on once a time while the others turned off, especially when the number of IRS elements is large. He~\emph{et al.}~\cite{He2020IRS} propose a three-state channel estimation algorithm for MIMO systems. The algorithm first obtains the channel matrices of the BS-IRS and IRS-UE links via matrix factorization. Then, by exploiting the IRS state matrix, the ambiguity of the solutions to matrix factorization is eliminated. In the final stage, the properties of channel matrices are utilized to recover the missing entries. The proposed estimation algorithm is also time consuming because the IRS units should also be turned on in turns to eliminate the ambiguity in the second stage. To address such a problem, Zheng~\emph{et al.}~\cite{Zheng2019IRS} propose a novel method to estimate the channel state information for IRS-enhanced OFDM systems. The authors first design a reflection pattern of the IRS for channel estimation by exploiting pilot signals, and then perform a joint channel estimation and reflection optimization based on the strongest signal path resolved in the first stage. This novel method avoids the requirement of large numbers of pilot signals and operation for each IRS units, therefore, reducing the complexity of the receiver for a large scale.

Besides the fundamental application as described above, there exist some other IRS-assisted communication systems or IRS technologies in the available references. For instance, Yang~\emph{et al.}~\cite{Yang2019IRS} discuss the application of the IRS in OFDM systems where multiple paths exist in the wireless communication environment. The authors establish an objective of maximizing the capacity with the maximum transmitting power limited, where the coefficients of the IRS and the power allocation for each subcarrier are unknown parameters. By exploiting the idea of alternating optimization, the optimal power allocation and the IRS coefficients can be derived. Hu~\emph{et al.}~\cite{Hu2018IRS} investigate the capacity performance with hardware impairments exist in the IRS. The authors also give a conclusion that by splitting the IRS into several subgroups with each subgroup consists of a number of the IRS' units, the performance degradation can be mitigated to a certain degree.
Basar~\cite{Basar2019IRS} investigate the bit-error rate (BER) performance of the IRS-aided SM systems. In the manuscript, optimal (exhaustive search) and suboptimal (greedy) detectors for the IRS-assisted SM or SSK schemes are formulated in detail, and their theoretical average bit error probabilities are derived as well, to give a benchmark of the IRS-IM/SSK wireless communication systems.

Furthermore, Huang~\emph{et al.}~\cite{huang2020holographic} propose the scheme where the IRS is combined together with MIMOS to form the HMIMOS systems. Because the interference is regarded as useful resources for developing holographic communication systems~\cite{zong20196g}, the multi-user interference can be decomposed into constructive and destructive parts using simple geometric relations~\cite{meng2020interference}. Constructive part is considered as beneficial communication resources, which can be used to improve QoS of 6G communication systems. To be more specific, active HMIMOS using the IRS are equipped with RF circuits and signal processing units, whereas passive HMIMOS only use the IRS for reflecting signals.


The above models all consider that the phase and amplitude of the IRS units are independent to each other. However, in some cases, the amplitudes of IRS units are relevant to their own phases. Di~\emph{et al.}~\cite{Di2020IRS} construct a reflecting coefficient model by exploiting the Lorentzian resonance response~\cite{Smith2017IRS}, where the coefficient is expressed as $\phi_i=(1j+e^{1j\eta_i})/2$ with $\eta_i \in [0,2\pi)$. In addition, Abeywickrama~\emph{et al.} consider a phase-dependent amplitude model in~\cite{Abeywickrama2020IRS,Abeywickrama2020IRS1} according to the lumped circuit model~\cite{Pozar2012IS,Koziel2013IS}. The amplitude of coefficient is represented as $\alpha_i=(1-\alpha_{\min})[(1+\sin(\theta_i-\eta))/2]^k+\alpha_{\min}$, where constant parameters $\alpha_{\min}\in[0,1]$, $\eta\ge 0$, and $k\ge 0$ are set depending on the specific circuit implementation. The recent development of the IRS technology in available references are listed in Table~\ref{table:IRS}.

\textbf{Challenges of IRS Technologies.} Although the advantages of the IRS is attractive, there still exists some challenges in the application of the IRS technology~\cite{Gong2020IRS} :
\begin{itemize}
\item Seeking for the optimal coefficients of the IRS and the beamforming vector is painstaking, it consumes a lot of time and hardware resources in solving the optimization problems.
\item Although available references refer to the methods for channel estimation, the derivation of CSI is still a complex work in engineering, especially when the number of IRS element is large.
\end{itemize}

\section{Index Modulation} \label{sec:IM}

\begin{figure*}[htbp]
    \centering
    \includegraphics[scale=0.9]{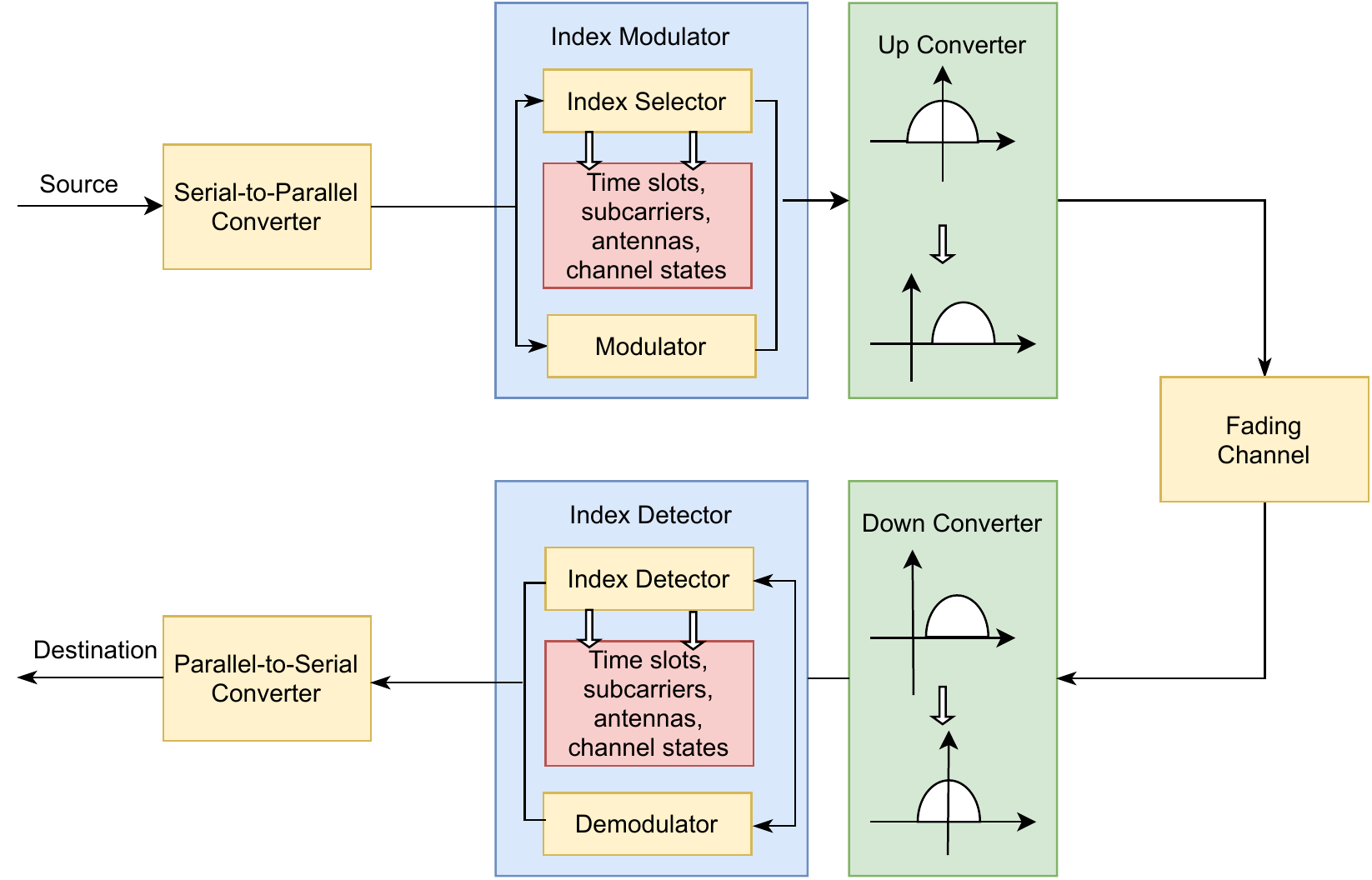}
    \caption{The structure of IM-aided systems.}
    \label{fig:IM}
\end{figure*}

Index Modulation (IM) is a kind of modulation scheme that sends extra information bits through resource entities' index. On the one hand, high capacity can be achieved due to the spare information bits' transmission. On the other hand, because the sending of these spare bits does not consume any power and spectrum resources, IM has both high spectral efficiency and high energy efficiency simultaneously compared with its non-IM-aided counterpart. The structure of the IM system is depicted in Fig.~(\ref{fig:IM}). As shown in the figure, at the transmitter, the information bits are divided into two parts after the operation of serial to parallel conversion, one part for the classical amplitude phase modulation (APM) such as phase-shift keying (PSK) or quadrature amplitude modulation (QAM), etc., and the other part for the activation of the index resources exploited for transmitting the modulated APM signals. After the up-conversion, passband modulated signals are transmitted through the activated resource entities. At the receiver, down-conversion is first performed to convert the received signals to baseband. An index demodulator is carried out to detect the information bits used for APM and index activation of resource entities. Finally, a parallel to serial conversion is conducted to recover the original information sequence. According to the categories of the resource entities used for index selection, IM can be classified into spatial-domain IM (SD-IM), frequency-domain IM (FD-IM), time-domain IM (TD-IM), and channel-domain IM (CD-IM). As stated in Table \ref{table}, IM is one of the main multiplexing architectures of 6G network~\cite{dang2020should}, and a lot of IM based technologies have been studied to satisfy requirements of 6G network.

By exploiting SM technology in VLC systems, the overall spectral efficiency can be improved to satisfy the requirements of high reliability, high power efficiency and high spectral efficiency for 6G network. Nahhal~\emph{et al.}~\cite{Nahhal2020IM} introduce a flexible generalized spatial modulation (FGSM) scheme for VLC systems. The proposed scheme changes the modulation sizes and the number of activated transmit LEDs adaptively  to improve the average SER under a predefined spectral efficiency threshold of VLC systems. Gao~\emph{et al.}~\cite{Gao2020IM} propose a light emitting dionode (LED) grouping SM scheme, where the channel correlation effect can be decreased and thus improve the SER performance significantly. Kumar~\emph{et al.}~\cite{Kumar2018IM}  optimize the collaborative constellation of GSM to enhance the power efficiency for VLC systems. And in~\cite{Wang2018IM}, by using different modulation sizes according to which LED is activated, adaptive SM is investigated to improve the spectral efficiency.

In addition, Huang~\emph{et al.}~\cite{Huang2020IM} combine IM with frequency diverse array (FDA) and propose a IM-FDA scheme. The proposed scheme can transmit information and sense target locations simultaneously. Such a feature makes IM-FDA an attractive technique for the future sensor network which requires the knowledge of each sensor’s location. And Nusenu~\emph{et al.}~\cite{Nusenu2020IM}  propose a space-frequency increment IM by utilizing the indices of the activated transmit FDA antennas and the corresponding frequency increments. A low-complexity decoding algorithm is investigated and performance is evaluated, in order to provide theoretical analysis for future 5G beyond network.

Purwita~\emph{et al.}~\cite{Purwita2020IM} introduce a signle carrier generalized time slot IM for single-carrier frequency domain equalization (SC-FDE) systems. By changing the number of activated time slots, the performance limitations due to the multipath optical wireless channel can be mitigated, leading to more than 3dB gain than its non-IM-aided SC-FDE technique. The proposed scheme is more attractable for future indoor optical wireless communications with limited dynamic range of LEDs.

The ability of serving huge numbers of UEs is also a key requirement for future network. To improve such an ability, Nguyen~\emph{et al.}~\cite{Nguyen2019IM} introduce a combined hybrid OFDM-based modulation scheme for narrowband IoT. By properly designing the constellation used in IM aided fast-OFDM systems, trade-offs between spectral efficiency and energy efficiency is obtained , thus being able to serve huge numbers of devices efficiently in 5G beyond networks. Althunibat~\emph{et al.}~\cite{Althunibat2019IM}  propose a quadrature index modulation multiple access (QIMMA), which allows each user to activate two of the orthogonal resources when communicating with the base station. Such an access scheme can serve larger numbers of users compared with its non quadrature counterpart, which is beneficial for future network where the number of users grows exponentially.

Security is another important requirement of 6G network. Yang~\emph{et al.}\cite{Yang2020IM} encrypt the SM signals through random mapping patterns to avoid being eavesdropped for SM assisted multi-hop wiretap ad-hoc networks. By activating the SM mapping patterns through random CQI patterns over the legitimate link, the security performance of the conceived scheme is enhanced a lot, and thus satisfying the security requirement of 6G network.

The error probability performance of IM relevant schemes is also studied by a lot of references to meet the requirement of high reliability. Shi~\emph{et al.}~\cite{Shi2019IM} investigate an OFDM scheme with all index modulation (OFDM-AIM). By replacing the PAM constellation modulator with subblock modulator of an OFDM symbol, the diversity order can be enhanced. Therefore, the BER performance is improved significantly. They also consider the legitimate subblock realizations in OFDM-AIM as the chromosomes of genetic algorithm~\cite{Shi2019IM1}. By utilizing the ABEP as the population fitness, subblock realizations with good BER performance can be obtained. Yu~\emph{et al.}~\cite{Yu2020IM} investigate a phase rotation based precoding SM scheme. By optimizing the phase rotation-based precoding matrix, the minimum Euclidean distance of the SM signal constellation is improved, and thus enhancing the BER performance to a large scale. In addition, Zhang~\emph{et al.}~\cite{Zhang2020IM} introduce a soft-input soft-output detector based on expectation propagation framework for SM schemes. By calculating the LLRs of the transmitted bits iteratively, the proposed detector can conduct trade-offs between BER performance and detector complexity. However, all the schemes metioned above need the knowledge of CSI when detecting the transmitted information bits, In~\cite{Katla2020IM}, a multi-set STSK scheme transmitted in mmWave channel is proposed, and a neural network is trained to jointly detect the transmitted source information. For the proposed detection algorithm, the channel estimation can be eliminated and outperforms the classical ML algorithm in the scenario with channel impairments. And Satyanarayana~\emph{et al.}~\cite{Satyanarayana2020IM} also propose a deep learning based soft detector. For the proposed scheme, a neural network is trained to obtain the soft values for transmitted signals without CSI information, which is promising for 6G wireless network where perfect CSI is difficult to evaluate. The summaries of IM-based technologies for future network are listed in Table~\ref{table:IM}.

\textbf{Challenges of IM Techniques.}
Although IM technology has the advantage of high data rate and low energy consumption, it still meets some challenges as follows~\cite{Mao2019IM,Wen2019IM} :
\begin{itemize}
\item For IM system, the non-activated resource entities are kept empty in transmission, and therefore the utilization efficiency is unsatisfactory.
\item The complexity of detection algorithm is very high.
\item Since the APM symbols are conveyed only through the selected resources, once the detection of the index is incorrect, the demodulation is almost destructive in the decoupled detection algorithm.
\item In MBM system, the number of mirrors is large to achieve high data rate, therefore, huge training sequence should be required for acquiring the channel states, which consumes a lot of resource in transmission.
\item Since hybrid IM technologies outperform its signal domain counterpart in error performance and transmission rate, current references are not enough in investigation into hybrid IM technologies.
\end{itemize}

\begin{table*}[!htp]
\centering
\caption{Summary of IM-based Technologies for future network.}
\label{table:IM}
\renewcommand\arraystretch{1.5}
\begin{tabular}{l|l}
\toprule
\textbf{IM-based technologies}               & \textbf{Features}  \\  \hhline{==}
SM in VLC systems   & \begin{tabular}[c]{@{}l@{}}- Change the modulation sizes and the number of activated transmit LEDs adaptively~\cite{Nahhal2020IM}\\ - LED grouping SM~\cite{Gao2020IM}\\ - Optimize the collaborative constellation of GSM~\cite{Kumar2018IM}\\ - Use different modulation sizes according to which LED is activated~\cite{Wang2018IM} \end{tabular} \\
\hline
IM combined FDA systems & \begin{tabular}[c]{@{}l@{}}- Transmit information and sense target locations simultaneously~\cite{Huang2020IM}\\ - Utilize the indices of the activated transmit FDA antennas and the corresponding frequency increments~\cite{Nusenu2020IM} \end{tabular} \\
\hline
IM assisted SC-FDE systems & Change the number of activated time slots in SC-FDE systems~\cite{Purwita2020IM} \\
\hline
Ability of serving huge numbers of UE & \begin{tabular}[c]{@{}l@{}} - Design a constellation properly used in IM aided fast-OFDM systems~\cite{Nguyen2019IM}\\ - Allow each user to activate two of the orthogonal resources~\cite{Althunibat2019IM} \end{tabular} \\
\hline
Security requirement & Encrypt the SM signals through random mapping patterns~\cite{Yang2020IM}\\
\hline
Error probability performance & \begin{tabular}[c]{@{}l@{}} - Replace the PAM constellation modulator with subblock modulator~\cite{Shi2019IM}\\ - Consider the legitimate subblock realizations in OFDM-AIM as the chromosomes of genetic algorithm~\cite{Shi2019IM1}\\ - Optimize the phase rotation-based precoding matrix~\cite{Yu2020IM}\\ - Soft-input soft-output detector based on expectation propagation framework for SM schemes~\cite{Zhang2020IM} \end{tabular} \\
\hline
Detector without CSI & \begin{tabular}[c]{@{}l@{}} - Jointly detect the transmitted source information exploiting neural network~\cite{Katla2020IM}\\ - Obtain the soft values for transmitted signals exploiting neural network~\cite{Satyanarayana2020IM} \end{tabular} \\
\bottomrule
\end{tabular}
\end{table*}

\section{Simultaneous Wireless Information and Power Transfer (SWIPT)}~\label{sec:swipt}

In the future 6G network, various mobile wireless devices will be accessed to the network. The high transmission rate leads to an increase in power consumption, and thus, requires the long lifetime of battery-powered devices. SWIPT is such a novel technology promisingly used in the 6G network where energy harvesting through RF signals can be made to fulfill the requirement of high power efficiency~\cite{SWIPT2018Gu,SWIPT2017Pan,Bahbaei2018SWIPT,Perera2018SWIPT}. The idea of energy harvesting can date back to 1969, when Brown \emph{et al.}~\cite{Brown1969EH} conduct a series of experiments and demonstrate that a helicopter at a height of 50 feet operating on 2.45GHz can provide a direct current power supply of about 270W. The energy is harvested either from the ambient wireless signals or from a specific controlled power source. In 2008, Varshney \emph{et al.}~\cite{Varshney2008SWIPT} first propose the concept of SWIPT, which deliver information and energy concurrently from a theoretical perspective. The performance is evaluated by assessing the reception reliability and information transmission rate.

\begin{figure}[htbp]
    \centering
    \includegraphics[scale=1]{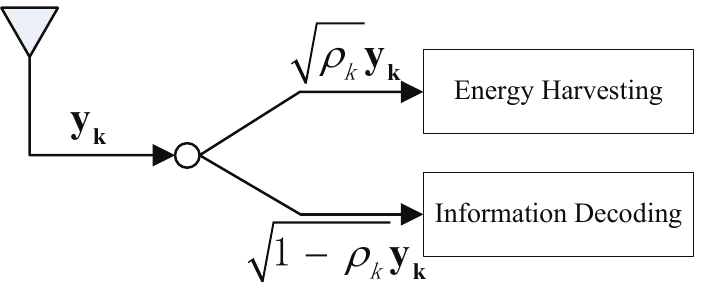}
    \caption{Architecture of SWIPT system.}
    \label{fig:SWIPT}
\end{figure}

Fig.~(\ref{fig:SWIPT}) depicts the architecture of SWIPT schemes. The generally-used structures for resource allocation are time switching (TS) structure, power splitting (PS) structure, and antenna switching (AS) structure. In TS structure, a frame consisting of several time slots are classified into information transfer slots and power transfer slots, and a switcher is used at the receiver to decide which state is at work~\cite{Tang2020SWIPT,Buckley2020SWIPT}. The receiver splits the received signal into two streams with power ratio $\rho(t) : 1-\rho(t)$, where the two parts are used for harvesting energy and decoding the information, respectively. The architecture of SWIPT system is depicted in Fig.~(\ref{fig:SWIPT}). Assume the received block consists of $N_f$ time slots, therefore the splitting vector can be defined as $\bm{\rho}=[\rho_1,...,\rho_{N_f}]^T$. For the TS structure, in the first $\alpha N_f$ time slots, the received signal power are used for energy harvesting overall, and vice versa, all the signal power are exploited for information decoding for the remaining $(1-\alpha)N_f$ time slots. Therefore, the element of the splitting vector for TS structure can be expressed as

\begin{equation}\label{splittingTS}
{\rho _k} = \left\{ \begin{array}{l}
1,k = 1,...,\alpha {N_f}\\
0,k = \alpha {N_f} + 1,...,{N_f}.
\end{array} \right.
\end{equation}

In PS structure, a power splitting factor is utilized to decide how much signal flows to the decoder circuit for signal processing or the battery circuit for energy harvesting. And subsequently, the two power streams can function separately but simultaneously through the overall frame duration~\cite{Wang2019SWIPT}. At the receiver, the ratio of the power used for harvesting energy and decoding information can be set as a constant $\rho$, leading to the element of the splitting vector be $\rho_k=\rho, (k=1,...,N_f)$. In addition, the power splitting factor can be designed to balance according to different system requirements.

In AS architecture, the RXs are divided into two subsets for energy harvesting and information decoding separately~\cite{Krikidis2014SWIPT}. Compared to TS and PS structures, AS is easier to implement and thus appealing for practical SWIPT deployment, however, the performance will degrade significantly in case of hardware impairments. In one frame duration, all splitting factors remain unchanged for a given received antenna, $\rho_k=\rho, (k=1,...,N_f)$ for energy harvesting RXs and $\rho_k=1-\rho, (k=1,...,N_f)$ for information decoding RXs. The overall performance can be increased by optimizing the combination of the signal power and the AS ratio.

As the goal of SWIPT scheme is to decide the splitting factor to balance the tradeoff between transmission rate and energy harvesting, the architecture of SWIPT is depicted in Fig.~(\ref{fig:SWIPT}). In most application scenarios, the research works focus on different resource allocation schemes considering data rate, harvested energy power, transmit power, and received SNR. By performing a joint optimization of resource allocation with power control and resource scheduling, both the capacity of the system and the harvested energy can be promoted considerably. A lot of references have discuss the energy efficiency of SWIPT-based technologies. Ojo \emph{et al.}~\cite{Ojo2019SWIPT} investigate the energy efficiency performance of a SWIPT-enabled cooperative relay network with interfering exists in the systems. Two different scenarios where the relay can harvest only from the source and that harvest energy from both the source and the interference are considered. A suboptimal PS factor is derived and a new energy-saving power splitting relaying protocol is investigated, which provides PS based relaying protocol for the future relay networks. Fang \emph{et al.}~\cite{Fang2019SWIPT} consider the SWIPT scheme with linear precoder MIMO system. By optimizing the precoder to maximize the harvested energy with the power constraint, high power efficiency is achieved, which is feasible for both separated and co-located receivers. Park \emph{et al.}~\cite{Park2020SWIPT} study the effect of high power amplifiers’ (HAPs’) nonlinearity on the multi-tone SWIPT systems.  Authors conclude that HAP has adverse effects on both energy harvesting and information decoding. To address this problem, Jang \emph{et al.}~\cite{Jang2020SWIPT} give a frequency-splitting SWIPT, which separates the power and transmitted signals in frequency domain. The proposed scheme can reduce the harmonic distortion of the high power amplifier, thus decreasing the effects amplifiers’ non-linearity. Frequency-splitting SWIPT outperforms its PS counterpart in terms of the harvested power, and is more beneficial to the 6G network where a lot of low-power wearable devices are accessed to the network. Zargari~\emph{et al.}~\cite{Zargari2020SWIPT} study an IRS-assisted SWIPT technology for MISO systems. By introducing an energy efficiency indicator to trade off between transmission rate and harvested energy, a joint optimization is performed to obtain the beamforming vector, the PS ratio of each user, and the phase shifts of IRS. The optimization problem is divided into two subproblems and a suboptimal solution is achieved for the system. The proposed scheme has the advantages of both low power consumption and high BER performance, which is attractive for the 6G network. 

Sun~\emph{et al.}~\cite{Sun2020SWIPT} also consider an IRS-assisted SWIPT technology, where the security issue is emphasized since the broadcast nature makes the system be vulnerable to eavesdropping. The secrecy rate is optimized by jointly designing the beamforming vector at the BS and reflective beamforming at the IRS, subject to the transmit power constraint and energy harvest constraints. The proposed scheme is useful for future network because both low power consumption and high security are guaranteed, which can satisfy the requirements of 6G network. Additionally, Ma~\emph{et al.}~\cite{Ma2020SWIPT} investigate the security issues of SWIPT networks. A SWIPT system enabled by full-duplex jamming is provided to improve the security, where the full-duplex energy-limited receiver is used to generate jamming signals. By optimizing the PS factor, the security throughput is maximized. Zhu~\emph{et al.}~\cite{Zhu2020SWIPT} employ artificial noise generation to improve the security of transmission. The minimum harvested energy among users is maximized with the constraints of secrecy rate and transmit power. Thakur~\emph{et al.}~\cite{Thakur2020SWIPT} investigate the security performance of SWIPT scheme for cognitive radio network. The intercept and secrecy outage probability is derived with imperfect CSI and the influence of power splitting factor. The secrecy performance is also studied in detail. All the SWIPT based scheme mentioned above consider security as one of the main problems in system design, and is beneficial for 6G network where high security is required to protect the individual data of each UE. The summarys of SWIPT relavant schemes for future network are listd in Table~\ref{table:SWIPT}.

\begin{table*}[!h]
\centering
\small \setlength{\tabcolsep}{1.1pt} \caption{Summary of SWIPT Technologies for future network.}
\label{table:SWIPT}
\renewcommand\arraystretch{2}
\begin{tabular}{l|l}
\toprule
\textbf{Developments}               & \textbf{Key SWIPT Technologies}            \\                   \hhline{==}
SWIPT   & \begin{tabular}[c]{@{}l@{}} - Energy harvesting and information perform decoding at different time slots~\cite{Tang2020SWIPT,Buckley2020SWIPT} \\ - A power splitting factor is used for energy harvesting and information decoding~\cite{Wang2019SWIPT}\\ - RXs are divided into two subsets for energy harvesting and information decoding~\cite{Krikidis2014SWIPT} \end{tabular} \\
\hline
SWIPT with efficiency & \begin{tabular}[c]{@{}l@{}} - Energy-saving power splitting relaying protocol is investigated~\cite{Ojo2019SWIPT} \\ - Optimize the precoder to maximize the harvested energy~\cite{Fang2019SWIPT} \\ - Separate the power and transmitted signals in frequency domain~\cite{Jang2020SWIPT} \\ - An energy efficiency indicator is used to trade off between transmission rate and harvested energy~\cite{Zargari2020SWIPT} \end{tabular} \\
\hline
SWIPT with security        & \begin{tabular}[c]{@{}l@{}}  - The secrecy rate is optimized by jointly designing the beamforming vector and the reflective beamforming~\cite{Sun2020SWIPT}\\ - A SWIPT system enabled by full-duplex jamming is provided to improve the security~\cite{Ma2020SWIPT}\\ - Employ artificial noise generation to improve the security of transmission~\cite{Zhu2020SWIPT}\\ - The intercept and secrecy outage probability is derived with imperfect CSI~\cite{Thakur2020SWIPT}  \end{tabular}\\
\hline
\end{tabular}
\end{table*}

\textbf{Challenges of SWIPT.}
The SWIPT-based systems metioned above exemplify the general usage of SWIPT technology in wireless communication networks. It can degrade the power consumption and takes advantage at scenarios where it is difficult to charge the battery such as volcano and marine areas. However, there still exist some challenges of SWIPT technology~\cite{Perera2018SWIPT,Hossain2019SWIPT}:
\begin{itemize}
\item Existing studies mainly focus on low-power relays and sensors due to the fact that the level of the harvested energy is too low to satisfy the requirements of large wireless devices.
\item In 6G, the dense deployment of wireless devices leads to complex electromagnetic interference, which is desirable to power transfer while adverse to information decoding.
\item Intense RF radiation has adverse impact on human health. However, to achieve high QoS of SWIPT-based system, denser sensors or relays should be deployed for energy harvesting, which causes more intense RF exposure.
\end{itemize}


\section{Space-Air-Ground-Sea Integrated Network}~\label{sec:sagsin}

\begin{figure}[!h]
    \centering
    \includegraphics[scale=0.6]{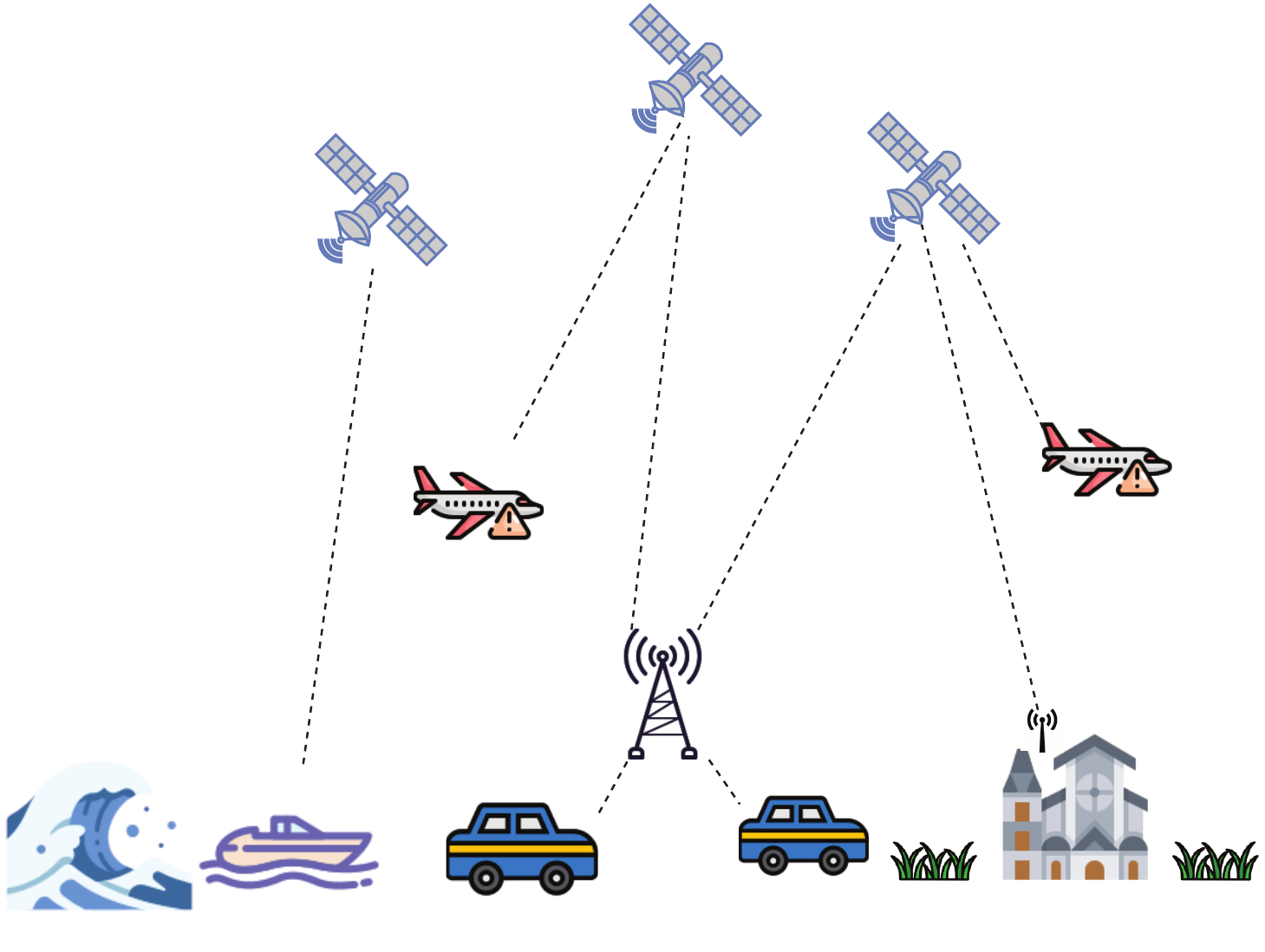}
    \caption{Space-Air-Ground-Sea Integrated Network.}
    \label{fig:satellite}
\end{figure}

In the 5G network, users only have access to a single wireless communication system, for example, the terrestrial wireless communication system, which is experiencing explosive growth in both the number of users and services available. To overcome 5G's bottleneck, 6G integrates satellite communication network, UAV communication network, terrestrial communication network, and maritime communication network to build a space–air–ground-sea integrated network (SAGSIN) to support the global coverage and ubiquitous connection as shown in Fig.~(\ref{fig:satellite}). These networks may work independently or collaboratively. 

In the previous  generations of communication systems, satellite-air-ground integrated network (SAGIN) has been hotly discussed by the academic. Specifically, SAGIN is essential to support applications like IoT, big data, and cloud computing~\cite{Kawamoto2013SAGIN}. Radhakrishnan~\emph{et al.}~\cite{Radhakrishnan2016SAGIN} summarize the inter-satellite communication from viewpoints of the physical and network layer. Besides, Niephaus~\emph{et al.}~\cite{Niephaus2016SAGIN} analyze the QoS provisioning in converged satellite and terrestrial networks. They emphasize the technical challenges related to the convergence of satellite and terrestrial networks in detail, where the satellite networks act as a complement to the existing terrestrial infrastructures. Hamdi~\emph{et al.}~\cite{Hamdi2008SAGIN} study a satellite-based hybrid sensor network in the perspective of detection and tracking in a mobile scenario. However, they only consider two networks, either space-ground or air-ground integrated networks.
Different from the space-air-ground integrated network in 5G, SAGSIN in 6G also provides coverage for the underwater and deep-sea communications.  In the following, we introduce each component of the SAGSIN in detail.

\subsection{Satellite Communication Network}

5G communication network focuses on the terrestrial coverage, which leaves some areas uncovered. Satellite communication in 6G is envisioned to integrate with terrestrial communication to provide full coverage and high throughput for the regions where terrestrial wireless communication cannot reach, such as rural areas. Satellite communication can be categorized into Geosynchronous Earth Orbit (GEO) and non-GEO based on the satellites' altitude. Chini~\emph{et al.}~\cite{chini2009} conclude that a GEO satellite is at an altitude of about 35,800 km. GEO satellite communication uses high frequencies, which aggravates the path loss. Non-GEO satellites have Low Earth Orbit (LEO) and Medium Earth Orbit (MEO). LEO is at an altitude of 500 to 2000 km, and MEO is at an altitude of 8,000 to 12,000 km. Non-GEO satellites have lower end-to-end delay compared to GEO satellites.  Different orbits decide that satellites are suitable for different scenarios. GEO is stationary to the Earth, while LEO is moving with the Earth. Therefore, GEO can provide  a designated region with continuous service. However, due to the long distance, it suffers a relatively high signal delay. Compared with GEO, LEO has a shorter delay because its orbit is much lower than GEO, which is more convenient for communicating with ground terminals, such as GPS communication and satellite phones.

\subsection{UAV Communication Network} 

A unmanned aerial vehicle (UAV) is an unmanned aircraft, which means it works automatically. Each UAV can be viewed as a node. Different UAVs construct a UAV network. Since the UAV communication network is distributed and federated learning is an emerging distributed learning framework, some studies leverage federated learning to assist UAV communication~\cite{lim2020towards, zeng2020federated}. In~\cite{lim2020towards}, Lim~\emph{et al.} propose a collaborative learning scheme based on federated learning and involve UAVs for applications in the IoV paradigm. In~\cite{zeng2020federated}, Zeng~\emph{et al.} apply federated learning algorithms to a UAV swarm and also present a joint power allocation and scheduling design, which increases the convergence rate of federated learning. 

\subsection{Maritime Communication Network}

It is envisioned that deep sea communication will be supported by the robust underwater link in 6G~\cite{zhang2020maritime}. That is, the underwater links will support communication among ships, submarines, and sensors. In a maritime communication network, it has shipped and involves the ashore stations and satellites. There are self-organized ad hoc networks which are consisted of vessels. This network's function is to accelerate the communication between mobile stations and aid navigation, and switch the route~\cite{zhang2020maritime}. Maritime communication also has several challenges to be overcome. Xia~\emph{et al.}~\cite{xia2020maritime} conclude eight existing challenges. They are ubiquitous connectivity $\&$ service continuity, capacity $\&$ scalability,  simplicity $\&$ reliability, device heterogeneity, traffic non-uniformity, service-centricity, interoperability, and radio spectrum internationality.





However, SAGSIN is still facing many challenges as listed in Table~\ref{tab:sum_challenge}.



\begin{table*}[]
\caption{A Summary of Challenges in SAGSIN~\cite{gupta2015survey, zhang2020maritime, xia2020maritime}.}
\label{tab:sum_challenge}
\begin{tabular}{l|l}
\toprule
\textbf{Networks} & \textbf{Challenges} \\ \hhline{==}
Satellite Communication Network & \begin{tabular}[c]{@{}l@{}} \\ - The design of the physical layer transmission to have accurate modelling of the satellite channels;\\ \\ - The application of massive MIMO;\\ \\ - Resource allocation in cooperative satellite transmission;\\ \\ - The physical layer transmission and media access control (MAC) protocols; \\ \\  \end{tabular} \\ \hline
UAV Communication Network & \begin{tabular}[c]{@{}l@{}}\\ - UAV deployment is too flexible to reach efficient result;\\ \\ - How to keep users' sessions if the UAV is out of service;\\ \\ - Designing low power consumption is tricky;\\ \\ - Deficiencies like node mobility, network partitioning, intermittent links are remained to be ameliorated;\\ \\ - Optimization of throughput, delay, etc. \\ \\ \end{tabular}  \\ \hline
Maritime Communication Network & \begin{tabular}[c]{@{}l@{}}\\- The services over the sea are limited so that ubiquitous connectivity and service continuity are not realized.\\ \\ - The distribution of maritime traffic is sparse on high seas. However, it is more dense close to the shore.\\ \\ - The maritime communication system should adapt to various applications.\\ \\ - The maritime communication network should be able to support different types of devices.\\ \\ - Maritime communication network should be deployed with low costs and high security.\\ \\- The maritime communication system should be extendible for the growing capacity.\\ \\ - Offer services for various information systems.\\ \\ - Standards and regulations of international frequency band are not addressed.\\ \\ \end{tabular} \\ \bottomrule
\end{tabular}
\end{table*}

\section{Other Potential Technologies} \label{sec:other-technologies}

We have  presented some key technologies  that will be enabled by 6G in detailed in the previous section. To meet the KPIs of 6G, we introduce more advanced technologies that may contribute to the paradigm shift. In this section, we briefly introduce some potential technologies in 6G, including terahertz communication, symbiotic radio, device-to-device communication, free-duplex, cell-free massive MIMO, blockchain-based network, and visible light communication.

\subsection{Terahertz Communications}~\label{sec:thz}

The terahertz (THz) frequency band, which ranges from 0.1 to 10 THz, is the unexplored span of radio spectrum~\cite{yuan2019potential, sarieddeen2019next}. Besides, THz communications provide new communication paradigms with the ultra-high bandwidth and ultra-low latency~\cite{sarieddeen2019next}. It is envisioned the data rate should be as high as Tbps to satisfy 6G applications' requirements of high throughput and low latency~\cite{yuan2019potential}.  

Traditionally, the THz frequency band limits the widespread use of THz. THz transceiver design is regarded as the most critical factor in facilitating THz communications~\cite{yuan2019potential}. Recent technology advancements in THz transceivers, such as photonics-based devices and electronics-based devices, overcome the THz gap, and enable some potential use cases in 6G~\cite{sarieddeen2019next}. The electronic technologies such as silicon-germanium BiCMOS, III-V semiconductor,  and standard silicon CMOS related technologies (III and V represents the old numbering of the periodic system groups), have been vastly advanced, such that mixers and amplifiers can operate at around 1THz frequency~\cite{yuan2019potential,han2019terahertz}. The photonic technologies are possiblely be used in the practical THz communication systems~\cite{yuan2019potential,han2019terahertz}. In addition, the combination of electronic-based transmitter and photonics-based receiver is feasible. Recent nanomaterials may help to develop novel devices that can used for THz communications~\cite{han2019terahertz}. In addition, a novel approach to generate the THz frequency is discovered by Chevalier~\emph{et al.}~\cite{chevalier2019widely}. They build a compact device that can use the nitrous oxide or laughing gas to produce a THz laser. The frequency of the laser can be tuned over a wide range at room temperature. 

Due to the high transmission RF, signals transmitted through THz frequency band suffer from a high pass loss. According to the Friis' law, the pass loss in free space increases quadratically with the operating frequency~\cite{HanTHz}. This feature limits the use of THz to short-distance transmission such as indoor communications~\cite{KhalidTHz}. Meanwhile, THz band can satisfy the requirement of ultra-high data rate; therefore, ultra-broadband applications, for example, virtual reality (VR) and wireless personal area networks can also exploit THz band to transmit signals~\cite{HanTHz}. THz technique can also be used in secure wireless communications. Since THz signals possess a narrow beam, it's very hard  to wiretap the information for the eavesdropper when locating outside the transmitter beam~\cite{KhalidTHz}.

\begin{figure}[!t]
     \centering
     \begin{subfigure}[b]{0.4\textwidth}
         \centering
         \includegraphics[width=\textwidth]{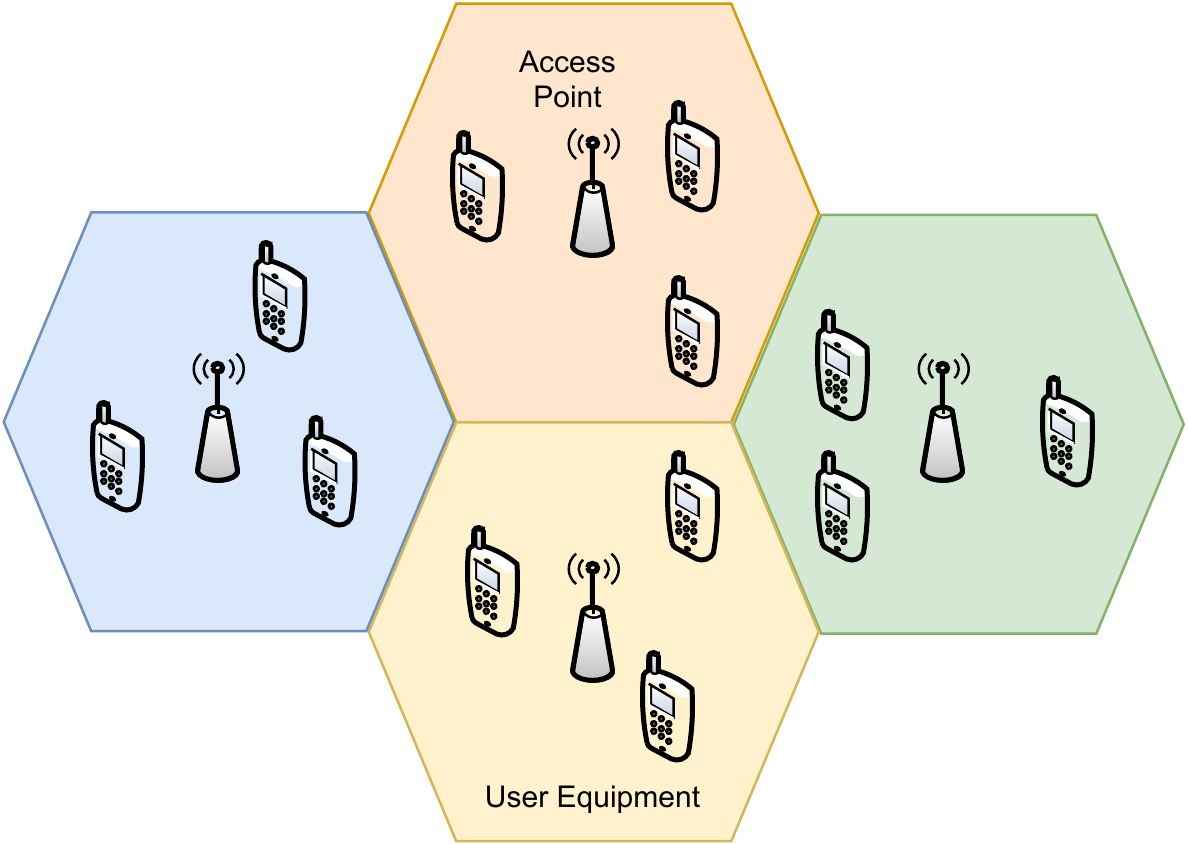}
         \caption{Traditional cellular system.}
         \label{fig:cellular}
     \end{subfigure}
     \hfill
     \hfill
     \begin{subfigure}[b]{0.4\textwidth}
         \centering
         \includegraphics[width=\textwidth]{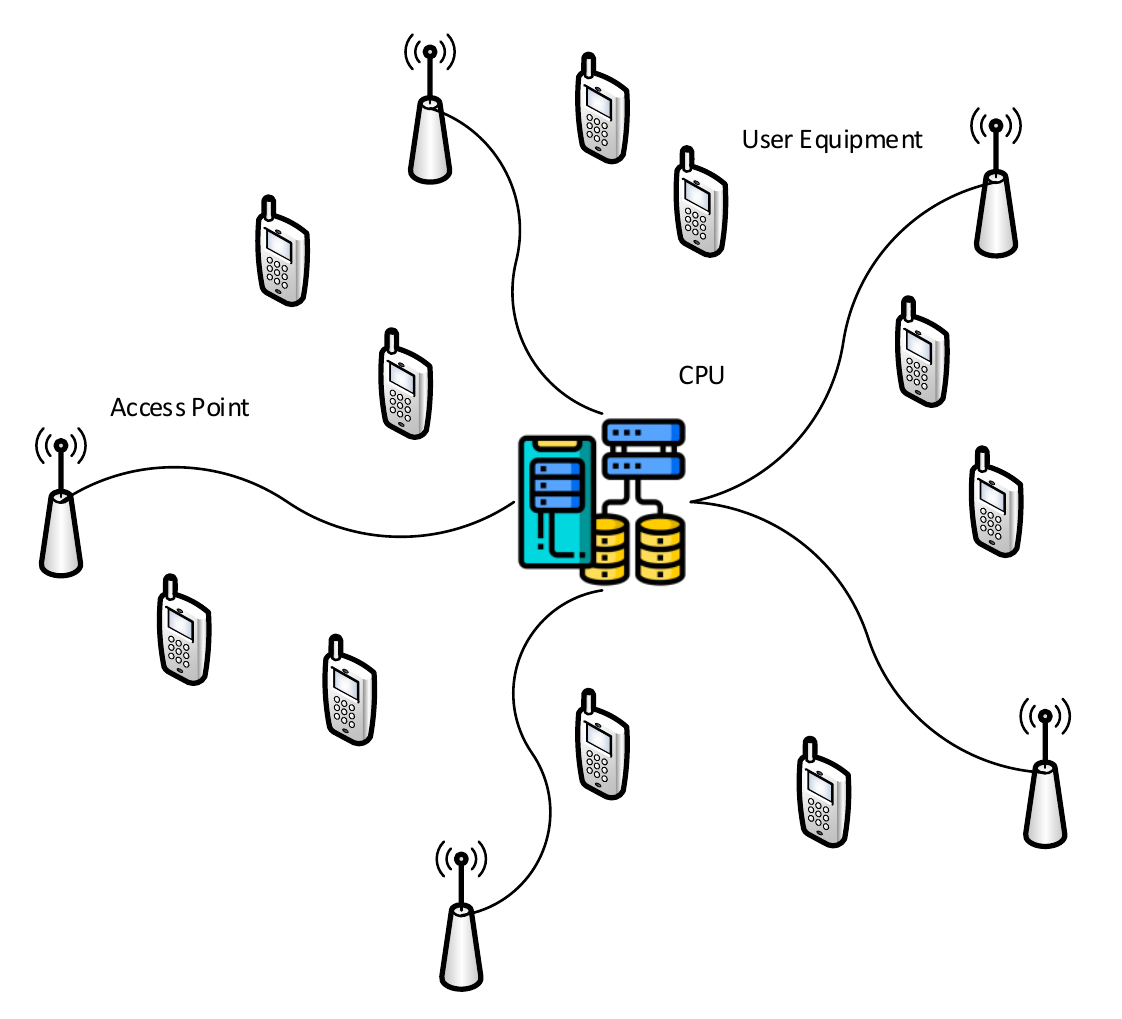}
         \caption{Cell-free massive MIMO.}
         \label{fig:cell-free}
     \end{subfigure}
        \caption{Compare the cellular system with the cell-free massive MIMO.}
        \label{fig:classic_mimo}
\end{figure}

\subsection{Cell-Free Massive MIMO}~\label{sec:cf}

The conventional cellular network divides the network into cells according to the topology theory. Each cell uses few access points which are equipped with large co-located antenna arrays~\cite{yang2014macro}. Due to the serious path loss, the cell center and the cell edge's performances are different. At the cell edge, the spectral efficiency and energy efficiency are very low. In 6G, cell-free massive multiple-input-multiple-output (CFmMM) is envisioned to be widely deployed. The CFmMM uses a large number of cheap access points (APs), and each of them is equipped with few antennas instead of the traditional way of deploying few access points with large co-located antenna arrays~\cite{rajatheva2020white}. The central processing units coordinate these APs. Since the number of APs increases, each user's equipment (UE) may use multiple APs and antennas nearby with reduced path loss. If the number of APs is large enough, there is no cell concept because UEs can connect to several neighbouring APs anywhere. Thus, there will be less path loss. Each UE can choose the cell with the strongest signals in the traditional cellular network, but it will try to access all the neighboring APs to seek support in the cell-free network. The proposed CFmMM systems ensure that UEs can achieve similar performance regardless of position and low-complexity signal processing. Thus, the number of APs should be more than that of users. Therefore, a CFmMM system is able to provide uniform service throughout its coverage area instead of being  partitioned into small cells. In order to illustrate their differences clearly, we compare the structures of conventional cellular system and the CFmMM system in Fig.~(\ref{fig:classic_mimo}).

As AI technologies will be widely utilized in 6G, for example, deep learning and federated learning, novel techniques and schemes based on CFmMM are proposed to solve machine learning problems~\cite{vu2020cell, bashar2020deep}. Instead of traditional optimization techniques, Bashar~\emph{et al.}~\cite{bashar2020deep} have developed a new deep learning based algorithm which uses a neural network to increase the achievable uplink sum rate of the CFmMM system. In addition, as the number of mobile phones increases, telecommunication operators start exploring ways to leverage the wireless computation to solve the machine learning problems. To conquer the federated learning problems, UEs compute  local updates with  local training data to train local models, and then they send trained models to the central server, while the central processor can be used to aggregate the local models to obtain the global model. Contrary to the traditional deep learning approaches, users can maintain their data locally to support training to prevent from the privacy leakage~\cite{vu2020cell}.

\begin{figure}[!t]
     \centering
     \begin{subfigure}[b]{0.25\textwidth}
         \centering
         \includegraphics[width=\textwidth]{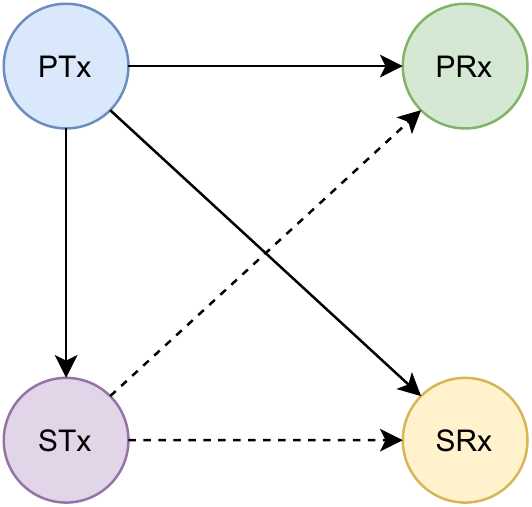}
         \caption{System model for SR. PTx uses active radio to transmit messages to PRx, and STx exploits backscattering radio to transmit messages to SRx riding over the RF signals from PTx.}
         \label{fig:sr}
     \end{subfigure}
     \hfill
     \hfill
     \begin{subfigure}[b]{0.25\textwidth}
         \centering
         \includegraphics[width=\textwidth]{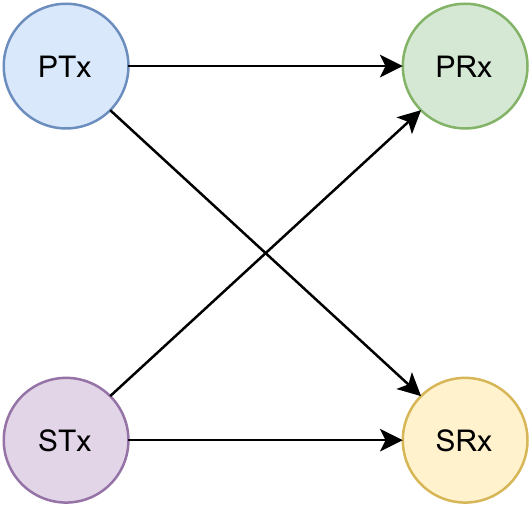}
         \caption{System model for CR. PTx transmits messages to PRx, and STx transmits messages to SRx by sharing the same radio spectrum with the primary system. STx refrains its transmit power to ensure that the caused interference level to PRx is below a tolerable threshold. Solid lines denote transmissions from active radios, while dash lines denote reflections from backscattering radios.}
         \label{fig:cr}
     \end{subfigure}
        \caption{Compare SR and CR's system models~\cite{liang2020symbiotic}.}
        \label{fig:sr-system}
\end{figure}

\subsection{Symbiotic Radio}

Symbiotic radio (SR) potentially enables the spectrum- and energy-efficient massive access technology of 6G~\cite{liang2020symbiotic, nawaz2020non, bariah2020prospective, chen2020vision, zhang20196g}. It is a cognitive backscattering communication system. SR reserves the advantages and solves the existing challenges  of cognitive radio (CR)~\cite{mitola1999cognitive} and ambient backscatter communications (AmBC)~\cite{liu2013ambient}. Specifically, CR means that the radio network is smart enough to learn from the environment and history, so that it may allocate spectrum resources dynamically to improve the efficiency of the radio spectrum utilization and avoid congestion. The secondary transmission and primary transmission exist at the same time, which may result in interfering with each other. AmBC enable that communications among smart devices using the ambient radio frequency instead of active RF transmission~\cite{liu2013ambient, van2018ambient}. In the wireless communication systems with low power, AmBC is very effective in enhancing the energy efficiency. However, the backscatter transmission may have direct-link interference due to the nature of the spectrum sharing, which may affect the performance of the transmission~\cite{8907447}.

SR can enhance the spectrum-, energy-, and cost-efficiency for wireless networks. Specifically, SR has two spectrum sharing systems, i.e., the primary system and the secondary system. Different from the interfering spectrum sharing in CR, SR achieves mutually beneficial spectrum sharing. By leveraging the joint decoding, SR obtains highly reliable backscattering communications, which makes up the disadvantage of the AmBC. The primary and secondary systems work collaboratively at both the transmitter and receiver's sides to improve the spectrum and energy efficiency further. Fig.~(\ref{fig:sr-system}) illustrates system models of SR and CR, respectively. SR utilizes the backscattering radio technology to support the transmission from the transmitter (STx) to the  receiver (SRx) in the secondary system, while the active RF chains are used by CR at both the primary receiver (PTx) and STx shown in Fig.~(\ref{fig:cr}). Fig.~(\ref{fig:sr}) illustrates that the backscattering link from PTx-STx-SRx has to go through the double fading, so the strength of the direct link from PTx to SRx is much stronger.  The secondary system's performance will be limited, and the primary system's performance will be improved. In addition, the backscattering radios can backscatter ambient signals within a specific frequency band to capture the RF signals that the secondary system uses.

As introduced in Section~\ref{sec:irs}, IRS is also a promising technology to improve the  efficiency of the energy in the wireless communication system. Specifically, IRS has the possibility to enhance  the backscattering link's signal, which contributes to improving the performance of the transmission. 
IRS-assisted SR is capable to adjust the reflecting elements intelligently, so SR system can capture the direction of the backscattring link to further improve the strength of the backscattering link. Also, the reflecting coefficients can be tuned to maximize the primary system's objectives. Besides, the number of reflecting elements can be increased as well to enhance the strength of the received SNR of the backscattering link. Thus, AmBC backscatters all the ambient signals; on the contrary, IRS-assisted SR is able to capture the intended PTx signal and avoid the undesired interference at the SRx.


\subsection{Free Duplex}~\label{sec:fd}

\begin{figure}[!h]
    \centering
    \includegraphics[scale=0.65]{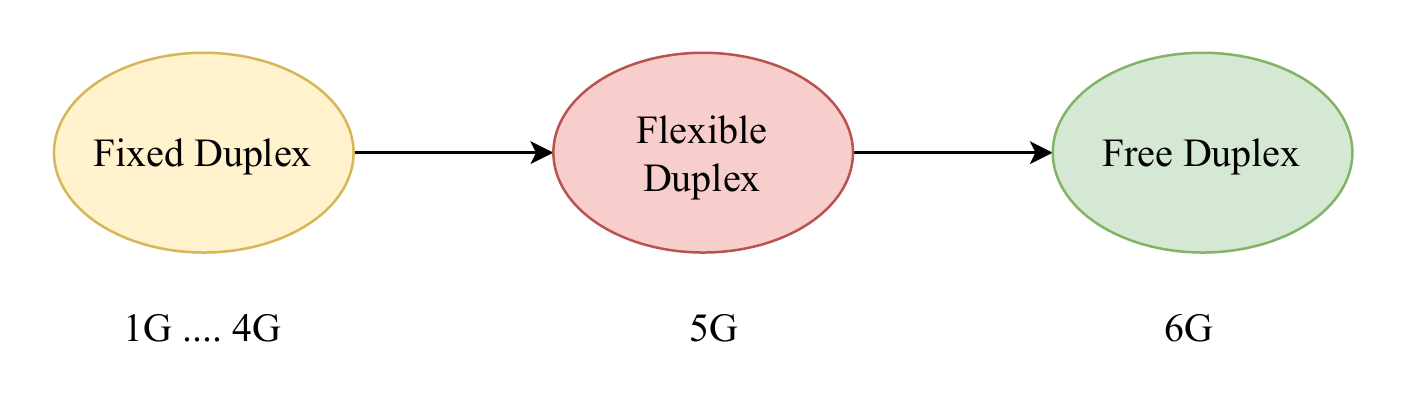}
    \caption{Evolution of Duplex Techniques.}
    \label{fig:free-duplex}
\end{figure}

To efficiently utilize the frequency resources, 6G is expected to implement the free duplex~\cite{lu20206g, zhao20196g, akhtar2020shift}. Free duplex means that there is no difference between the time division duplex (TDD) and frequency division duplex (FDD), so that the utilization of spectrum resources is free in terms of time, frequency, and space. In the previous generations of communication systems, the frequency is allocated based on TDD or FDD. Specifically, before 5G, the frequency resources are allocated statically according to TDD or FDD. Then, transmission resources are allocated dynamically in 5G, which is called flexible duplex.  Fig.~(\ref{fig:free-duplex}) shows the evolution of the duplex technologies. Free duplex is built on the full duplex technology. In the following, we present a detailed explanation on the development of the duplex technology.

\begin{figure}[!h]
    \centering
    \includegraphics[scale=0.4]{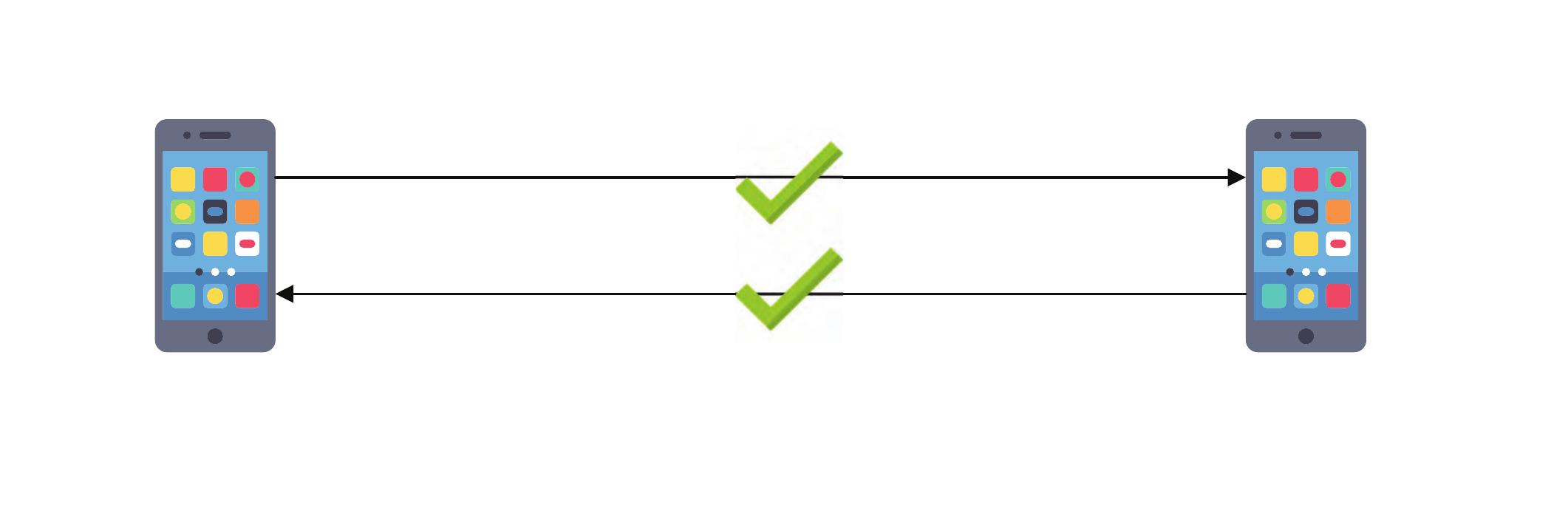}
    \caption{Full-Duplex.}
    \label{fig:fd}
\end{figure}

\begin{figure}[!h]
    \centering
    \includegraphics[scale=0.4]{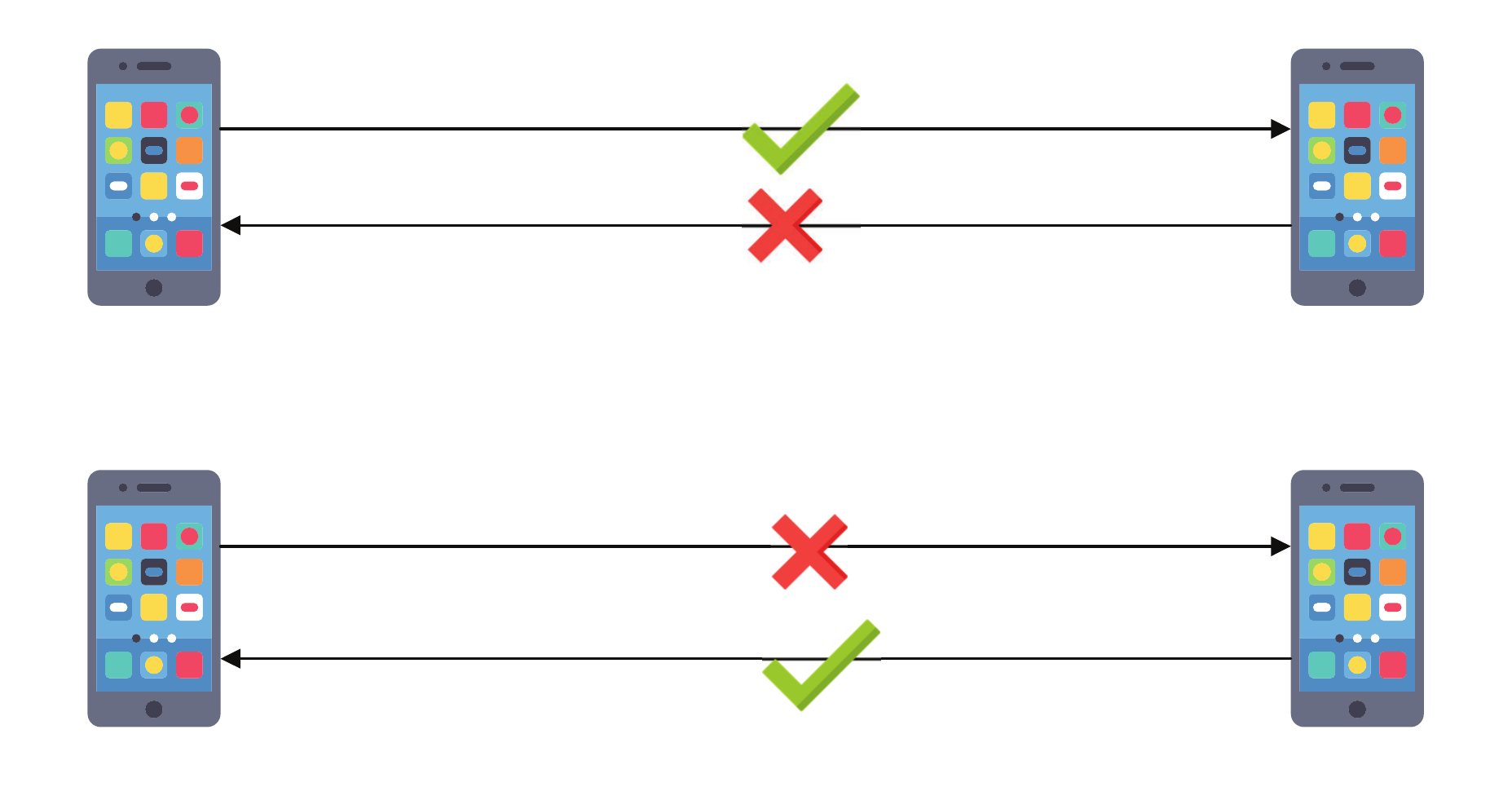}
    \caption{Half-Duplex.}
    \label{fig:hd}
\end{figure}

Duplex represents the ability of communication supported by two systems, including transmission and reception~\cite{amjad2017full}. Based on the capability of systems' data flow, the transmission and reception can be simultaneous or asynchronous, which are called full-duplex (FD) and half-duplex (HD), respectively. To be more specific, if systems are in FD mode, they can transit and receive simultaneously; otherwise, they choose to transmit or receive in different time slots. 
That means that the system using HD has to spend half of time  transmitting and the other half of time receiving, which greatly reduces the throughput and efficiency compared with leveraging FD. Fig.~(\ref{fig:fd}) and Fig.~(\ref{fig:hd}) illustrate differences between full-duplex and half-duplex.

In the 4G/5G wireless systems, transmission and reception cannot be done at the same frequency or a same time interval, because HD does not support performing the transmission and reception at the same time. Currently, 5G's spectrum is limited to the time division duplex (TDD) or frequency division duplex (FDD), and most of spectral resources are TDD. TDD and FDD are orthogonal transmission which may decrease the efficiency of the utilization of spectrum. FD technique has been enabled by 5G, but it has not been adopted by 3GPP yet. 

FD will be utilized completely in 6G wireless systems. FD technologies have the possibility to double current efficiency in sharing spectrum and increasing the throughput of the networks and communication systems. Both FD and its related techniques such as in-band full-duplex (IBFD) technologies improve the efficiency of communication by allowing devices to transmit and receive a signal in the same frequency band~\cite{rajatheva2020white}. Compared with HD, FD technology leverages self-interference cancellation technology to increase the utility of spectral resources, improve the throughput, and reduce the transmission delay between transceiver and receiver links. The difference between TDD and FDD will be eliminated, and the true FD mode based on the communication requirements. The arrival of data packets follows Poisson distribution, so that the utility of resources fluctuates dynamically. 

By using FD technology, devices can transmit and receive signals at the same time. Yuan~\emph{et al.}~\cite{yuan2020high} propose to improve the receiver with self-interference cancellation to realize self-interference cancellation, which uses $20\%$ time-frequency spectrum resources of traditional solution. However, the hard part is to eliminate the self-interference and transmit signal generates over 100dB higher noise than the receiver noise floor. Thus, the new scheduling algorithms and cost saving circuits should be designed for 6G networks. Currently, three types of self-interference cancellation techniques are proposed, which include digital cancellation, analog cancellation, and passive suppression.

Except for self-cancellation techniques, Shen~\emph{et al.}~\cite{xu2020resource} and Xu~\emph{et al.}~\cite{shen2020beamforming} propose to utilize intelligent reflecting surface (IRS) introduced in Section~\ref{sec:irs} to assist FD to improve the system performance and mitigate the interference. The cost for IRS is much cheaper than relay and it does not consume energy by using its soft-controlled functionalities of electromagnetic (EM) waves. IRS can assist FD application in two ways : (1) IRS acts as a bridge to facilitate FD transmission. Different from RF relays, IRS supports co-time and co-frequency FD transmission scenarios, including line-of-sight (LOS) and non-LOS. (2) Employ IRS to generate reflect/transmission waves of FD signals for specific purposes, including WPT, artificial noise, and cooperative jamming.

By integrating IRS and FD, the energy and cost saving FD-enabled IRS systems are proposed. Besides, the self-interference cancellation is unnecessary because IRS does not require any RF components, and then IRS is free of interference~\cite{pan2020full}. IRS enables electromagnetic (EM) functions such as anomalous reflection, frequency shifting, absorption, wavefront shaping, nonreciprocity, and focusing, etc. These functionalities can contribute to FD-enabled wireless transmission as follows :

\begin{itemize}
    \item Anomalous Reflection.
    \item Frequency Shifting.
    \item Absorption.
    \item Wavefront Shaping.
    \item Nonreciprocity.
    \item Focusing.
\end{itemize}

Many applications can leverage FD, including : (1) Application which are in low transmission power scenarios such as device-to-device (D2D) (introduced in Section~\ref{sec:d2d})  and vehicle-to-vehicle (V2V) applications. By using FD, V2V communication can be more reliable with low latency. (2) Sensing-based semi-persistent scheduling (SPS) achieves a relatively better performance. (3) Scenarios equipped transceiver devices with unlimited complexity and cost, such as wireless relay and wireless Backhaul. Scenarios which use spatial freedom and narrow beams.

\subsection{Blockchain-based Network}~\label{sec:blockchain}

Blockchain is a chain of blocks that constitute a distributed database. It is initially designed for cryptocurrencies (e.g., bitcoin). However, nowadays, blockchain can do more than just in cryptocurrencies but run Turing-complete programs such as smart contracts in a distributed way (e.g., Ethereum)~\cite{wood2014ethereum}. Blockchain provides a secure and distributed database for storing records of transactions, and each node includes the previous block's cryptographic hash, a time stamp, and transaction data~\cite{mollah2020blockchain, zhang20196g}.  Besides, blockchain-like mechanisms are expected to provide the distributed authentication and control by leveraging digital actions provided by smart contracts~\cite{strinati20196g}.

\begin{figure}[!h]
    \centering
    \includegraphics[scale=0.65]{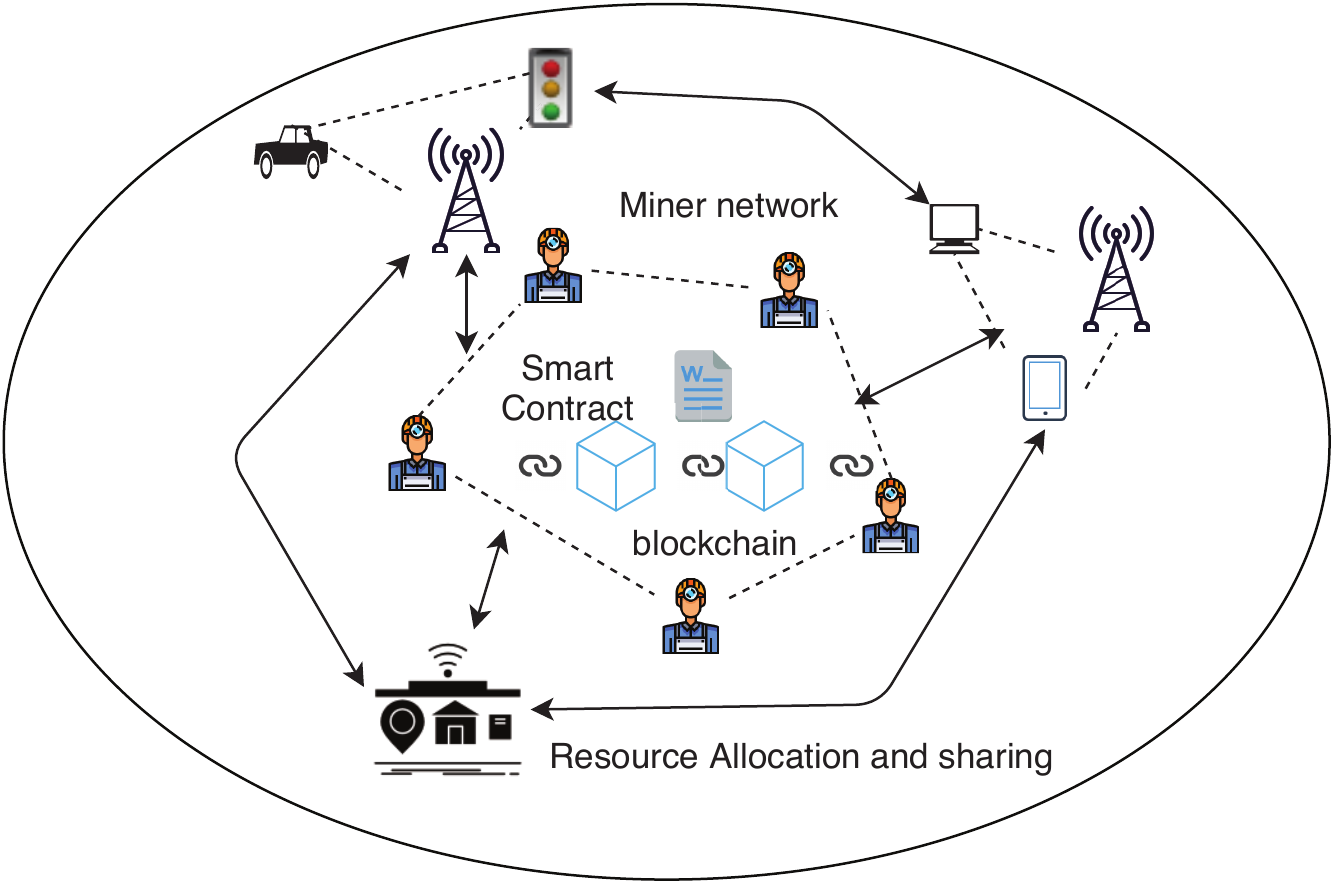}
    \caption{Blockchain-based network.}
    \label{fig:blockchain}
\end{figure}

With these unique features, blockchain is envisioned to support numerous applications in 6G. In particular, by combining with federated learning, blockchain-based AI architectures are shifting AI processing to the edge~\cite{porambagesec}.  Recently, a blockchain radio access network (B-RAN) has been proposed with the prototype~\cite{ling2019blockchain, le2019prototype}. Thus, blockchain can help to form a secure and decentralized environment in 6G. Blockchain can provide a secure architecture for 6G wireless networks~\cite{dai2019blockchain}. 
Blockchain makes the consensus through miners instead of a central authority as shown in Fig.~(\ref{fig:blockchain}), and it includes a wide range of applications in 6G. Blockchain can help enhance authentication security by the approach of distributed ledger technologies~\cite{strinati20196g}. Besides, blockchain has applications in solving the problem of the low spectrum utilization and spectrum monopoly when deployed in spectrum sharing system~\cite{zhang20196g}.  By integrating wireless networks and blockchain, the central administrator can be eliminated, improving network security and reducing costs. As D2D and IoE are gaining popularity in 6G, the cooperation among devices is getting more frequent, so that the distributed way of resource management and spectrum control is required. Since blockchain guarantees transparency, it is easy to track the spectrum's real-time utilization, which enables to allocate the spectrum dynamically and efficiently. Thus, if 6G wireless network couples with blockchain when building, it will be simple to track the resource management and spectrum sharing. IoT and D2D enable applications such as smart farming, healthcare, and machine-to-machine communication, etc.

\subsection{Visible Light Communication}~\label{sec:vlc}

The visible light communication (VLC) is considered as one of the techniques that will be used in 6G, and it operates at the rarely explored THz frequencies~\cite{ariyanti2020visible}. Specifically, 6G moves to higher frequencies because of the spectral congestion in frequencies that 5G uses and the increasing requirements for higher data rates.  
VLC has high data rates, a large frequency spectrum, high-speed transmission, and robustness against interference~\cite{arfaoui2020physical}. Hence, VLC contributes to the development of short-range communications in 6G~\cite{6gwhite}. For the short-range communication, either data-modulated white laser diodes or light-emitting diodes are used as transmitters, while photodetectors are utilized as receivers. Besides, VLC is considered as a complementary technology for radio frequency communication because it can utilize an unlicensed spectrum for communication~\cite{giordani2019towards}.

\begin{figure}[!htbp]
    \centering
    \includegraphics[scale=0.6]{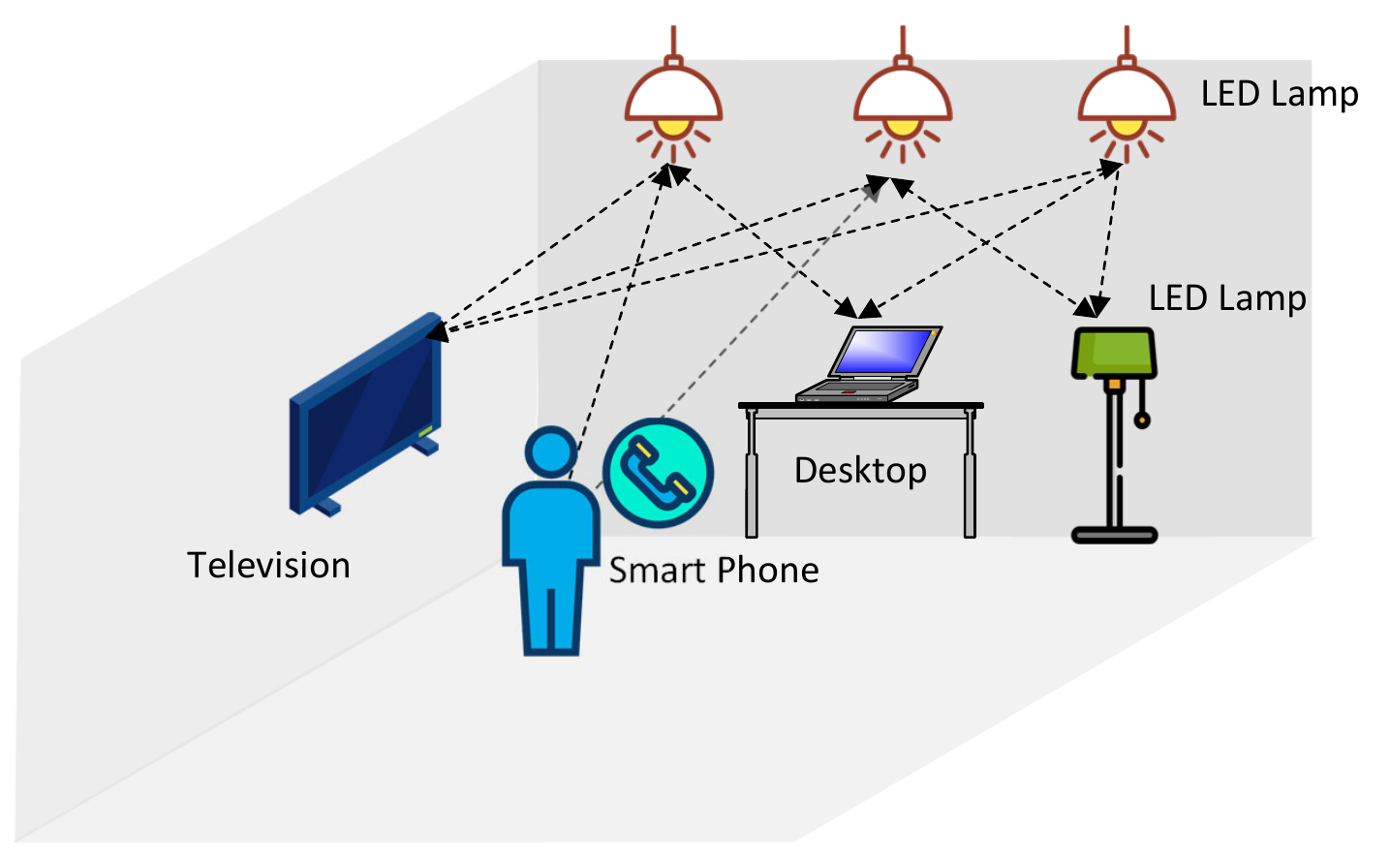}
    \caption{Visible Light Communication.}
    \label{fig:vlc}
\end{figure}

The laser diode (LD)-phosphor conversion lighting technology can provide better performance in efficiency, brightness, and larger illumination range compared with traditional lighting techniques~\cite{yuan2019potential}. Thus, it is considered as the most promising technology for 6G. The speed LD-based VLC system is possible to reach 100Gbps, which meets the requirements of  ultra-high data density (uHDD) services in 6G. Besides, the upcoming new light sources based on microLED will overcome the limitation of low speed in the short-range communication~\cite{strinati20196g}. As massive parallelization of microLED arrays, spatial multiplexing techniques, CMOS driver arrays, and THz communications develop, VLC's data rate is expected to reach Tbps in the short-range indoor scenario by the year 2027~\cite{strinati20196g, sarieddeen2019next}. Fig.~(\ref{fig:vlc}) illustrates the indoor VLC scenario.

VLC is envisioned to be utilized in various applications in 6G. By integrating with SAGSIN,  short-range network, and cellular communication, VLC can be used to provide a better coverage~\cite{nawaz2019quantum}. Also, traditional electromagnetic-wave signals cannot achieve high data transmission speed using laser beams in the free space and underwater. Still, VLC has a ultra-high bandwidth and high data transmission speed~\cite{chowdhury2019role}. Therefore, VLC is useful in cases where traditional RF communication is less active, for example, indoor communication, underwater communication, underground communication, and in-cabin internet service~\cite{tariq2019speculative, katz2020opportunities}. Furthermore, VLC is envisioned to be widely used in vehicle-to-vehicle communications, which depend on the head and tail lights of cars for communications~\cite{tariq2019speculative, chowdhury2019role, yuan2019potential}. Besides, VLC serves as a potential solution to build gigabit wireless networks underwater.

However, VLC is still facing many challenges. For example, the light source's bandwidth restricts the speed of VLC~\cite{ariyanti2020visible}, so new materials and mechanisms should be developed to increase the light source's bandwidth. Besides, Si-based detectors used by VLC systems are more sensitive to infrared waves than visible light. Moreover, no application-specific integrated circuits for VLC baseband processing. Furthermore, the data processing for future systems will be a much more complex to process.

\subsection{Device-to-Device Communication}~\label{sec:d2d}

As the Internet of Everything (IoE) will be introduced in 6G, the number of edge devices' booming increase imposes unpredictable pressure on the communication between centralized servers and edge devices. Device-to-Device (D2D) communication enables devices to communicate directly without going through infrastructures like base stations or access points, which guarantees ultra-low latency, speeds up the transmission, improves communication quality, and offloads traffic from the conventional cellular network in end-to-end communication~\cite{kar2020critical, malik2020survey}.  Thus, D2D communication will gain much more attention in increasing data transmission speed, decreasing end-to-end latency, and reducing the cost of the communication~\cite{zhang2020envisioning}.

Specifically, D2D communication can operate in licensed bands and unlicensed bands (i.e., Bluetooth and Wi-Fi). 
Because of the decentralized features that D2D possesses, D2D obtains its advantages in multiple areas, such as mobile edge computing, IoT, IoE, etc., which also share decentralized features.  With the aid of D2D communication, coordination between the centralized server and user’s devices has significantly changed. Users who are close to each other exchange data without forwarding through the BS nearby~\cite{chen2020reconfigurable}. Hence, D2D communication can reduce energy consumption and upgrade its quality of service (QoS) demands.

D2D is encountering challenges, including severe interference, prohibitively high cost, a vast amount of signaling, complex resource management,   and energy consumption. To facilitate the D2D communication, more advanced and state-of-the-art technologies such as AI and IRS can be utilized~\cite{zhang2020envisioning}. To be more specific, D2D communication will become AI-Driven and intelligent in 6G. One application is to leverage AI to manage resources intelligently. AI-driven D2D communication will enable three kinds of applications, including  NOMA-based D2D cognitive networking, intelligent D2D-enhanced mobile edge computing, and D2D-enabled intelligent network slicing. Network slicing (NS) helps to manage and share resources in 5G network. A large number of D2D clusters can extend the network's flexibility, which enables to provide services according to users' needs. In 6G, the large number of D2D clusters can provide both physical and virtual resources, which are critical to building NS. AI will be employed to support and manage D2D clusters.  The process of distributing resources will be intelligent and automatic, which means that AI will monitor the underlying resource and network slices to intelligently achieve resource mapping. Combined with NOMA, D2D will support the cognitive network. Delta-orthogonal multiple access (D-OMA) technique is proposed to obtain large scale concurrent access. D-OMA which serves as the new multiple access scheme for 6G is leveraged to tackle terminal devices, including the high complexity and increased energy consumption at terminal devices.

\subsection{Network in Box}~\label{sec:nib}

More and more kinds of technologies will be embedded in 6G, such as the autonomous vehicles, factory automation, etc. To satisfy the real-time and reliable features of the network, Network in Box (NIB) technique has attracted much more attention from the industrial automation because NIB offers a device that can provide seamless connectivity between different services. On the other hand, NIB covers the wireless environments, including ground, air, and marine well agreed to the vision of the IoE in 6G.

\section{Security and Privacy}~\label{sec:security}

In the 6G wireless communication systems, edge devices with the high mobility are getting popular. A huge amount of data are generated by mobile applications containing sensitive information, which closely relates to users' privacy. Because of the wireless communication system's broadcast nature, eavesdroppers may monitor and steal users' confidential information during the transmission. To avoid the necessary but complex key sharing in traditional cryptography, physical layer security methods that can guarantee wireless communication security are proposed.

\subsection{Attack Types}

The attacks to the physical layer of the wireless communication system can be classified into two categories: the passive attack and the active attack.

\begin{figure}[htbp]
    \centering
    \includegraphics[scale=0.85]{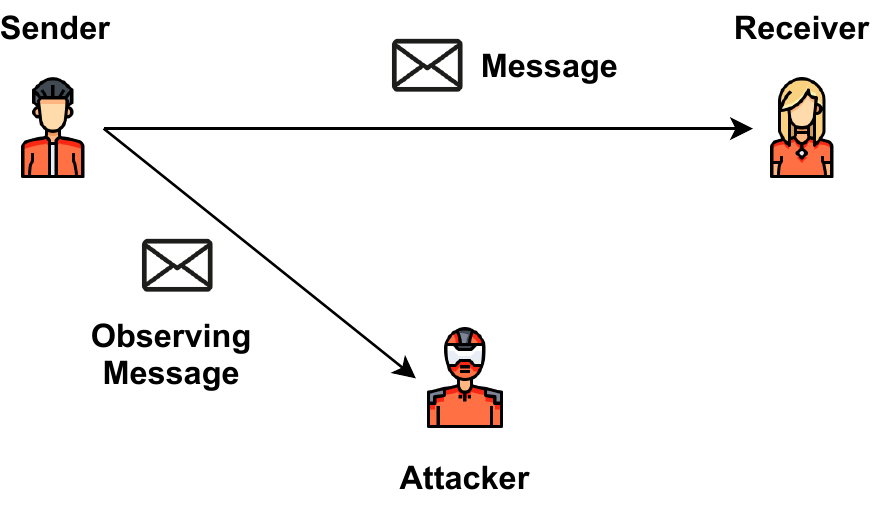}
    \caption{Passive Attack.}
    \label{fig:passive_attack}
\end{figure}

\textbf{Passive Attack.} The passive attack means that attackers attempt to learn the information from the transmission without affecting the system resources, as shown in Fig.~(\ref{fig:passive_attack}). It can be divided into two distinct categories: one is to obtain the content of the message directly; the other one is to analyze the data flow. The eavesdropping attack is a way to obtain the effective data sent from the original station to the destination station without affecting the normal data communication. By monitoring the data transmission, the eavesdropping attack damages data confidentiality and results in privacy leakage. Suppose some approaches, such as encryption, make the attacker unable to obtain the true content of the message from the intercepted parts. In that case, the attacker may obtain the message format, determine the location and identity of both sides of the communication, the number of communication times, and the length of the message, which may also be sensitive to both sides of the communication. Since the passive attack does not make any modification to the message, it is difficult to detect.

\begin{figure}[htbp]
    \centering
    \includegraphics[scale=0.85]{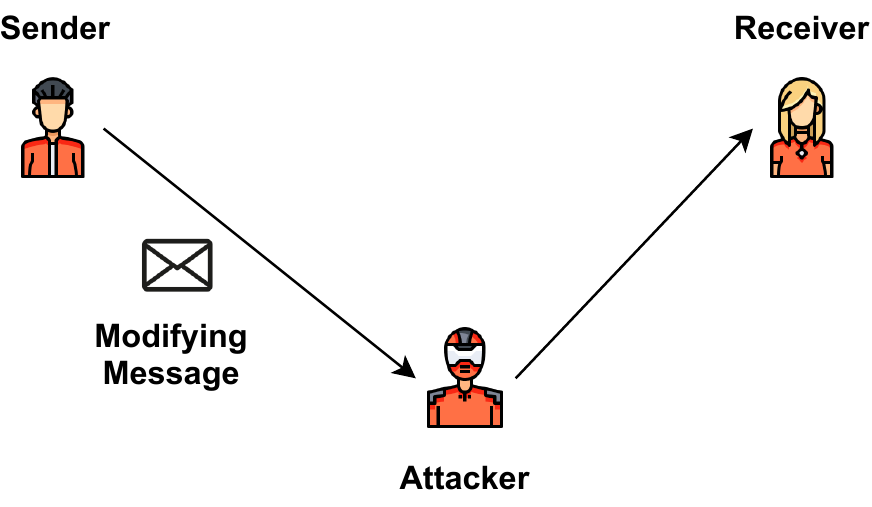}
    \caption{Active Attack.}
    \label{fig:active_attack}
\end{figure}

\textbf{Active Attack.} The active attack is a deliberate action of actively accessing the required data, causing a direct impact on legitimate users as illustrated in Fig.~(\ref{fig:active_attack}). The active attack can be divided into three categories:  The first one is to intercept the data sent by the original station, interrupting the effective data so that the destination station cannot receive the data transmitted by the original station. It affects data availability. The second one is to tamper with the data sent from the original station to the destination station, thereby affecting the information received by the destination station. It affects data integrity. The third one is to forge data and send it to the destination station even if the original station does not send data, which affects data authenticity. The primary approach to deal with the active attack is to detect and recover the damage caused by it.

\subsection{Technologies for 6G Physical Layer Security}
In 6G, two basic requirements need to be satisfied when it comes to the physical layer security: confidentiality and authentication. Confidentiality means that eavesdroppers have no access to the effective data. The authentication ensures that attackers cannot forge data and send them to the destination station. Although traditional cryptographic technologies such as the public key cryptography can enforce security, they are all implemented in the upper layer. Wyner shows that secure communication can be achieved in the physical layer just by technologies adopted in noise and interference~\cite{wyner1975wire}, which becomes the foundation of research on the security of the wireless communication theory. Csisz{\'a}r and Köner~\cite{csiszar1978broadcast} generalize Wyner's results by adopting a non-degraded discrete memoryless broadcast channel. Since then, more and more researchers focus on this field and propose a large number of approaches. To be more specific, existing methods can be categorized into channel approaches, power allocation approaches, and signal processing approaches. The intelligent reflecting surface (IRS) is a kind of metasurfaces that can improve the efficiency of the spectrum utilization by reconfiguring the reflection angle of signals regardless of the incidence angle. However, attacks may happen in IRS, and many have been investigated~\cite{shen2019secrecy, hong2020artificial, yu2020robust}.

\textbf{Channel Approaches.}
In recent years, many researchers focus on the fundamental issues of secure channel capacity. Their work's main idea is to distinguish the quality of signals received by legitimate users and by unauthorized receivers. Wyner has shown that reliable and secure transmission can be achieved in degraded broadcast channels. The perfect secrecy capacity is the gap between the attacker's capacity and legitimate user's capacity for discrete memoryless channels~\cite{wyner1975wire}.~\cite{yamamoto1991coding, leung1978gaussian} introduce the Gaussian channels based on Wyner's work and generalize the conclusion to Gaussian channels. Klinc~\emph{et al.}~\cite{klinc2011ldpc} propose an effective coding scheme for Gaussian wiretap channel based on LOPC codes, which is encodable in linear time. It can combine with cryptography techniques, providing improved data security protection in communication channels. The radio frequency recognition system proposed by Sperandio~\emph{et al.}~\cite{sperandio2002wireless} can recognize the identities of transmitters from the received signal. Each participant has its intrinsic physical properties, so the system aims to process the extracted features and obtain a fingerprint for each party. Cobb~\emph{et al.}~\cite{cobb2011intrinsic} use a Bayesian classifier to analyze amplitude, phase, and frequency, thus providing authentication for the communication system. Dan~\emph{et al.}~\cite{shan2013phy} propose a method using the radio frequency as physical fingerprints to authenticate a Wi-Fi device's identity. The approach of radio frequency fingerprint authentication is effective in the aspect of preventing network intrusion, especially when cryptography based authentication techniques are challenging to implement in some specific systems.

\textbf{Power Allocation Approaches.}
In Wyner's wiretap channel model, the eavesdropper's channel's quality is worse than that of the legitimate user's to achieve communication confidentiality. But in some cases, the eavesdropper is close to the original station than the legitimate user, which means that the eavesdropper has better channel quality than the legitimate user. Hence, the eavesdropper may monitor and receive information that is confidential to the legitimate users. To prevent the information leakage, Goel~\emph{et al.}~\cite{goel2008guaranteeing} propose to add the artificial noise to the channel to deteriorate the eavesdropper's channel, thus achieving minimum guaranteed secrecy capacity. Using this method, if the original station has more antennas than the eavesdropper, the transmitter can use part of the power to generate the artificial noise and inject it into the channel with multiple antennas. The information signal is transmitted in the range space in the legitimate channel. At the same time, the artificial noise is generated in the null space in the eavesdropper's channel, so the artificial noise only impairs the eavesdropper's channel but not the intended receiver’s channel. However, this design heavily depends on the obtainment of accurate channel knowledge.

\textbf{Signal Processing Approaches.}
In the 6G wireless communication systems, edge devices with high mobility are getting popular. A huge amount of data are generated by mobile applications containing sensitive information, which may closely relate to users' privacy. Also, technologies are proposed, for example, the intelligent reflecting surfaces. Because of the wireless communication systems' broadcast nature, eavesdroppers may monitor and steal users' confidential information during data transmission. To avoid the necessary but complex key sharing in traditional cryptography, physical layer security methods that can guarantee the security of the wireless communication are proposed.

\section{Applications}~\label{sec:applications}

Generally, the Internet of Things (IoT) refers to a network of connected devices (also know as smart objects) that can collect and exchange data over the Internet~\cite{Fortino14}. Over the past few years, there is an emerging trend of employing artificial intelligence (AI) for IoT as AI has made lots of remarkable achievements in the big-data era~\cite{Poniszewska15}. To date, artificial intelligence of things (AIoT) has been widely used in various kinds of areas. As AIoT becomes ubiquitous, 6G will play a vital role in AIoT applications. First, 6G will support a large number of smart devices to work cooperatively in an AIoT system while maintaining the low latency. Second, by integrating space-air-ground-sea networks, 6G provides a wider coverage for AIoT applications. Third, advanced AI technologies will be implemented in 6G-based AIoT. Last but not least, 6G enables AIoT to better protect users' security and privacy. In a word, with the help of 6G wireless network, both device-device communication  and device-server communication will be greatly enhanced. For example, autonomous cars with high running speeds reply on low-latency communication to react quickly when avoiding of collision. To this end, automobiles exchange information (e.g. speed and location) to each other rapidly through vehicle-to-vehicle communication (V2V communication), a typical application of device-device communication. The example of device-server communication is smart healthcare. The wearable and in-body sensors will continuously collect patients' information and upload the data to the cloud serve for real time monitoring  through low-latency communication~\cite{Mucchi20}. In fact, many researchers have demonstrated the versatility of 6G-based AIoT systems on a wide range of scenarios. In the following, we focus on several typical applications for further illustration and summarize them in Table~\ref{tab:aiot}.

\subsection{AI in Network Management} As 6G network becomes complex, it may utilize deep learning instead of human operators to improve the flexibility and efficiency in the network management~\cite{piran2019learning}. AI technologies are applicable to both the physical and network layers. In physical layer, AI techniques have involved in design and resource allocation in wireless communications~\cite{ho2019next}.  For example, unsupervised learning are applicable to interference cancellation, optimal modulation, channel-aware feature-extraction, and channel estimation, etc.~\cite{piran2019learning}. Deep reinforcement learning is possible to be employed for link preservation, scheduling,  transmission optimization, on-demand beamforming, and energy harvesting~\cite{piran2019learning, shafin2019artificial}. In addition, AI technologies can be used to the network layer as well. Supervised learning techniques can tackle problems such as resource allocation, fault prediction~\cite{piran2019learning}. Besides, unsupervised learning algorithms can help in routing, traffic control, parameter prediction, resource allocations, etc.~\cite{piran2019learning}. Reinforcement learning can be important for traffic prediction, packet scheduling, multi-objective routing, security, and classification~\cite{piran2019learning, shafin2019artificial}.

\subsection{AI in Autonomy}
AI technologies are potential to enable 6G wireless systems to be autonomous~\cite{loven2019edgeai,gacanin2019autonomous,zhang20196g}. Agents with intelligence can detect and resolve network issues actively and autonomously. AI-based network management contributes to  monitoring network status in real-time and keep network health. Also, AI techniques can provide intelligence at the edge devices and edge computing, which enables edge devices and edge computing to learn to solve security problems autonomously~\cite{porambagesec, mollah2017secure, loven2019edgeai}. In addition, autonomous applications such as autonomous aerial vehicles and autonomous robots are envisioned to be available in 6G~\cite{ho2019next}.

\begin{table*}[!h]
\centering
\caption{Summarization of Applications in Artificial Intelligence of Things.}
\renewcommand\tabcolsep{20.0pt}
\begin{tabular}{l|l|l}
\toprule
\textbf{Typical applications}                        & \textbf{Reference}                       & \textbf{Requirements}                                                                                                                                                           \\ \hhline{===}
Smart Healthcare                  &~\cite{Nayak2020Healthcare},~\cite{tariq2019speculative},~\cite{dang2020should},~\cite{nayak20206g},~\cite{Hewa20} & \begin{tabular}[c]{@{}l@{}}Ultra-low latency\\ High bandwidth\\ High security\end{tabular}                                                                             \\ \hline
Smart Manufacturing               &~\cite{rajatheva2020white},~\cite{giordani2019towards},~\cite{Berardinelli18},~\cite{strinati20196g},~\cite{Sekaran20},~\cite{Peltonen20} & \begin{tabular}[c]{@{}l@{}}Ultra-low latency\\ Ultra-high reliability\\ Ultra-high bandwidth\\ Very high intelligence\end{tabular}                                     \\ \hline
Smart Home                        &~\cite{Nayak20Communications},~\cite{Noury07},~\cite{Peltonen20},~\cite{Mao20AI} & \begin{tabular}[c]{@{}l@{}}Ultra-low latency\\ Ultra-high security\\ Very high intelligence\end{tabular}                                                               \\ \hline
Intelligent Transportation System &~\cite{Nayak20Communications},~\cite{Zheng19},~\cite{rajatheva2020white},~\cite{chowdhury2019role},~\cite{Wang20Security},~\cite{Sun20},~\cite{Zhang18},~\cite{Dizdar20},~\cite{chowdhury2019role}    & \begin{tabular}[c]{@{}l@{}}Ultra-low latency\\ Ultra-high bandwidth\\ Ultra-high security\\ High intelligence\\ High mobility\\ Long distance communication\end{tabular} \\ \hline
Smart Grid                        &~\cite{Fang12},~\cite{ho2019next}                              & \begin{tabular}[c]{@{}l@{}}Ultra-low latency\\ Ultra-high security\end{tabular}                                                                                        \\ \hline
Unmanned Aerial Vehicle           &~\cite{Wang20Security},~\cite{Nayak20Communications},~\cite{tariq2019speculative},~\cite{Mehta20},~\cite{chowdhury20196g} & \begin{tabular}[c]{@{}l@{}}Ultra-low latency\\ Ultra-high bandwidth\\ Ultra-high mobility\\ Ultra-long distance communication\\  \end{tabular} \\ \bottomrule
\end{tabular}
\label{tab:aiot}
\end{table*}

\subsection{Smart Healthcare}
\begin{figure}[htbp]
    \centering
    \includegraphics[scale=0.62]{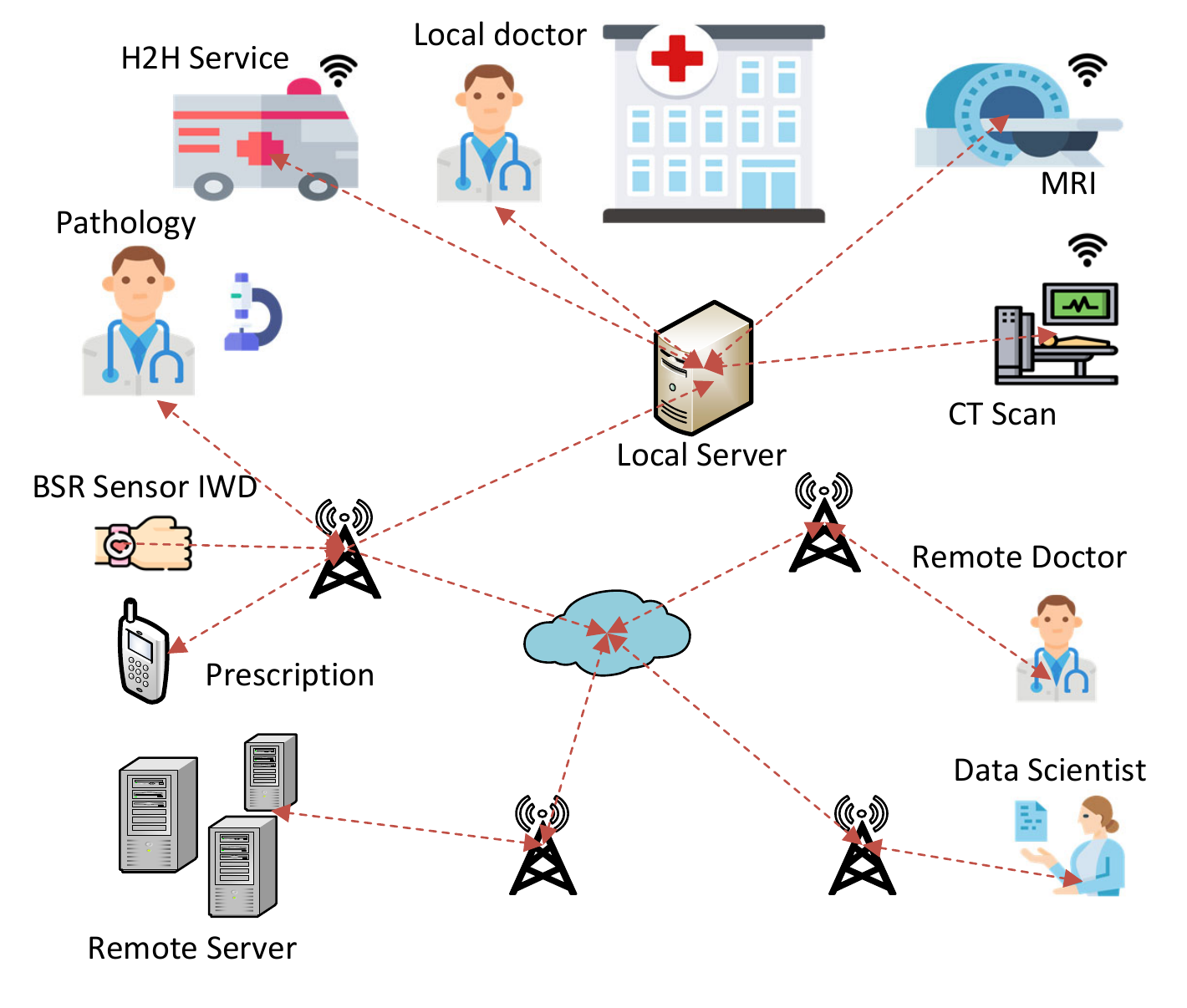}
    \caption{Smart Healthcare.}
    \label{fig:smart_healthcare}
\end{figure}

The number of Chronic patients and the aging of population are increasing dramatically year by year~\cite{Mucchi20}. Besides, traditional healthcare systems  require patients to visit the hospitals, which are time-consuming and labor-intensive~\cite{Aygun19}. Therefore, healthcare systems improvement is significant for people's wellbeing. An efficient healthcare system is expected to carry out health monitoring, disease diagnosis, and medical treatment remotely with high efficiency. Therefore, researchers have resorted to the smart healthcare. Smart healthcare is a healthcare service system that combines a variety of technologies such as wearable sensors and AIoT~\cite{Tian19}. Smart healthcare can not only provide convenience for the people but also save lives in emergency. Even though time and space are barriers of current healthcare systems, 6G wireless network enables smart healthcare to overcome these barriers. Therefore, 6G allows healthcare systems to complete more useful and sophisticated tasks as illustrated in Fig.~(\ref{fig:smart_healthcare}). That is, patients can be accurately diagnosed and treated by professional doctors even if they are at home. 

\textbf{Health Monitoring and Disease Detection.} Continuous health status monitoring is crucial in healthcare service, especially for those patients who suffer from sudden disease (e.g., cardiovascular diseases). In 6G-based healthcare systems, data-driven models that can detect abnormal health status will be uploaded to the cloud in advance. Patients' physiological data can be collected continuously by noninvasive sensors and will be transmitted to the cloud. The models on the cloud will analyse the data and send a warning message if the diseases occur suddenly. 6G technology will play a critical role as both real-time monitoring and disease notification require high data rates to react quickly. What's more, 6G will prompt the development of the hospital-to-home (H2H) service to replace the traditional ambulance service~\cite{Nayak2020Healthcare}.

\textbf{Medical Treatment.} The merits of 6G will support doctors to perform more effective medical treatment. For example, augmented reality (AR) and virtual reality (VR) can be applied in medical treatment because of high data rates and low latency in 6G communication. To be more specific, AR enables doctors to observe the inside of the patient's body clearly without making any incision, and doctors can try out various surgical plans in a simulated environment before real surgeries~\cite{Nayak2020Healthcare}. 

6G also revitalizes the medical robots~\cite{tariq2019speculative}. For example, the medical robots will take care of the patients and provide timely help for them when the hospital is too busy and most nurses are unavailable. Besides, the medical robots will help doctors in surgeries as the 6G wireless network allows the robots to carry out complex tasks with high precision. Specifically, the size of the medical robots can be very small so that the doctors will control them to enter the human's body to take pictures, deliver drugs, or remove diseased tissues. 

\textbf{Privacy and Security in Healthcare.} Security and privacy are challenges of the 6G technology and the key concerns for patients~\cite{dang2020should}. Edge computing can be used to protect patients' privacy~\cite{Nayak2020Healthcare}. The data will be delivered to different edge nodes because the memory of edge nodes is small. Therefore, healthcare data do not need to be stored in only one place, hence increasing communication security. Besides, selectively uploading the data to the cloud helps to improve security as it is easier for the cloud to protect lesser data~\cite{Nayak2020Healthcare}. Last but not least, blockchains or federated learning are also possible approaches to address the privacy problem in healthcare systems in the future~\cite{nayak20206g, Hewa20}. 

\subsection{Smart Manufacturing}
The Industry 4.0 has envisioned a digital transformation of manufacturing through cyber physical systems and IoT services, and its main goal is to reduce the human intervention in industrial processes by using efficient control approaches and communication technologies~\cite{strinati20196g}. 6G will finally realize this revolution by investigating smart manufacturing~\cite{rajatheva2020white}. Smart manufacturing refers to a IoT-connected manufacturing system that applies a variety of control and data analytics approaches to improve manufacturing performance. The advantages of 6G will boost the communication and computing capabilities of the connected sensors and machinery, thus leading to a preciser and smarter manufacturing system~\cite{giordani2019towards}. 

\textbf{High-precision Manufacturing.} It is vital for the manufacturing system to maintain high precision during operations. Numerically speaking, implementing high-precision manufacturing requires very high reliability (up to about 10$^9$) and extremely low latency (0.1 to 1ms round trip time)~\cite{Berardinelli18}. In addition, massive amounts of data and transmissions are involved in industrial control networks, hence requiring a very low delay jitter (about 1$\mu s$)~\cite{Berardinelli18}. 

The above requirements can hardly be met if using 5G technology. In contrast, the emergence of 6G paves the way for high-precision manufacturing because of its superior features. For example, a new 6G-based architecture that integrates different resources has been proposed to satisfy the tight physical constraints~\cite{strinati20196g}. Some researchers also resort to other advanced IoT approaches, such as blockchain~\cite{Sekaran20} and edge computing~\cite{Peltonen20}, to improve the performance of the manufacturing systems.

\textbf{Intelligent Robots.} It is very common in a modern manufacturing system to apply intelligent robots to deal with dull and tedious work. Robots can also replace humans to carry out dangerous industrial operations or tasks requiring extremely high precision. According to Industry 4.0, the robots are required to react quickly when interacting with humans and machinery in a dynamic environment~\cite{Torfs17}. To this end, it is essential to apply 6G technology into robotics communication~\cite{Nayak2020Healthcare}. 

The intelligent robots connected by 6G wireless network robots will be able to conduct complex cooperative operations~\cite{Peltonen20, Monserrat20}. For example, intelligent robots on the edge side can take videos of the industrial process and then upload the data to the cloud, while the learning algorithms on the cloud will make decisions to control the robots. Aided by 6G technology, robots will be competent enough even though the whole control loop requires ultrahigh data rates. As a result, senors, robots, machinery, and 6G will form an efficient distributed intelligent network which has terabytes of computing capacity~\cite{zong20196g}. Furthermore, there are some manufacturing processes which are hazardous but still require high precision, such as nuclear power plants and oil pipelines. In this case, nano-robots can be used in those dangerous environments~\cite{Akyildiz20}. 

\subsection{Smart Home}
A smart home contains different kinds of IoT devices and AI-driven in-network services to remotely control the household systems like lighting, furniture, and thermostats~\cite{Peltonen20}. Current smart homes have been able to control the furniture and house environment according to people's commands. In the future, 6G will allow the household systems to be smarter, like providing adaptive real-time control without much human intervention. In addition to the convenience, residents' safety and privacy will also be well protected using 6G technologies such as federated learning.

\textbf{Intelligent Furniture.} Intelligent furniture in the smart home will facilitate people's life as well as save energy. To begin with, an intelligent light will switch off when nobody occupies the room, and the light intensity can be continuously tuned according to sunlight intensity~\cite{Nayak20Communications}. Over time, this will save substantial electrical energy while people do not have to pay extra attention to this matter. Similarly, the air conditioner can work based on the indoor temperature and occupancy detection. Furthermore, the running data of furniture can be recorded on the cloud side so that individual preferences can be learned by AI algorithms. Under this circumstance, the intelligent furniture will be tuned with the considering of individual preferences, thus leading to a comfortable home. Noted that real-time control, occupancy detection, and individual preferences estimation will generate a large amount of data and requires higher capacity requirements. Therefore, 6G is essential for the implementation of intelligent furniture.

\textbf{Emergency Detection.} 6G will aid the smart homes in emergency detection so as to keep residents' safe. For example, fall detection of the elderly is a major public health challenge~\cite{Noury07}. If an old man suddenly falls down in a smart home, the data collected by embedded intelligent sensors and video surveillance will be sent to the cloud by 6G wireless network immediately. The well-trained prediction model on the cloud side should detect this emergency by analysing the data and then send distress signals to the man's relatives and the ambulance. Other kinds of emergencies, such as forced entry and fire, can be detected easily by a similar method.

\textbf{Privacy Protection.} Privacy sensitive data are frequently transmitted in the smart home, hence requiring a reliable privacy protection approach. Some researchers tackled the privacy problems through edge-native solutions~\cite{Peltonen20}, which means data storage and processing will be done within the residents' premises. Mao~\emph{et al.}~\cite{Mao20AI} propose an AI based adaptive security specification method for 6G IoT networks to address the privacy problems, and the proposed method has been evaluated in a smart room.

\subsection{Intelligent Transportation System}
The intelligent transportation system (ITS) utilizes advanced communication, control, and sensing technologies to provide safer and efficient traffic and traffic management. In a ITS, autonomous driving vehicles require reliability above 99.99999$\%$ and latency below 1 ms, while the vehicle speed in some cases can be as high as to 1000 km/h~\cite{giordani2019towards,strinati20196g}. However, the ITS fails to meet these requirements in such high mobility scenarios because of the insufficient capability of the current communication technologies. In contrast, 6G network will significantly improve the capability of the ITS and make it satisfy the strict requirements. 

\textbf{Traffic Management.} An effective traffic management approach can force down the traffic jams, reduce passengers' waiting time, and preserve the road security. To provide real time transportation planning, the traffic information has to be collected in high data rates. Besides, the global optimal solution will be obtained only if the coverage of the mobile communication network is large enough. Thus, it is necessary to apply 6G technology in traffic management owing to its high speed Internet, low latency, and extensive coverage~\cite{Nayak20Communications}. The ITS empowered by 6G will keep guiding the drivers so as to minimize the travel time. It is also promising to investigate 6G for traffic signal control problems since traffic signal control involves a massive amount of real time traffic data and requires sophisticated algorithms to make decisions~\cite{Zheng19}. Last but not least, 6G technology will enhance the public traffic security. Police can utilize vehicle surveillance to track a suspected vehicle~\cite{Nayak20Communications}. The parameters and components of the vehicles will be continuously monitored to ensure safe driving. 

\textbf{Autonomous Vehicles.} Autonomous vehicles are key applications in the ITS~\cite{Bizon14}. Compared with the traffic management, the implementation of autonomous vehicles requires even higher data rates~\cite{rajatheva2020white}. 6G will help autonomous vehicles overcome the physical barriers and realize full automation. Advanced communication methods, such as dedicated short-range communication (DSRC),  vehicle-to-network (V2N), vehicle-to-infrastructure (V2I), vehicle-to-Pedestrian (V2P), Vehicle-to-Home (V2H), and vehicle-to-everything (V2X), have great potential to formulate a comprehensive autonomous vehicle network~\cite{chowdhury2019role}. Besides, novel AI approaches, like real-time intelligent edge, are indispensable for vehicle networks implementation as they enable the autonomous vehicles to react to the unfamiliar environment in real time~\cite{Wang20Security}. In addition to improving the speed, users' privacy will be better protected in the next generation autonomous vehicles. For example, a Efficient and Privacy-preserving Truth Discovery (EPTD) method is developed to strengthen the privacy protection~\cite{Sun20}. Also, researchers solve the security problems of vehicular Ad-hoc Networks (VANETs) by designing a privacy preserving machine learning-based collaborative intrusion detection system~\cite{Sun20, Zhang18}.

\begin{figure}[!h]
    \centering
    \includegraphics[scale=0.45]{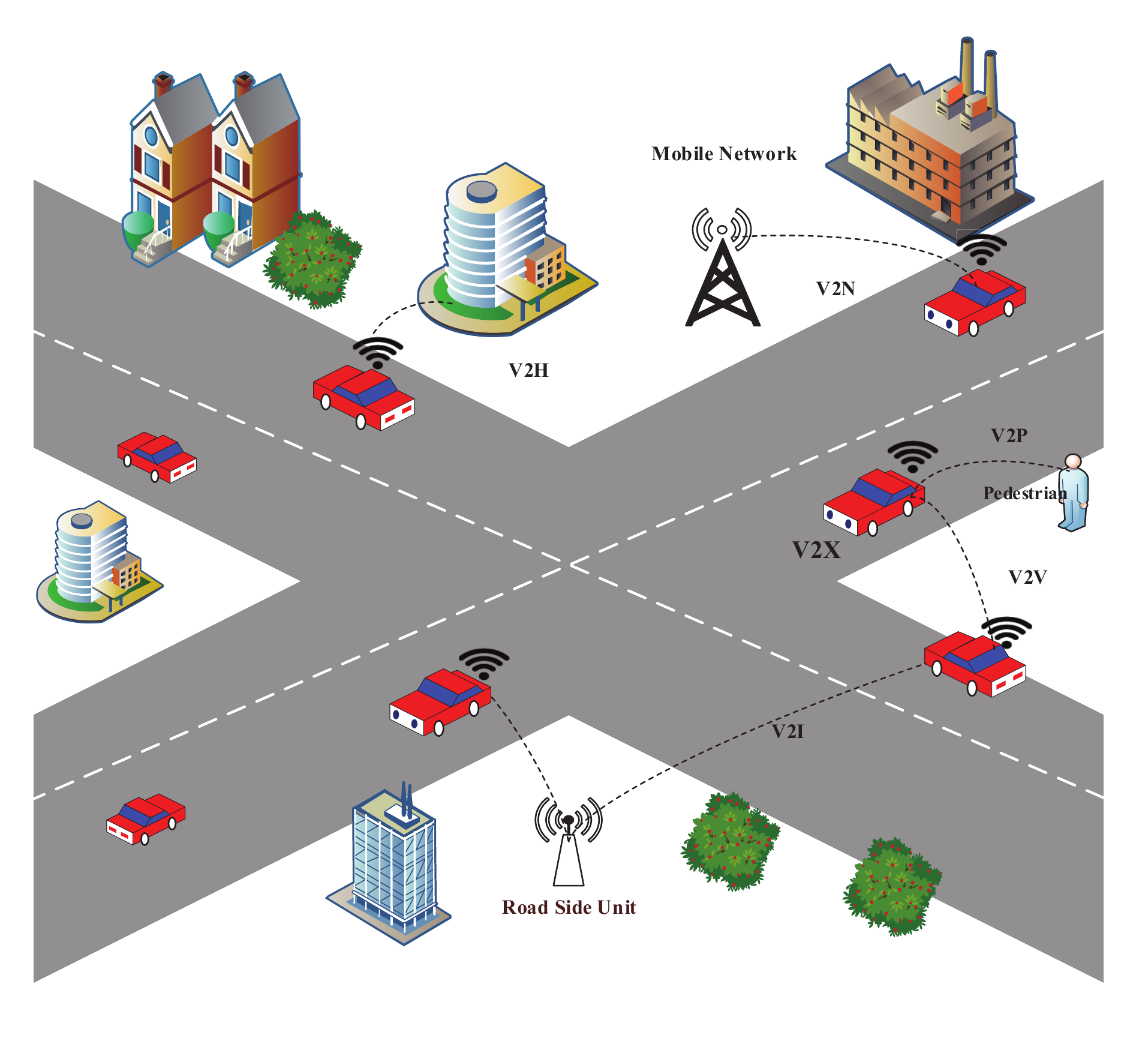}
    \caption{Vehicle-to-Everything (V2X).}
    \label{fig:v2v}
\end{figure}

\textbf{Airport and Waterway Transportation.} In addition to the land transportation, the applicability of 6G technology will expand into airport and waterway transportation. By integrating with satellite communication, 6G can provide localization services, broadcast, Internet connectivity, and weather information to cellular users~\cite{piran2019learning}. Satellite communication has potential benefits, such as providing readily connection to moving objects, and it is expected to be used in wireless network architectures in the future. Specifically,  the global coverage provided by 6G enables the ships and airplanes to be connected with a variety of IoT devices, which leads to a more intelligent transportation system. With the ubiquitous connectivity, 6G can keep updating the weather condition to the captains, which ensures the transportation safety~\cite{Nayak20Communications}. 
Besides, the satellite communication can be applied to connect the land, air, and sea together into an integrated 6G system~\cite{Dizdar20}. For instance, AANETs, which is proposed in~\cite{Zhang19}, demonstrated the feasibility of satellite communication in transport network. Moreover, optical wireless communication, such as free space optics (FSO)~\cite{chowdhury2019role} and visible light communication (VLC)~\cite{rajatheva2020white}, will also be useful in airport and waterway transportation.

\subsection{Smart Grid}
Smart grid is an IoT-based electricity network that utilizes advanced communication and AI methods to deliver power in more efficient ways~\cite{Fang12}. Researchers have been working on integrating 5G into the smart grid~\cite{ho2019next}, yet few people have considered applying 6G communication into this area. In the near future, however, 6G will be an indispensable technology for the further development of smart grid due to the fact that smart grid systems will demand more extensive computations and higher data rates.

In a smart grid system, all activities and electrical equipment should be supervised in order to make sure the system runs smoothly and safely. 6G enables smart grid systems that contain a great number of IoT devices to conduct real-time remote monitoring and control. Moreover, because of ultra reliability and low latency of 6G, the electricity network will be able to detect the fault quickly and then take actions in time. In addition, by employing 6G technology, the scale of the smart grid can be greatly extended without sacrificing the control precision and increasing the communication latency.

\subsection{Unmanned Aerial Vehicle}

The operation of unmanned aerial vehicle (UAV) requires either manual control or autonomous control by intelligent algorithms. Both two control approaches need exchange large amounts of data every second. Moreover, UAV is usually expected to carry out tasks in a long distance at high altitudes, which requires the wireless network to have high data rates and large coverage. Even though UAV cannot be applied successfully in the 5G network, 6G technology will facilitate the implementation of UAV due to the high capabilities of 6G~\cite{Wang20Security}.

UAV will be used as an aerial base station (BS) in 6G wireless communication owing to its aerial superiority. By using  Drone-to-Infrastructures (D2I) and Drone-to-Drone (D2D)  communication, the 6G network will maintain high data rate while extending the wireless coverage~\cite{Nayak20Communications}. Besides the coverage, UAV has several advantages over fixed BS infrastructures, such as  strong line-of-sight links, easy deployment, and high mobility~\cite{tariq2019speculative, Mehta20}. Therefore, UAV can be used as a flexible and low-latency BS in the areas where the infrastructures are absent or heavily loaded. For instance, natural disasters can destroy the ground communication infrastructures, and it is very difficult to establish new infrastructures in such a hazardous place~\cite{chowdhury20196g}. In this case, UAV-supported aerial base station will provide stable wireless connectivity.

\subsection{Smart and Autonomous Communication Systems}

In this section, we present some potential use cases of AI in 6G such as AI in network management and AI in autonomy. As 6G network becomes complex, it may utilize deep learning instead of human operators to improve the flexibility and efficiency in the network management~\cite{piran2019learning}. AI technologies are applicable to both the physical and network layers. In physical layer, AI techniques have involved in design and resource allocation in wireless communication~\cite{ho2019next}.  For example, unsupervised learning are applicable to channel-aware feature-extraction, optimal modulation, interference cancellation, and channel estimation, etc.~\cite{piran2019learning}. Deep reinforcement learning is possible to be employed for link preservation, scheduling,  transmission optimization, on-demand beamforming, and energy harvesting, etc.~\cite{piran2019learning, shafin2019artificial}. In addition, AI technologies can be used to the network layer as well. Supervised learning techniques can tackle problems such as resource allocation, fault prediction, etc.~\cite{piran2019learning}. Besides, unsupervised learning algorithms can help in routing, traffic control, parameter prediction, resource allocations, etc.~\cite{piran2019learning}. Reinforcement learning can be important for traffic prediction, packet scheduling, multi-objective routing, security, and classification, etc.~\cite{piran2019learning, shafin2019artificial}.

In addition, AI technologies have potentials to enable 6G wireless systems to be autonomous~\cite{loven2019edgeai,gacanin2019autonomous,zhang20196g}. Agents with intelligence can detect and resolve network issues actively and autonomously. AI-based network management contributes to  monitoring network status in real-time and keep network healthy. Also, AI techniques can provide intelligence at the edge devices and edge computing, which enables edge devices and edge computing to learn to solve security problems autonomously~\cite{porambagesec, mollah2017secure, loven2019edgeai}. Besides, autonomous applications such as autonomous aerial vehicles and autonomous robots are envisioned to be available in 6G~\cite{ho2019next}. 

\subsection{Intelligent Vehicle-to-Everything Communications}

The vehicular network builds the bridge between human beings and transportation~\cite{tang2019future}, for example, vehicle-to-vehicle shown in Fig.~(\ref{fig:v2v}). As the number of vehicles is increasing rapidly, the over crowded vehicular network fails to achieve high latency and low reliability. While traditional vehicular networks attach great attention to vehicle-to-vehicle (V2V) and vehicle-to-infrastructure (V2I) communications, the 6G vehicular network will realize space-air-ground-sea even underwater vehicles.

Although 5G technology has spanned over  cognitive radio (CR), network function virtualization (NFV), and reactive vehicular network control, they cannot meet the requirements of 6G communications. 6G requires to evolve to network intelligentization, intelligent radio, and self-learning with proactive exploration. Aided by advanced automation techniques and sensitive collision avoidance ability,  vehicular networks's performance can be significantly enhanced in 6G era.

\subsection{Radio Access Network}
A radio access network (RAN) connects a device to other parts of a network by radio access technology. RAN is of utmost importance as it can be applied to different kinds of AIoT applications such as smart healthcare, connected vehicles, and public surveillance. To integrate RAN into 6G network, Lee~\emph{et al.}~\cite{Lee19arx} elaborate three critical performance requirements for 6G RAN, flexibility, massive interconnectivity, and energy efficiency.

Moreover, several variants of RAN have great application potential in the 6G era. For example, open-radio access network (O-RAN) alliance, which combines extensible RAN (xRAN) forum and cloud RAN (C-RAN) alliance, is proposed in~\cite{Solmaz20}. O-RAN adopts virtualized network elements and open interfaces to integrate intelligence into RAN, thus enabling itself to support 6G technology. Fadlullah~\emph{et al.}~\cite{Fadlullah20} propose an all-photonic RAN that contains two key elements: a photonic engine (PE) and an all-photonic arrayed antenna unit (AAU). All-photonic RAN is envisioned to tackle sensing tasks and services when deploying in 6G networks. 

\subsection{Energy-efficient Networks}

As a wide range of IoT devices will be involved in the future, the implementation of 6G networks will further exacerbate e-waste management challenges. Therefore, it is vital to consider energy consumption for 6G networks.

Ali~\emph{et al.}~\cite{ali20206g} demonstrates several ways to build energy-efficient 6G networks. For example, the MAC layer is the best layer of the system to perform energy conversation. Besides, long battery life is also a crucial factor for designing energy-efficient 6G Networks. In addition, transmit power control or retransmissions is a promising method to enhance 6G networks' energy efficiency. In~\cite{Karan20}, the authors tackle the energy conversation problem in 6G networks through intelligent resource management. In~\cite{strinati20196g}, the authors anticipate that effective energy-efficient communication strategies will be developed in 6G networks. The strategies are expected to achieve battery-free communications, targeting communication efficiency in the order of 1 pJ/bit~\cite{Mudasar19}. Last but not least, Dean Bubley envisions that "energy budget" should be tied closely to costs (including externalities) in a wide variety of fields~\cite{Deanweb}.

\subsection{Holography Radio}

Holography radio is the highest level of interference exploitation and it improves spectrum efficiency and network capacity by controlling  the entire physical space and the full loop of the electromagnetic field through spatial spectral holography and spatial wave field synthesis~\cite{huang2020holographic, williams2020communication, saad2019vision}. Specifically, unwanted signals are treated as noises, and people try to reduce the interference caused by these noises. However, in 6G, the interference is regarded as useful resources for developing holographic communication systems~\cite{zong20196g}. Interference exploitation is that communication system  obtains gains through decomposing interference.  According to~\cite{meng2020interference}, the multi-user interference can be decomposed into constructive and destructive parts using simple geometric relations. Constructive part is considered as beneficial communication resources, which  can be used to improve QoS of 6G communication systems.  In recent years, MIMO is gaining its popularity because of its high throughput. Thus, 6G is expected to realize holographic MIMOs (HMIMOS) by combining MIMOS with LIS or IRS. HMIMOS can be categorized as active HMIMOS and passive HMIMOS based on the power consumption, which are supported by LIS and IRS, respectively~\cite{huang2020holographic}. To be more specific, active HMIMOS using LIS are equipped with  RF circuits and signal processing units, whereas passive HMIMOS only use IRS for reflecting signals.

\section{Conclusion}~\label{sec:conclusion}
In this paper, we highlight some promising technologies in 6G networks. We present a detailed explanation of artificial intelligence, intelligent reflecting surfaces, SWIFT,  THz communications, blockchain, space-air-ground-sea integrated network, free-duplex technologies and how these technologies will be applied in 6G. In addition, we discuss the potential security and privacy problems brought by these technologies. Moreover, we envision that 6G will enable a large sum of new application in facilitating our daily life.

\bibliographystyle{IEEEtran}
\bibliography{related}

\begin{thebibliography}{100}
\providecommand{\url}[1]{#1}
\csname url@samestyle\endcsname
\providecommand{\newblock}{\relax}
\providecommand{\bibinfo}[2]{#2}
\providecommand{\BIBentrySTDinterwordspacing}{\spaceskip=0pt\relax}
\providecommand{\BIBentryALTinterwordstretchfactor}{4}
\providecommand{\BIBentryALTinterwordspacing}{\spaceskip=\fontdimen2\font plus
\BIBentryALTinterwordstretchfactor\fontdimen3\font minus
  \fontdimen4\font\relax}
\providecommand{\BIBforeignlanguage}[2]{{%
\expandafter\ifx\csname l@#1\endcsname\relax
\typeout{** WARNING: IEEEtran.bst: No hyphenation pattern has been}%
\typeout{** loaded for the language `#1'. Using the pattern for}%
\typeout{** the default language instead.}%
\else
\language=\csname l@#1\endcsname
\fi
#2}}
\providecommand{\BIBdecl}{\relax}
\BIBdecl

\bibitem{5Gcountries}
J.~Wills, ``5{G} technology: Which country will be the first to adapt?''
  \url{https://www.investopedia.com/articles/markets-economy/090916/5g-technology-which-country-will-be-first-adapt.asp},
  April 23, 2020.

\bibitem{matti2019key}
L.~Matti and L.~Kari, ``Key drivers and research challenges for 6{G} ubiquitous
  wireless intelligence,'' \emph{6{G} Flagship, Oulu, Finland, White Paper},
  2019.

\bibitem{chowdhury20196g}
M.~Z. Chowdhury, M.~Shahjalal, S.~Ahmed, and Y.~M. Jang, ``{6G} wireless
  communication systems: Applications, requirements, technologies, challenges,
  and research directions,'' \emph{IEEE Open Journal of the Communications
  Society}, vol.~1, pp. 957--975, 2020.

\bibitem{saad2019vision}
W.~Saad, M.~Bennis, and M.~Chen, ``A vision of {6G} wireless systems:
  Applications, trends, technologies, and open research problems,'' \emph{IEEE
  network}, vol.~34, no.~3, pp. 134--142, 2019.

\bibitem{ho2019next}
T.~M. Ho, T.~D. Tran, T.~T. Nguyen, S.~Kazmi, L.~B. Le, C.~S. Hong, and
  L.~Hanzo, ``Next-generation wireless solutions for the smart factory, smart
  vehicles, the smart grid and smart cities,'' \emph{arXiv preprint
  arXiv:1907.10102}, 2019.

\bibitem{mollah2019emerging}
M.~B. Mollah, S.~Zeadally, and M.~A.~K. Azad, ``Emerging wireless technologies
  for {I}nternet of {T}hings applications: Opportunities and challenges,'' in
  \emph{Encyclopedia of Wireless Networks}.\hskip 1em plus 0.5em minus
  0.4em\relax Springer International Publishing Cham, 2019, pp. 1--11.

\bibitem{loven2019edgeai}
L.~Lov{\'e}n, T.~Lepp{\"a}nen, E.~Peltonen, J.~Partala, E.~Harjula,
  P.~Porambage, M.~Ylianttila, and J.~Riekki, ``Edge{AI}: A vision for
  distributed, edge-native artificial intelligence in future 6{G} networks,''
  \emph{The 1st 6{G} Wireless Summit}, pp. 1--2, 2019.

\bibitem{david2019defining}
K.~David, J.~Elmirghani, H.~Haas, and X.-H. You, ``Defining 6{G}: Challenges
  and opportunities [from the guest editors],'' \emph{IEEE Vehicular Technology
  Magazine}, vol.~14, no.~3, pp. 14--16, 2019.

\bibitem{tariq2019speculative}
F.~Tariq, M.~R. Khandaker, K.-K. Wong, M.~A. Imran, M.~Bennis, and M.~Debbah,
  ``A speculative study on {6G},'' \emph{IEEE Wireless Communications},
  vol.~27, no.~4, pp. 118--125, 2020.

\bibitem{giordani2019towards}
M.~Giordani, M.~Polese, M.~Mezzavilla, S.~Rangan, and M.~Zorzi, ``Toward {6G}
  networks: Use cases and technologies,'' \emph{IEEE Communications Magazine},
  vol.~58, no.~3, pp. 55--61, 2020.

\bibitem{xiao2017millimeter}
M.~Xiao, S.~Mumtaz, Y.~Huang, L.~Dai, Y.~Li, M.~Matthaiou, G.~K. Karagiannidis,
  E.~Bj{\"o}rnson, K.~Yang, I.~Chih-Lin, and A.~Ghosh, ``Millimeter wave
  communications for future mobile networks,'' \emph{IEEE Journal on Selected
  Areas in Communications}, vol.~35, no.~9, pp. 1909--1935, 2017.

\bibitem{andrews2016modeling}
J.~G. Andrews, T.~Bai, M.~N. Kulkarni, A.~Alkhateeb, A.~K. Gupta, and R.~W.
  Heath, ``Modeling and analyzing millimeter wave cellular systems,''
  \emph{IEEE Transactions on Communications}, vol.~65, no.~1, pp. 403--430,
  2016.

\bibitem{zhu2019millimeter}
L.~Zhu, Z.~Xiao, X.-G. Xia, and D.~O. Wu, ``Millimeter-wave communications with
  non-orthogonal multiple access for {B}5{G}/6{G},'' \emph{IEEE Access},
  vol.~7, pp. 116\,123--116\,132, 2019.

\bibitem{nawaz2019quantum}
S.~J. Nawaz, S.~K. Sharma, S.~Wyne, M.~N. Patwary, and M.~Asaduzzaman,
  ``Quantum machine learning for 6{G} communication networks: State-of-the-art
  and vision for the future,'' \emph{IEEE Access}, vol.~7, pp.
  46\,317--46\,350, 2019.

\bibitem{corre2019sub}
Y.~Corre, G.~Gougeon, J.-B. Dor{\'e}, S.~Bica{\"\i}s, B.~Miscopein,
  E.~Faussurier, M.~Saad, J.~Palicot, and F.~Bader, ``{Sub-THz} spectrum as
  enabler for 6{G} wireless communications up to 1 {Tbit/s},'' 2019.

\bibitem{giordani2020satellite}
M.~Giordani and M.~Zorzi, ``Satellite communication at millimeter waves: {A}
  key enabler of the 6{G} era,'' in \emph{2020 International Conference on
  Computing, Networking and Communications (ICNC)}.\hskip 1em plus 0.5em minus
  0.4em\relax IEEE, 2020, pp. 383--388.

\bibitem{underwood2016blockchain}
S.~Underwood, ``Blockchain beyond bitcoin,'' 2016.

\bibitem{piran2019learning}
M.~J. Piran and D.~Y. Suh, ``Learning-{D}riven wireless communications, towards
  6{G},'' in \emph{2019 International Conference on Computing, Electronics and
  Communications Engineering (ICCECE)}.\hskip 1em plus 0.5em minus 0.4em\relax
  IEEE, 2019, pp. 219--224.

\bibitem{dang2020should}
S.~Dang, O.~Amin, B.~Shihada, and M.-S. Alouini, ``What should 6{G} be?''
  \emph{Nature Electronics}, vol.~3, no.~1, pp. 20--29, 2020.

\bibitem{nayak20206g}
S.~Nayak and R.~Patgiri, ``6{G}: Envisioning the key issues and challenges,''
  \emph{arXiv preprint arXiv:2004.04024}, 2020.

\bibitem{kato2020ten}
N.~Kato, B.~Mao, F.~Tang, Y.~Kawamoto, and J.~Liu, ``Ten challenges in
  advancing machine learning technologies toward 6{G},'' \emph{IEEE Wireless
  Communications}, 2020.

\bibitem{shafin2020artificial}
R.~Shafin, L.~Liu, V.~Chandrasekhar, H.~Chen, J.~Reed, and J.~C. Zhang,
  ``Artificial intelligence-enabled cellular networks: A critical path to
  beyond-5{G} and 6{G},'' \emph{IEEE Wireless Communications}, vol.~27, no.~2,
  pp. 212--217, 2020.

\bibitem{tan2020thz}
J.~Tan and L.~Dai, ``{TH}z precoding for 6{G}: Applications, challenges,
  solutions, and opportunities,'' \emph{arXiv preprint arXiv:2005.10752}, 2020.

\bibitem{elmeadawy2020enabling}
S.~Elmeadawy and R.~M. Shubair, ``Enabling technologies for 6{G} future
  wireless communications: Opportunities and challenges,'' \emph{arXiv preprint
  arXiv:2002.06068}, 2020.

\bibitem{gui20206g}
G.~Gui, M.~Liu, F.~Tang, N.~Kato, and F.~Adachi, ``6{G}: Opening new horizons
  for integration of comfort, security and intelligence,'' \emph{IEEE Wireless
  Communications}, 2020.

\bibitem{chen2020vision}
S.~Chen, Y.-C. Liang, S.~Sun, S.~Kang, W.~Cheng, and M.~Peng, ``Vision,
  requirements, and technology trend of 6{G}: how to tackle the challenges of
  system coverage, capacity, user data-rate and movement speed,'' \emph{IEEE
  Wireless Communications}, vol.~27, no.~2, pp. 218--228, 2020.

\bibitem{xiaohutowards}
\BIBentryALTinterwordspacing
X.~You, C.~Wang, J.~Huang, X.~Gao, Z.~Zhang, M.~Wang, Y.~Huang, C.~Zhang,
  Y.~Jiang, J.~Wang, M.~Zhu, B.~Sheng, D.~Wang, Z.~Pan, P.~Zhu, Y.~Yang,
  Z.~Liu, P.~Zhang, X.~Tao, S.~Li, Z.~Chen, X.~Ma, C.~I, S.~Han, K.~Li, C.~Pan,
  Z.~Zheng, L.~Hanzo, X.~Shen, Y.~J. Guo, Z.~Ding, H.~Haas, W.~Tong, P.~Zhu,
  G.~Yang, J.~Wang, E.~G. Larsson, H.~Ngo, W.~Hong, H.~Wang, D.~Hou, J.~Chen,
  Z.~Chen, Z.~Hao, G.~Li, R.~Tafazolli, Y.~Gao, V.~Poor, G.~Fettweis, and
  Y.~Liang, ``Towards 6{G} wireless communication networks: Vision, enabling
  technologies, and new paradigm shifts,'' \emph{SCIENCE CHINA Information
  Sciences}. [Online]. Available:
  \url{https://engine.scichina.com/publisher/Science China
  Press/journal/SCIENCE CHINA Information Sciences///10.1007/s11432-020-2955-6}
\BIBentrySTDinterwordspacing

\bibitem{Hu2018LIS}
S.~Hu, F.~Rusek, and O.~Edfors, ``Beyond massive {MIMO}: the potential of data
  transmission with large intelligent surfaces,'' \emph{IEEE Transactions on
  Signal Processing}, vol.~66, no.~10, pp. 2746--2758, 2018.

\bibitem{Hu2017LIS}
------, ``The potential of using large antenna arrays on intelligent
  surfaces,'' in \emph{IEEE 2017 Vehicular Technology Conference (VTC)}.\hskip
  1em plus 0.5em minus 0.4em\relax IEEE, 2017, pp. 1--6.

\bibitem{Wu2019IRS}
Q.~Wu and R.~Zhang, ``Intelligent reflecting surface enhanced wireless network
  via joint active and passive beamforming,'' \emph{IEEE Transactions on
  Wireless Communications}, vol.~18, no.~11, pp. 5394--5409, 2016.

\bibitem{sarieddeen2019next}
H.~Sarieddeen, N.~Saeed, T.~Y. Al-Naffouri, and M.-S. Alouini, ``Next
  generation terahertz communications: A rendezvous of sensing, imaging, and
  localization,'' \emph{IEEE Communications Magazine}, vol.~58, no.~5, pp.
  69--75, 2020.

\bibitem{liang2020symbiotic}
Y.-C. Liang, Q.~Zhang, E.~G. Larsson, and G.~Y. Li, ``Symbiotic radio:
  Cognitive backscattering communications for future wireless networks,''
  \emph{IEEE Transactions on Cognitive Communications and Networking}, vol.~6,
  no.~4, pp. 1242--1255, 2020.

\bibitem{lu20206g}
Y.~Lu and X.~Zheng, ``6{G}: A survey on technologies, scenarios, challenges,
  and the related issues,'' \emph{Journal of Industrial Information
  Integration}, p. 100158, 2020.

\bibitem{zhao20196g}
Y.~Zhao, G.~Yu, and H.~Xu, ``6{G} mobile communication network: vision,
  challenges and key technologies,'' \emph{arXiv preprint arXiv:1905.04983},
  2019.

\bibitem{akhtar2020shift}
M.~W. Akhtar, S.~A. Hassan, R.~Ghaffar, H.~Jung, S.~Garg, and M.~S. Hossain,
  ``The shift to 6g communications: vision and requirements,''
  \emph{Human-centric Computing and Information Sciences}, vol.~10, no.~1, pp.
  1--27, 2020.

\bibitem{White2020}
W.~paper, ``{5G} evolution and {6G},'' in \emph{NTT DOCOMO, INC.}, 2020.

\bibitem{xu2020blockchain}
H.~Xu, P.~V. Klainea, O.~Oniretia, B.~Caob, M.~Imrana, and L.~Zhang,
  ``Blockchain-enabled resource management and sharing for {6G}
  communications,'' \emph{arXiv preprint arXiv:2003.13083}, 2020.

\bibitem{fan2017blockchain}
K.~Fan, Y.~Ren, Y.~Wang, H.~Li, and Y.~Yang, ``Blockchain-based efficient
  privacy preserving and data sharing scheme of content-centric network in
  {5G},'' \emph{IET Communications}, vol.~12, no.~5, pp. 527--532, 2017.

\bibitem{kotobi2018secure}
K.~Kotobi and S.~G. Bilen, ``Secure blockchains for dynamic spectrum access:
  {A} decentralized database in moving cognitive radio networks enhances
  security and user access,'' \emph{ieee vehicular technology magazine},
  vol.~13, no.~1, pp. 32--39, 2018.

\bibitem{yang2017blockchain}
H.~Yang, H.~Zheng, J.~Zhang, Y.~Wu, Y.~Lee, and Y.~Ji, ``Blockchain-based
  trusted authentication in cloud radio over fiber network for {5G},'' in
  \emph{2017 16th International Conference on Optical Communications and
  Networks (ICOCN)}.\hskip 1em plus 0.5em minus 0.4em\relax IEEE, 2017, pp.
  1--3.

\bibitem{Chau2001SSK}
Y.~A. Chau and S.~H. Yu, ``Space modulation on wireless fading channels,'' in
  \emph{IEEE 54th Vehicular Technology Conference. VTC Fall 2001.
  Proceedings}.\hskip 1em plus 0.5em minus 0.4em\relax IEEE, 2001, pp.
  1668--1671.

\bibitem{Wang2014SWIPT}
H.~Wang, W.~Wang, and X.~Chen, ``Wireless information and energy transfer in
  interference aware massive {MIMO} systems,'' in \emph{2014 IEEE Global
  Communications Conference}.\hskip 1em plus 0.5em minus 0.4em\relax IEEE,
  2014, pp. 2556--2561.

\bibitem{Zhang2015SWIPT}
R.~Zhang, R.~G. Maunder, and L.~Hanzo, ``Wireless information and power
  transfer: from scientific hypothesis to engineering practice,'' \emph{IEEE
  Communications Magazine}, vol.~53, no.~8, pp. 99--105, 2015.

\bibitem{Liu2016SWIPT}
Y.~Liu, Z.~Ding, and M.~Elkashlan, ``Cooperative non-orthogonal multiple access
  with simultaneous wireless information and power transfer,'' \emph{IEEE
  Journals on Selected Areas in Communications}, vol.~34, no.~4, pp. 938--953,
  2016.

\bibitem{Yang2017SWIPT}
Z.~Yang, Z.~Ding, P.~Fan, and N.~Al-Dhahir, ``The impact of power allocation on
  cooperative non-orthogonal multiple access networks with {SWIPT},''
  \emph{IEEE Transactions on Wireless Communications}, vol.~16, no.~7, pp.
  4332--4343, 2017.

\bibitem{Gong2017SWIPT}
J.~Gong and X.~Chen, ``Achievable rate region of non-orthogonal multiple access
  systems with wireless powered decoder,'' \emph{IEEE Journals on Selected
  Areas in Communications}, vol.~35, no.~12, pp. 2846--2859, 2017.

\bibitem{Alsaba2018SWIPT}
Y.~Alsaba, C.~Y. Leow, and S.~K.~A. Rahim, ``Full-duplex cooperative
  non-orthogonal multiple access with beamforming and energy harvesting,''
  \emph{IEEE Access}, vol.~6, pp. 19\,726--19\,738, 2018.

\bibitem{Bariah2019SWIPT}
L.~Bariah, S.~Muhaidat, and A.~A. Dweik, ``Error probability analysis of
  {NOMA}-based relay networks with {SWIPT},'' \emph{IEEE Access}, vol.~23,
  no.~7, pp. 1223--1226, 2019.

\bibitem{Li2019SWIPT}
S.~Li, L.~Bariah, S.~Muhaidat, P.~Sofotasios, J.~Liang, and A.~Wang, ``Error
  analysis of {NOMA}-based user cooperation with {SWIPT},'' in \emph{2019 15th
  International Conference on Distributed Computing in Sensor Systems
  (DCOSS)}.\hskip 1em plus 0.5em minus 0.4em\relax IEEE, 2019, pp. 507--513.

\bibitem{haas2018lifi}
H.~Haas, ``{LiFi} is a paradigm-shifting {5G} technology,'' \emph{Reviews in
  Physics}, vol.~3, pp. 26--31, 2018.

\bibitem{strinati20196g}
E.~C. Strinati, S.~Barbarossa, J.~L. Gonzalez-Jimenez, D.~Ktenas, N.~Cassiau,
  L.~Maret, and C.~Dehos, ``{6G}: The next frontier: From holographic messaging
  to artificial intelligence using subterahertz and visible light
  communication,'' \emph{IEEE Vehicular Technology Magazine}, vol.~14, no.~3,
  pp. 42--50, 2019.

\bibitem{kar2020critical}
U.~N. Kar and D.~K. Sanyal, ``A critical review of {3GPP} standardization of
  device-to-device communication in cellular networks,'' \emph{SN Computer
  Science}, vol.~1, no.~1, p.~37, 2020.

\bibitem{malik2020survey}
P.~K. Malik, D.~S. Wadhwa, and J.~S. Khinda, ``A survey of device to device and
  cooperative communication for the future cellular networks,''
  \emph{International Journal of Wireless Information Networks}, pp. 1--22,
  2020.

\bibitem{asadi2014survey}
A.~Asadi, Q.~Wang, and V.~Mancuso, ``A survey on device-to-device communication
  in cellular networks,'' \emph{IEEE Communications Surveys \& Tutorials},
  vol.~16, no.~4, pp. 1801--1819, 2014.

\bibitem{zhang2014social}
Y.~Zhang, E.~Pan, L.~Song, W.~Saad, Z.~Dawy, and Z.~Han, ``Social network aware
  device-to-device communication in wireless networks,'' \emph{IEEE
  Transactions on Wireless Communications}, vol.~14, no.~1, pp. 177--190, 2014.

\bibitem{zhang2020beyond}
S.~Zhang, H.~Zhang, and L.~Song, ``Beyond {D2D}: Full dimension
  {UAV}-to-everything communications in {6G},'' \emph{IEEE Transactions on
  Vehicular Technology}, 2020.

\bibitem{sun2020machine}
Y.~Sun, J.~Liu, J.~Wang, Y.~Cao, and N.~Kato, ``When machine learning meets
  privacy in {6G}: A survey,'' \emph{IEEE Communications Surveys \& Tutorials},
  vol.~22, no.~4, pp. 2694--2724, 2020.

\bibitem{akhtar2018threat}
N.~Akhtar and A.~Mian, ``Threat of adversarial attacks on deep learning in
  computer vision: {A} survey,'' \emph{IEEE Access}, vol.~6, pp.
  14\,410--14\,430, 2018.

\bibitem{mcmahan2017communication}
B.~McMahan, E.~Moore, D.~Ramage, S.~Hampson, and B.~A. y~Arcas,
  ``Communication-efficient learning of deep networks from decentralized
  data,'' in \emph{Artificial Intelligence and Statistics}.\hskip 1em plus
  0.5em minus 0.4em\relax PMLR, 2017, pp. 1273--1282.

\bibitem{zhou2018robust}
Z.~Zhou, H.~Liao, B.~Gu, K.~M.~S. Huq, S.~Mumtaz, and J.~Rodriguez, ``Robust
  mobile crowd sensing: When deep learning meets edge computing,'' \emph{IEEE
  Network}, vol.~32, no.~4, pp. 54--60, 2018.

\bibitem{sattiraju2019ai}
R.~Sattiraju, A.~Weinand, and H.~D. Schotten, ``{AI}-assisted {PHY}
  technologies for 6{G} and beyond wireless networks,'' \emph{arXiv preprint
  arXiv:1908.09523}, 2019.

\bibitem{Letaief2019IEEE}
K.~B. Letaief, W.~Chen, Y.~Shi, J.~Zhang, , and Y.~A. Zhang, ``The roadmap to
  {6G}: {AI} empowered wireless networks,'' \emph{IEEE Communications
  Magazine}, vol.~57, no.~8, pp. 84--90, 2019.

\bibitem{Nayak2020Healthcare}
S.~Nayak and R.~Patgiri, ``{6G} communication technology: A vision on
  intelligent healthcare,'' \emph{arXiv preprint arXiv:2005.07532}, 2020.

\bibitem{rajatheva2020white}
N.~Rajatheva, I.~Atzeni, E.~Bjornson, A.~Bourdoux, S.~Buzzi, J.-B. Dore,
  S.~Erkucuk, M.~Fuentes, K.~Guan, Y.~Hu, X.~Huang, J.~Hulkkonen, J.~M. Jornet,
  M.~Katz, R.~Nilsson, E.~Panayirci, K.~Rabie, N.~Rajapaksha, M.~Salehi,
  H.~Sarieddeen, T.~Svensson, O.~Tervo, A.~Tolli, Q.~Wu, and W.~Xu, ``White
  paper on broadband connectivity in 6{G},'' \emph{arXiv preprint
  arXiv:2004.14247}, 2020.

\bibitem{Nayak20Communications}
S.~Nayak and R.~Patgiri, ``{6G} communications: A vision on the potential
  applications,'' \emph{arxiv}, 04 2020.

\bibitem{Fang12}
X.~Fang, S.~Misra, G.~Xue, and D.~Yang, ``Smart grid — the new and improved
  power grid: A survey,'' \emph{Communications Surveys \& Tutorials, IEEE},
  vol.~14, pp. 944--980, 01 2012.

\bibitem{Wang20Security}
M.~Wang, T.~Zhu, T.~Zhang, J.~Zhang, S.~Yu, and W.~Zhou, ``Security and privacy
  in {6G} networks: New areas and new challenges,'' \emph{Digital
  Communications and Networks}, vol.~6, 07 2020.

\bibitem{zhang2020towards}
S.~Zhang and D.~Zhu, ``Towards artificial intelligence enabled 6g: State of the
  art, challenges, and opportunities,'' \emph{Computer Networks}, p. 107556,
  2020.

\bibitem{elsayed2019ai}
M.~Elsayed and M.~Erol-Kantarci, ``A{I}-{E}nabled future wireless networks:
  Challenges, opportunities, and open issues,'' \emph{IEEE Vehicular Technology
  Magazine}, vol.~14, no.~3, pp. 70--77, 2019.

\bibitem{mao2018deep}
Q.~Mao, F.~Hu, and Q.~Hao, ``Deep learning for intelligent wireless networks: A
  comprehensive survey,'' \emph{IEEE Communications Surveys \& Tutorials},
  vol.~20, no.~4, pp. 2595--2621, 2018.

\bibitem{zappone2019model}
A.~Zappone, M.~Di~Renzo, M.~Debbah, T.~T. Lam, and X.~Qian, ``Model-aided
  wireless artificial intelligence: Embedding expert knowledge in deep neural
  networks for wireless system optimization,'' \emph{IEEE Vehicular Technology
  Magazine}, vol.~14, no.~3, pp. 60--69, 2019.

\bibitem{he2019model}
H.~He, S.~Jin, C.-K. Wen, F.~Gao, G.~Y. Li, and Z.~Xu, ``Model-driven deep
  learning for physical layer communications,'' \emph{IEEE Wireless
  Communications}, 2019.

\bibitem{zong20196g}
B.~Zong, C.~Fan, X.~Wang, X.~Duan, B.~Wang, and J.~Wang, ``6{G} technologies:
  Key drivers, core requirements, system architectures, and enabling
  technologies,'' \emph{IEEE Vehicular Technology Magazine}, vol.~14, no.~3,
  pp. 18--27, 2019.

\bibitem{letaief2019roadmap}
K.~B. Letaief, W.~Chen, Y.~Shi, J.~Zhang, and Y.-J.~A. Zhang, ``The roadmap to
  {6G}: {AI} empowered wireless networks,'' \emph{IEEE Communications
  Magazine}, vol.~57, no.~8, pp. 84--90, 2019.

\bibitem{gacanin2019autonomous}
H.~Gacanin, ``Autonomous wireless systems with artificial intelligence: A
  knowledge management perspective,'' \emph{IEEE Vehicular Technology
  Magazine}, vol.~14, no.~3, pp. 51--59, 2019.

\bibitem{stoica20196g}
R.-A. Stoica and G.~T.~F. de~Abreu, ``6{G}: the wireless communications network
  for collaborative and {AI} applications,'' \emph{arXiv preprint
  arXiv:1904.03413}, 2019.

\bibitem{zhang20196g}
Z.~Zhang, Y.~Xiao, Z.~Ma, M.~Xiao, Z.~Ding, X.~Lei, G.~K. Karagiannidis, and
  P.~Fan, ``6{G} wireless networks: Vision, requirements, architecture, and key
  technologies,'' \emph{IEEE Vehicular Technology Magazine}, vol.~14, no.~3,
  pp. 28--41, 2019.

\bibitem{zhangLin20196g}
L.~Zhang, Y.-C. Liang, and D.~Niyato, ``6{G} visions: Mobile ultra-broadband,
  super {I}nternet-of-{T}hings, and artificial intelligence,'' \emph{China
  Communications}, vol.~16, no.~8, pp. 1--14, 2019.

\bibitem{tang2019future}
F.~Tang, Y.~Kawamoto, N.~Kato, and J.~Liu, ``Future intelligent and secure
  vehicular network toward 6{G}: Machine-learning approaches,''
  \emph{Proceedings of the IEEE}, vol. 108, no.~2, pp. 292--307, 2019.

\bibitem{xiao2020towards}
Y.~Xiao, G.~Shi, Y.~Li, W.~Saad, and H.~V. Poor, ``Towards self-learning edge
  intelligence in {6G},'' \emph{arXiv preprint arXiv:2010.00176}, 2020.

\bibitem{jameel2020machine}
F.~Jameel, N.~Sharma, M.~A. Khan, I.~Khan, M.~M. Alam, G.~Mastorakis, and C.~X.
  Mavromoustakis, ``Machine learning techniques for wireless-powered ambient
  backscatter communications: Enabling intelligent {IoT} networks in {6G}
  era,'' in \emph{Convergence of Artificial Intelligence and the Internet of
  Things}.\hskip 1em plus 0.5em minus 0.4em\relax Springer, 2020, pp. 187--211.

\bibitem{guo2020explainable}
W.~Guo, ``Explainable artificial intelligence for {6G}: Improving trust between
  human and machine,'' \emph{IEEE Communications Magazine}, vol.~58, no.~6, pp.
  39--45, 2020.

\bibitem{du2020machine}
J.~Du, C.~Jiang, J.~Wang, Y.~Ren, and M.~Debbah, ``Machine learning for {6G}
  wireless networks: Carrying forward enhanced bandwidth, massive access, and
  ultrareliable/low-latency service,'' \emph{IEEE Vehicular Technology
  Magazine}, vol.~15, no.~4, pp. 122--134, 2020.

\bibitem{han2020artificial}
S.~Han, T.~Xie, I.~Chih-Lin, L.~Chai, Z.~Liu, Y.~Yuan, and C.~Cui,
  ``{Artificial-Intelligence-Enabled} air interface for {6G}: Solutions,
  challenges, and standardization impacts,'' \emph{IEEE Communications
  Magazine}, vol.~58, no.~10, pp. 73--79, 2020.

\bibitem{liu2019machine}
Y.~Liu, S.~Bi, Z.~Shi, and L.~Hanzo, ``When machine learning meets big data: A
  wireless communication perspective,'' \emph{IEEE Vehicular Technology
  Magazine}, vol.~15, no.~1, pp. 63--72, 2019.

\bibitem{liu2020data}
R.~W. Liu, J.~Nie, S.~Garg, Z.~Xiong, Y.~Zhang, and M.~S. Hossain,
  ``Data-driven trajectory quality improvement for promoting intelligent vessel
  traffic services in {6G}-enabled maritime {IoT} systems,'' \emph{IEEE
  Internet of Things Journal}, 2020.

\bibitem{yang2020artificial}
H.~Yang, A.~Alphones, Z.~Xiong, D.~Niyato, J.~Zhao, and K.~Wu,
  ``Artificial-intelligence-enabled intelligent {6G} networks,'' \emph{IEEE
  Network}, vol.~34, no.~6, pp. 272--280, 2020.

\bibitem{huang2020general}
Z.~Huang and X.~Cheng, ``A general 3d space-time-frequency non-stationary model
  for {6G} channels,'' \emph{IEEE Transactions on Wireless Communications},
  2020.

\bibitem{giordani2020toward}
M.~Giordani, M.~Polese, M.~Mezzavilla, S.~Rangan, and M.~Zorzi, ``Toward {6G}
  networks: Use cases and technologies,'' \emph{IEEE Communications Magazine},
  vol.~58, no.~3, pp. 55--61, 2020.

\bibitem{dai2019blockchain}
Y.~Dai, D.~Xu, S.~Maharjan, Z.~Chen, Q.~He, and Y.~Zhang, ``Blockchain and deep
  reinforcement learning empowered intelligent 5{G} beyond,'' \emph{IEEE
  Network}, vol.~33, no.~3, pp. 10--17, 2019.

\bibitem{tang2020deep}
F.~Tang, Y.~Zhou, and N.~Kato, ``Deep reinforcement learning for dynamic
  uplink/downlink resource allocation in high mobility {5G} hetnet,''
  \emph{IEEE Journal on Selected Areas in Communications}, vol.~38, no.~12, pp.
  2773--2782, 2020.

\bibitem{xiong2019deep}
Z.~Xiong, Y.~Zhang, D.~Niyato, R.~Deng, P.~Wang, and L.-C. Wang, ``Deep
  reinforcement learning for mobile {5G} and beyond: Fundamentals,
  applications, and challenges,'' \emph{IEEE Vehicular Technology Magazine},
  vol.~14, no.~2, pp. 44--52, 2019.

\bibitem{zhao2020deep}
R.~Zhao, X.~Wang, J.~Xia, and L.~Fan, ``Deep reinforcement learning based
  mobile edge computing for intelligent {Internet of Things},'' \emph{Physical
  Communication}, vol.~43, p. 101184, 2020.

\bibitem{al2020multiple}
Y.~Al-Eryani, M.~Akrout, and E.~Hossain, ``Multiple access in cell-free
  networks: Outage performance, dynamic clustering, and deep reinforcement
  learning-based design,'' \emph{IEEE Journal on Selected Areas in
  Communications}, 2020.

\bibitem{he2020beamspace}
H.~He, R.~Wang, S.~Jin, C.-K. Wen, and G.~Y. Li, ``Beamspace channel estimation
  in terahertz communications: A model-driven unsupervised learning approach,''
  \emph{arXiv preprint arXiv:2006.16628}, 2020.

\bibitem{huang2020reconfigurable}
C.~Huang, R.~Mo, C.~Yuen \emph{et~al.}, ``Reconfigurable intelligent surface
  assisted multiuser {MISO} systems exploiting deep reinforcement learning,''
  \emph{arXiv preprint arXiv:2002.10072}, 2020.

\bibitem{zhang2020uav}
Y.~Zhang, Z.~Mou, F.~Gao, J.~Jiang, R.~Ding, and Z.~Han, ``{UAV}-enabled secure
  communications by multi-agent deep reinforcement learning,'' \emph{IEEE
  Transactions on Vehicular Technology}, vol.~69, no.~10, pp. 11\,599--11\,611,
  2020.

\bibitem{rout20206g}
S.~P. Rout, ``6{G} wireless communication: Its vision, viability, application,
  requirement, technologies, encounters and research,'' in \emph{2020 11th
  International Conference on Computing, Communication and Networking
  Technologies (ICCCNT)}.\hskip 1em plus 0.5em minus 0.4em\relax IEEE, 2020,
  pp. 1--8.

\bibitem{du2020mec}
J.~Du, F.~R. Yu, G.~Lu, J.~Wang, J.~Jiang, and X.~Chu, ``{MEC}-assisted
  immersive {VR} video streaming over terahertz wireless networks: A deep
  reinforcement learning approach,'' \emph{IEEE Internet of Things Journal},
  vol.~7, no.~10, pp. 9517--9529, 2020.

\bibitem{wang2020multi}
L.~Wang, K.~Wang, C.~Pan, W.~Xu, N.~Aslam, and L.~Hanzo, ``Multi-agent deep
  reinforcement learning based trajectory planning for multi-{UAV} assisted
  mobile edge computing,'' \emph{IEEE Transactions on Cognitive Communications
  and Networking}, 2020.

\bibitem{lu2020low}
Y.~Lu, X.~Huang, K.~Zhang, S.~Maharjan, and Y.~Zhang, ``Low-latency federated
  learning and blockchain for edge association in digital twin empowered {6G}
  networks,'' \emph{IEEE Transactions on Industrial Informatics}, 2020.

\bibitem{aledhari2020federated}
M.~Aledhari, R.~Razzak, R.~M. Parizi, and F.~Saeed, ``Federated learning: A
  survey on enabling technologies, protocols, and applications,'' \emph{IEEE
  Access}, vol.~8, pp. 140\,699--140\,725, 2020.

\bibitem{yang2020federated}
K.~Yang, T.~Jiang, Y.~Shi, and Z.~Ding, ``Federated learning via over-the-air
  computation,'' \emph{IEEE Transactions on Wireless Communications}, vol.~19,
  no.~3, pp. 2022--2035, 2020.

\bibitem{khan2020federated}
L.~U. Khan, W.~Saad, Z.~Han, E.~Hossain, and C.~S. Hong, ``Federated learning
  for {Internet of Things}: Recent advances, taxonomy, and open challenges,''
  \emph{arXiv preprint arXiv:2009.13012}, 2020.

\bibitem{chen2020joint}
M.~Chen, Z.~Yang, W.~Saad, C.~Yin, H.~V. Poor, and S.~Cui, ``A joint learning
  and communications framework for federated learning over wireless networks,''
  \emph{IEEE Transactions on Wireless Communications}, 2020.

\bibitem{khan20206g}
L.~U. Khan, I.~Yaqoob, M.~Imran, Z.~Han, and C.~S. Hong, ``{6G} wireless
  systems: A vision, architectural elements, and future directions,''
  \emph{IEEE Access}, vol.~8, pp. 147\,029--147\,044, 2020.

\bibitem{qu2020empowering}
Y.~Qu, C.~Dong, J.~Zheng, Q.~Wu, Y.~Shen, F.~Wu, and A.~Anpalagan, ``Empowering
  the edge intelligence by air-ground integrated federated learning in {6G}
  networks,'' \emph{arXiv preprint arXiv:2007.13054}, 2020.

\bibitem{fadlullah2020hcp}
Z.~M. Fadlullah and N.~Kato, ``{HCP:} heterogeneous computing platform for
  federated learning based collaborative content caching towards {6G}
  networks,'' \emph{IEEE Transactions on Emerging Topics in Computing}, 2020.

\bibitem{zhao2020federated}
Z.~Zhao, C.~Feng, H.~H. Yang, and X.~Luo, ``Federated-learning-enabled
  intelligent fog radio access networks: Fundamental theory, key techniques,
  and future trends,'' \emph{IEEE Wireless Communications}, vol.~27, no.~2, pp.
  22--28, 2020.

\bibitem{yang2020energy}
Z.~Yang, M.~Chen, W.~Saad, C.~S. Hong, and M.~Shikh-Bahaei, ``Energy efficient
  federated learning over wireless communication networks,'' \emph{IEEE
  Transactions on Wireless Communications}, 2020.

\bibitem{bai2019deep}
Q.~Bai, J.~Wang, Y.~Zhang, and J.~Song, ``Deep learning-based channel
  estimation algorithm over time selective fading channels,'' \emph{IEEE
  Transactions on Cognitive Communications and Networking}, vol.~6, no.~1, pp.
  125--134, 2019.

\bibitem{qin2019deep}
Z.~Qin, H.~Ye, G.~Y. Li, and B.-H.~F. Juang, ``Deep learning in physical layer
  communications,'' \emph{IEEE Wireless Communications}, vol.~26, no.~2, pp.
  93--99, 2019.

\bibitem{wang2017deep}
T.~Wang, C.-K. Wen, H.~Wang, F.~Gao, T.~Jiang, and S.~Jin, ``Deep learning for
  wireless physical layer: Opportunities and challenges,'' \emph{China
  Communications}, vol.~14, no.~11, pp. 92--111, 2017.

\bibitem{doan2018content}
K.~N. Doan, T.~Van~Nguyen, T.~Q. Quek, and H.~Shin, ``Content-aware proactive
  caching for backhaul offloading in cellular network,'' \emph{IEEE
  Transactions on Wireless Communications}, vol.~17, no.~5, pp. 3128--3140,
  2018.

\bibitem{rajendran2019crowdsourced}
S.~Rajendran, V.~Lenders, W.~Meert, and S.~Pollin, ``Crowdsourced wireless
  spectrum anomaly detection,'' \emph{IEEE Transactions on Cognitive
  Communications and Networking}, vol.~6, no.~2, pp. 694--703, 2019.

\bibitem{yang2020machine}
H.~Yang, X.~Xie, and M.~Kadoch, ``Machine learning techniques and a case study
  for intelligent wireless networks,'' \emph{IEEE Network}, vol.~34, no.~3, pp.
  208--215, 2020.

\bibitem{huang2018unsupervised}
H.~Huang, W.~Xia, J.~Xiong, J.~Yang, G.~Zheng, and X.~Zhu, ``Unsupervised
  learning-based fast beamforming design for downlink {MIMO},'' \emph{IEEE
  Access}, vol.~7, pp. 7599--7605, 2018.

\bibitem{nikbakht2020unsupervised}
R.~Nikbakht, A.~Jonsson, and A.~Lozano, ``Unsupervised learning for parametric
  optimization,'' \emph{IEEE Communications Letters}, 2020.

\bibitem{ali20206g}
S.~Ali, W.~Saad, N.~Rajatheva, K.~Chang, D.~Steinbach, B.~Sliwa, C.~Wietfeld,
  K.~Mei, H.~Shiri, H.-J. Zepernick \emph{et~al.}, ``6{G} white paper on
  machine learning in wireless communication networks,'' \emph{arXiv preprint
  arXiv:2004.13875}, 2020.

\bibitem{konevcny2016federated}
J.~Kone{\v{c}}n{\`y}, H.~B. McMahan, F.~X. Yu, P.~Richt{\'a}rik, A.~T. Suresh,
  and D.~Bacon, ``Federated learning: Strategies for improving communication
  efficiency,'' \emph{arXiv preprint arXiv:1610.05492}, 2016.

\bibitem{shafin2019artificial}
R.~Shafin, L.~Liu, V.~Chandrasekhar, H.~Chen, J.~Reed, and J.~Zhang,
  ``Artificial intelligence-enabled cellular networks: A critical path to
  beyond-5{G} and 6{G},'' \emph{arXiv preprint arXiv:1907.07862}, 2019.

\bibitem{yang2021federated}
Z.~Yang, M.~Chen, K.-K. Wong, H.~V. Poor, and S.~Cui, ``Federated learning for
  6{G}: Applications, challenges, and opportunities,'' \emph{arXiv preprint
  arXiv:2101.01338}, 2021.

\bibitem{cousik2019cogrf}
T.~Cousik, R.~Shafin, Z.~Zhou, K.~Kleine, J.~Reed, and L.~Liu, ``{CogRF}: A new
  frontier for machine learning and artificial intelligence for 6{G} {RF}
  systems,'' \emph{arXiv preprint arXiv:1909.06862}, 2019.

\bibitem{guo2019explainable}
W.~Guo, ``Explainable artificial intelligence ({XAI}) for 6{G}: Improving trust
  between human and machine,'' \emph{arXiv preprint arXiv:1911.04542}, 2019.

\bibitem{faisal2019ultra}
A.~Faisal, H.~Sarieddeen, H.~Dahrouj, T.~Y. Al-Naffouri, and M.-S. Alouini,
  ``Ultra-massive {MIMO} systems at terahertz bands: Prospects and
  challenges,'' \emph{arXiv preprint arXiv:1902.11090}, 2019.

\bibitem{hu2018beyond}
S.~Hu, F.~Rusek, and O.~Edfors, ``Beyond massive {MIMO}: The potential of data
  transmission with large intelligent surfaces,'' \emph{IEEE Transactions on
  Signal Processing}, vol.~66, no.~10, pp. 2746--2758, 2018.

\bibitem{hu2017potential}
------, ``The potential of using large antenna arrays on intelligent
  surfaces,'' in \emph{2017 IEEE 85th Vehicular Technology Conference (VTC
  Spring)}.\hskip 1em plus 0.5em minus 0.4em\relax IEEE, 2017, pp. 1--6.

\bibitem{nadeem2019large}
Q.-U.-A. Nadeem, A.~Kammoun, A.~Chaaban, M.~Debbah, and M.-S. Alouini, ``Large
  intelligent surface assisted {MIMO} communications,'' \emph{arXiv preprint
  arXiv:1903.08127}, 2019.

\bibitem{jung2019performance}
M.~Jung, W.~Saad, and G.~Kong, ``Performance analysis of large intelligent
  surfaces ({LISs}): Uplink spectral efficiency and pilot training,''
  \emph{arXiv preprint arXiv:1904.00453}, 2019.

\bibitem{de2019non}
E.~De~Carvalho, A.~Ali, A.~Amiri, M.~Angjelichinoski, and R.~W. Heath~Jr,
  ``Non-stationarities in extra-large scale massive {MIMO},'' \emph{arXiv
  preprint arXiv:1903.03085}, 2019.

\bibitem{hu2018user}
S.~Hu, K.~Chitti, F.~Rusek, and O.~Edfors, ``User assignment with distributed
  large intelligent surface {(LIS)} systems,'' in \emph{2018 IEEE 29th Annual
  International Symposium on Personal, Indoor and Mobile Radio Communications
  (PIMRC)}.\hskip 1em plus 0.5em minus 0.4em\relax IEEE, 2018, pp. 1--6.

\bibitem{hu2017cramer}
S.~Hu, F.~Rusek, and O.~Edfors, ``Cram{\'e}r-rao lower bounds for positioning
  with large intelligent surfaces,'' in \emph{2017 IEEE 86th Vehicular
  Technology Conference (VTC-Fall)}.\hskip 1em plus 0.5em minus 0.4em\relax
  IEEE, 2017, pp. 1--6.

\bibitem{gong2020toward}
S.~Gong, X.~Lu, D.~T. Hoang, D.~Niyato, L.~Shu, D.~I. Kim, and Y.-C. Liang,
  ``Toward smart wireless communications via intelligent reflecting surfaces: A
  contemporary survey,'' \emph{IEEE Communications Surveys \& Tutorials},
  vol.~22, no.~4, pp. 2283--2314, 2020.

\bibitem{basar2019wireless}
E.~Basar, M.~Di~Renzo, J.~De~Rosny, M.~Debbah, M.-S. Alouini, and R.~Zhang,
  ``Wireless communications through reconfigurable intelligent surfaces,''
  \emph{IEEE Access}, vol.~7, pp. 116\,753--116\,773, 2019.

\bibitem{ozdogan2019intelligent}
{\"O}.~{\"O}zdogan, E.~Bj{\"o}rnson, and E.~G. Larsson, ``Intelligent
  reflecting surfaces: Physics, propagation, and pathloss modeling,''
  \emph{IEEE Wireless Communications Letters}, vol.~9, no.~5, pp. 581--585,
  2019.

\bibitem{liang2019large}
Y.-C. Liang, R.~Long, Q.~Zhang, J.~Chen, H.~V. Cheng, and H.~Guo, ``Large
  intelligent surface/antennas {(LISA)}: Making reflective radios smart,''
  \emph{Journal of Communications and Information Networks}, vol.~4, no.~2, pp.
  40--50, 2019.

\bibitem{liaskos2018new}
C.~Liaskos, S.~Nie, A.~Tsioliaridou, A.~Pitsillides, S.~Ioannidis, and
  I.~Akyildiz, ``A new wireless communication paradigm through
  software-controlled metasurfaces,'' \emph{IEEE Communications Magazine},
  vol.~56, no.~9, pp. 162--169, 2018.

\bibitem{qingqing2019towards}
Q.~Wu and R.~Zhang, ``Towards smart and reconfigurable environment: Intelligent
  reflecting surface aided wireless network,'' \emph{IEEE Communications
  Magazine}, vol.~58, no.~1, pp. 106--112, 2019.

\bibitem{He2019LIM}
Z.-Q. He and X.~Yuan, ``Cascaded channel estimation for large intelligent
  metasurface assisted massive {MIMO},'' \emph{IEEE Wireless Communications
  Letters}, vol.~9, no.~2, pp. 210--214, 2019.

\bibitem{Tan2016SRA}
X.~Tan, Z.~Sun, and J.~M. Jornet, ``Increasing indoor spectrum sharing capacity
  using smart reflect-array,'' in \emph{2016 IEEE International Conference on
  Communications (ICC)}.\hskip 1em plus 0.5em minus 0.4em\relax IEEE, 2016, pp.
  1--6.

\bibitem{Tan2018SRA}
X.~Tan, Z.~Sun, and D.~Koutsoni, ``Enabling indoor mobile millimeter-wave
  networks based on smart reflect-arrays,'' in \emph{2018 IEEE 29th Annual
  International Symposium on Personal, Indoor and Mobile Radio Communications
  (PIMRC)}.\hskip 1em plus 0.5em minus 0.4em\relax IEEE, 2018, pp. 1--6.

\bibitem{Nie2019SRA}
S.~Nie, J.~M. Jornet, and I.~F. Akyildiz, ``Intelligent environments based on
  ultra-massive {MIMO} platforms for wireless communication in millimeter wave
  and terahertz bands,'' in \emph{2019 IEEE International Conference on
  Acoustics, Speech and Signal Processing (ICASSP)}.\hskip 1em plus 0.5em minus
  0.4em\relax IEEE, 2019, pp. 7849--7853.

\bibitem{Basar2019IRS}
E.~Basar, ``Reconfigurable intelligent surface-based index modulation: A new
  beyond {MIMO} paradigm for {6G},'' \emph{IEEE Transactions on
  Communications}, vol.~68, no.~5, pp. 3187--3196, 2020.

\bibitem{Liaskos2019IRS}
C.~Liaskos, A.~Tsioliaridou, and S.~Nie, ``An interpretable neural network for
  configuring pro-grammable wireless environments,'' in \emph{2019 IEEE 20th
  International Workshop on Signal Processing Advances in Wireless
  Communications (SPAWC)}.\hskip 1em plus 0.5em minus 0.4em\relax IEEE, 2019,
  pp. 1--6.

\bibitem{Yuille2003Convex}
A.~L. Yuille and A.~Rangarajan, ``The concave-convex procedure,'' \emph{Neural
  Computation}, vol.~15, no.~4, pp. 915--936, 2003.

\bibitem{Lipp2016Convex}
T.~Lipp and S.~Boyd, ``Variations and extension of the convex-concave
  procedure,'' \emph{Optimization and Engineering}, vol.~17, no.~2, pp.
  263--287, 2016.

\bibitem{Boyd2004Convex}
S.~Boyd and L.~Vandenberghe, ``Convex optimization,'' \emph{Cabridge University
  Press}, 2014.

\bibitem{Yang2019IRS}
Y.~Yang, S.~Zhang, and R.~Zhang, ``Irs-enhanced {OFDM}: Power allocation and
  passive array optimization,'' in \emph{2019 IEEE Global Communications
  Conference (GLOBECOM)}.\hskip 1em plus 0.5em minus 0.4em\relax IEEE, 2019,
  pp. 1--6.

\bibitem{Hu2018IRS}
S.~Hu, F.~Rusek, and O.~Edfors, ``Capacity degradation with modeling hardware
  impairment in large intelligent surface,'' in \emph{2018 IEEE Global
  Communications Conference (GLOBECOM)}.\hskip 1em plus 0.5em minus 0.4em\relax
  IEEE, 2018, pp. 1--6.

\bibitem{Nadeem2020IRS}
Q.~U.~A. Nadeem, H.~Alwazani, and A.~Kammoun, ``Intelligent reflecting surface
  assisted multi-user {MISO} communication: channel estimation and beamforming
  design,'' \emph{IEEE Open Journal of the Communications Society}, vol.~1, pp.
  661--680, 2020.

\bibitem{He2020IRS}
Z.~He and X.~Yuan, ``Cascaded channel estimation for large intelligent
  metasurface assisted massive {MIMO},'' \emph{IEEE Wireless Communication
  Letters}, vol.~9, no.~2, pp. 6210--6214, 2020.

\bibitem{Zheng2019IRS}
B.~Zheng and R.~Zhang, ``Intelligent reflecting surface-enhanced {OFDM}:
  Channel estimation and reflection optimization,'' \emph{IEEE Wireless
  Communications Letters}, vol.~9, no.~4, pp. 518--522, 2019.

\bibitem{huang2020holographic}
C.~Huang, S.~Hu, G.~C. Alexandropoulos, A.~Zappone, C.~Yuen, R.~Zhang,
  M.~Di~Renzo, and M.~Debbah, ``Holographic {MIMO} surfaces for {6G} wireless
  networks: Opportunities, challenges, and trends,'' \emph{IEEE Wireless
  Communications}, vol.~27, no.~5, pp. 118--125, 2020.

\bibitem{Di2020IRS}
B.~Di, H.~Zhang, L.~Song, Y.~Li, Z.~Han, and H.~V. Poor, ``Hybrid beamforming
  for reconfigurable intelligent surface based multi-user communications:
  Achievable rates with limited discrete phase shifts,'' \emph{IEEE Journal on
  Selected Areas in Communications}, vol.~38, no.~8, pp. 1809--1822, 2020.

\bibitem{Pozar2012IS}
D.~M. Pozar, \emph{Microwave engineering}.\hskip 1em plus 0.5em minus
  0.4em\relax John wiley \& sons, 2011.

\bibitem{Koziel2013IS}
S.~Koziel, ``Surrogate-based modeling and optimization: applications in
  engineering,'' \emph{New York, NY: Springer}, 2013.

\bibitem{meng2020interference}
X.~Meng, F.~Liu, J.~Zhou, and S.~Yang, ``Interference exploitation precoding
  for intelligent reflecting surface aided communication system,'' \emph{IEEE
  Wireless Communications Letters}, 2020.

\bibitem{Smith2017IRS}
D.~R. Smith, O.~Yurduseven, L.~P. Mancera, P.~Bowen, and N.~B. Kundtz,
  ``Analysis of a waveguide-fed metasurface antenna,'' \emph{Physical Review
  Applied}, vol.~8, no.~5, p. 054048, 2017.

\bibitem{Abeywickrama2020IRS}
S.~Abeywickrama, R.~Zhang, Q.~Wu, and C.~Yuen, ``Intelligent reflecting
  surface: practical phase shift model and beamforming optimization,'' in
  \emph{2020 IEEE International Conference on Communications (ICC)}.\hskip 1em
  plus 0.5em minus 0.4em\relax IEEE, 2020, pp. 1--6.

\bibitem{Abeywickrama2020IRS1}
------, ``Intelligent reflecting surface: practical phase shift model and
  beamforming optimization,'' \emph{IEEE Transactions on Communications},
  vol.~68, no.~9, pp. 5849--5863, 2020.

\bibitem{Gong2020IRS}
S.~Gong, X.~Lu, D.~T. Hoang, and \emph{et al.}, ``Toward smart wireless
  communications via intelligent reflecting surfaces: A contemporary survey,''
  \emph{IEEE Communications Surveys Tutorials}, vol.~22, no.~4, pp. 2283--2314,
  2020.

\bibitem{Nahhal2020IM}
M.~Al-Nahhal, E.~Basar, and U.~Uysal, ``Flexible generalized spatial modulation
  for visible light communications,'' \emph{IEEE Transactions on Vehicular
  Technology}, pp. 1--1, 2020.

\bibitem{Gao2020IM}
X.~Gao, Z.~Bai, P.~Gong, and D.~O. Wu, ``Design and performance analysis of
  led-grouping based spatial modulation in the visible light communication
  system,'' \emph{IEEE Transactions on Vehicular Technology}, vol.~69, no.~7,
  pp. 7317--7324, 2020.

\bibitem{Kumar2018IM}
C.~R. Kumar and R.~K. Jeyachitra, ``Dual-mode generalized spatial modulation
  mimo for visible light communications,'' \emph{IEEE Communications Letters},
  vol.~22, no.~2, pp. 280--283, 2018.

\bibitem{Wang2018IM}
J.~Wang, J.~Zhu, S.~Lin, and J.~Wang, ``Adaptive spatial modulation based
  visible light communications: Ser analysis and optimization,'' \emph{IEEE
  Photonics Journal}, vol.~10, no.~3, pp. 1--14, 2018.

\bibitem{Huang2020IM}
G.~Huang, S.~Ouyang, Y.~Ding, and V.~Fusco, ``Index modulation for frequency
  diverse array,'' \emph{IEEE Antennas and Wireless Propagation Letters},
  vol.~19, no.~1, pp. 49--53, 2020.

\bibitem{Nusenu2020IM}
S.~Y. Nusenu, S.~Huaizong, Y.~Pan, and A.~Basit, ``Space-frequency increment
  index modulation approach for fifth generation and beyond wireless
  communication systems,'' \emph{IEEE Transactions on Vehicular Technology},
  vol.~69, no.~6, pp. 6286--6298, 2020.

\bibitem{Purwita2020IM}
A.~A. Purwita, A.~Yesilkaya, M.~Safari, and H.~Haas, ``Generalized time slot
  index modulation for optical wireless communications,'' \emph{IEEE
  Transactions on Communications}, vol.~68, no.~6, pp. 3706--3719, 2020.

\bibitem{Nguyen2019IM}
N.~H. Nguyen, B.~Berscheid, and H.~H. Nguyen, ``Fast-ofdm with index modulation
  for nb-iot,'' \emph{IEEE Communications Letters}, vol.~23, no.~7, pp.
  1157--1160, 2019.

\bibitem{Althunibat2019IM}
S.~Althunibat, R.~Mesleh, and K.~Qaraqe, ``Quadrature index modulation based
  multiple access scheme for 5g and beyond,'' \emph{IEEE Communications
  Letters}, vol.~23, no.~12, pp. 2257--2261, 2019.

\bibitem{Yang2020IM}
Y.~Yang, M.~Ma, S.~Aïssa, and L.~Hanzo, ``Physical-layer secret key generation
  via cqi-mapped spatial modulation in multi-hop wiretap ad-hoc networks,''
  \emph{IEEE Transactions on Information Forensics and Security}, vol.~16, pp.
  1322--1334, 2020.

\bibitem{Shi2019IM}
Y.~Shi, X.~Lu, K.~Gao, J.~Zhu, and S.~Wang, ``Subblocks set design aided
  orthogonal frequency division multiplexing with all index modulation,''
  \emph{IEEE Access}, vol.~7, pp. 52\,659--52\,668, 2019.

\bibitem{Shi2019IM1}
------, ``Genetic algorithm aided ofdm with all index modulation,'' \emph{IEEE
  Communications Letters}, vol.~23, no.~12, pp. 2192--2195, 2019.

\bibitem{Yu2020IM}
Z.~Yu, Z.~Bai, K.~Pang, X.~Hao, X.~Yang, and R.~Li, ``Optimization of phase
  rotation-based precoding for spatial modulation system,'' in \emph{2020 IEEE
  20th International Conference on Communication Technology (ICCT)}, 2020, pp.
  236--240.

\bibitem{Zhang2020IM}
Z.~Zhang, C.~Gong, H.~Li, Y.~Dong, X.~Wang, and X.~Dai, ``Soft-input
  soft-output detection via expectation propagation for massive spatial
  modulation mimo systems,'' \emph{IEEE Communications Letters}, pp. 1--1,
  2020.

\bibitem{Katla2020IM}
S.~Katla, L.~Xiang, Y.~Zhang, E.~El-Hajjar, and \emph{et al.}, ``Deep learning
  assisted detection for index modulation aided mmwave systems,'' \emph{IEEE
  Access}, vol.~8, pp. 202\,738--202\,754, 2020.

\bibitem{Satyanarayana2020IM}
K.~Satyanarayana, M.~El-Hajjar, A.~A.~M. Mourad, and \emph{et al.},
  ``Soft-decoding for multi-set space-time shift-keying mmwave systems: A deep
  learning approach,'' \emph{IEEE Access}, vol.~8, pp. 49\,584--49\,595, 2020.

\bibitem{Mao2019IM}
T.~Mao, Q.~Wang, Z.~Wang, and S.~Chen, ``Novel index modulation techniques: A
  survey,'' \emph{IEEE Communications Surveys \& Tutorials}, vol.~21, no.~1,
  pp. 315--348, 2018.

\bibitem{Wen2019IM}
M.~Wen, B.~Zheng, K.~J. Kim, and \emph{et al.}, ``A survey on spatial
  modulation in emerging wireless systems: Research progresses and
  applications,'' \emph{IEEE Journal on Selected Areas in Communications},
  vol.~37, no.~9, pp. 1949--1972, 2019.

\bibitem{SWIPT2018Gu}
Q.~Gu, G.~Wang, and R.~Fan, ``Rate-energy tradeoff in simultaneous wireless
  information and power transfer over fading channels with uncertain
  distribution,'' \emph{IEEE Transactions on Vehicular Technology}, vol.~67,
  no.~4, pp. 3663--3668, 2018.

\bibitem{SWIPT2017Pan}
G.~Pan, H.~Lei, and Y.~Yuan, ``Performance analysis and optimization for
  {SWIPT} wireless wensor networks,'' \emph{IEEE Transactions on
  Communications}, vol.~65, no.~5, pp. 2291--2302, 2017.

\bibitem{Bahbaei2018SWIPT}
M.~Babaei, U.~Aygolu, and E.~Basar, ``Ber analysis of dual-hop relaying with
  energy harvesting in {Nakagami}-m fading channel,'' \emph{IEEE Transactions
  on Communications}, vol.~17, no.~7, pp. 4352--4361, 2018.

\bibitem{Perera2018SWIPT}
T.~D.~P. Perera, D.~N.~K. Jayakody, S.~K. Sharma, S.~Chatzinotas, and J.~Li,
  ``Simultaneous wireless information and power transfer ({SWIPT}): Recent
  advances and future challenges,'' \emph{IEEE Communications Surveys \&
  Tutorials}, vol.~20, no.~1, pp. 264--302, 2017.

\bibitem{Brown1969EH}
W.~C. Brown, ``Experiments involving a microwave beam to power and position a
  helicopter,'' \emph{IEEE Transactions on Aerospace and Electronic Systems},
  vol. AES-5, no.~5, pp. 692--702, 1969.

\bibitem{Varshney2008SWIPT}
L.~R. Varshney, ``Transporting information and energy simultaneously,'' in
  \emph{2008 IEEE International Symposium on Information Theory (ISIT)}.\hskip
  1em plus 0.5em minus 0.4em\relax IEEE, 2008, pp. 1612--1616.

\bibitem{Tang2020SWIPT}
J.~Tang, J.~Luo, J.~Ou, and \emph{et al.}, ``Decoupling or learning: Joint
  power splitting and allocation in mc-noma with swipt,'' \emph{IEEE
  Transactions on Communications}, vol.~68, no.~9, pp. 5834--5848, 2020.

\bibitem{Buckley2020SWIPT}
R.~F. Buckley and R.~W. Heath, ``System and design for selective ofdm swipt
  transmission,'' \emph{IEEE Transactions on Green Communications and
  Networking}, pp. 1--1, 2020.

\bibitem{Wang2019SWIPT}
J.~Wang, G.~Wang, Z.~Lin, and e.~, ``Swipt in mimo af relay systems with direct
  link,'' in \emph{2019 IEEE 89th Vehicular Technology Conference
  (VTC2019-Spring)}, 2019, pp. 1--6.

\bibitem{Krikidis2014SWIPT}
I.~Krikidis, S.~Sasaki, S.~Timotheou, and Z.~Ding, ``A low complexity antenna
  switching for joint wireless information and energy transfer in mimo relay
  channels,'' \emph{IEEE Transactions on Communications}, vol.~62, no.~5, pp.
  1577--1587, 2014.

\bibitem{Ojo2019SWIPT}
F.~K. Ojo and M.~F. Mohd~Salleh, ``Energy efficiency optimization for
  swipt-enabled cooperative relay networks in the presence of interfering
  transmitter,'' \emph{IEEE Communications Letters}, vol.~23, no.~10, pp.
  1806--1810, 2019.

\bibitem{Fang2019SWIPT}
B.~Fang and X.~Zhu, ``Linear precoder design maximizing energy harvesting in
  swipt systems with finite-alphabet inputs,'' in \emph{2019 IEEE 90th
  Vehicular Technology Conference (VTC2019-Fall)}, 2019, pp. 1--5.

\bibitem{Park2020SWIPT}
J.~J. Park, J.~H. Moon, H.~H. Jang, and D.~I. Kim, ``Performance analysis of
  power amplifier nonlinearity on multi-tone swipt,'' \emph{IEEE Wireless
  Communications Letters}, pp. 1--1, 2020.

\bibitem{Jang2020SWIPT}
H.~H. Jang, K.~W. Choi, and D.~I. Kim, ``Novel frequency-splitting swipt for
  overcoming amplifier nonlinearity,'' \emph{IEEE Wireless Communications
  Letters}, vol.~9, no.~6, pp. 826--829, 2020.

\bibitem{Zargari2020SWIPT}
S.~Zargari, A.~Khalili, and R.~Zhang, ``Energy efficiency maximization via
  joint active and passive beamforming design for multiuser miso irs-aided
  swipt,'' \emph{IEEE Wireless Communications Letters}, pp. 1--1, 2020.

\bibitem{Sun2020SWIPT}
W.~Sun, Q.~Song, L.~Guo, and J.~Zhao, ``Secrecy rate maximization for
  intelligent reflecting surface aided swipt systems,'' in \emph{2020 IEEE/CIC
  International Conference on Communications in China (ICCC)}, 2020, pp.
  1276--1281.

\bibitem{Ma2020SWIPT}
R.~Ma, H.~Wu, J.~Ou, S.~Yang, and Y.~Gao, ``Power splitting-based swipt systems
  with full-duplex jamming,'' \emph{IEEE Transactions on Vehicular Technology},
  vol.~69, no.~9, pp. 9822--9836, 2020.

\bibitem{Zhu2020SWIPT}
Z.~Zhu, N.~Wang, W.~Hao, Z.~Wang, and I.~Lee, ``Robust beamforming designs in
  secure mimo swipt iot networks with a non-linear channel model,'' \emph{IEEE
  Internet of Things Journal}, pp. 1--1, 2020.

\bibitem{Thakur2020SWIPT}
A.~Thakur, A.~Kumar, N.~Gupta, and P.~Chatterjee, ``Secrecy analysis of
  reconfigurable underlay cognitive radio networks with swipt and imperfect
  csi,'' \emph{IEEE Transactions on Network Science and Engineering}, pp. 1--1,
  2020.

\bibitem{Hossain2019SWIPT}
M.~A. Hossain, R.~Md~Noor, K.~A. Yau, I.~Ahmedy, and S.~S. Anjum, ``A survey on
  simultaneous wireless information and power transfer with cooperative relay
  and future challenges,'' \emph{IEEE Access}, vol.~7, pp. 19\,166--19\,198,
  2019.

\bibitem{Kawamoto2013SAGIN}
Y.~Kawamoto, H.~Nishiyama, N.~Kato, and N.~Kadowaki, ``A traffic distribution
  technique to minimize packet delivery delay in multilayered satellite
  networks,'' \emph{IEEE Transactions on Vehicular Technology}, vol.~62, no.~7,
  pp. 3315--3324, 2013.

\bibitem{Radhakrishnan2016SAGIN}
R.~Radhakrishnan, W.~W. Edmonson, F.~Afghah, R.~M. Rodriguez-Osorio, F.~Pinto,
  and S.~C. Burleigh, ``Survey of inter-satellite communication for small
  satellite systems: Physical layer to network layer view,'' \emph{IEEE
  Communications Surveys \& Tutorials}, vol.~18, no.~4, pp. 2442--2473, 2016.

\bibitem{Niephaus2016SAGIN}
C.~Niephaus, M.~Kretschmer, and G.~Ghinea, ``{QoS} provisioning in converged
  satellite and terrestrial networks: A survey of the state-of-the-art,''
  \emph{IEEE Communications Surveys \& Tutorials}, vol.~18, no.~4, pp.
  2415--2441, 2016.

\bibitem{Hamdi2008SAGIN}
M.~Hamdi, N.~Boudriga, and M.~S. Obaidat, ``Bandwidth-effictive design of a
  satellite-based hybrid wireless sensor network for mobile target detection
  and tracking,'' \emph{IEEE Systems Journal}, vol.~2, no.~1, pp. 74--82, 2008.

\bibitem{chini2009}
P.~Chini, G.~Giambene, and S.~Kota, ``A survey on mobile satellite systems,''
  \emph{Int. J. Satellite Communications Networking}, vol.~28, pp. 29--57, 01
  2009.

\bibitem{lim2020towards}
W.~Y.~B. Lim, J.~Huang, Z.~Xiong, J.~Kang, D.~Niyato, X.-S. Hua, C.~Leung, and
  C.~Miao, ``Towards federated learning in {UAV}-enabled {Internet of
  Vehicles}: A multi-dimensional contract-matching approach,'' \emph{arXiv
  preprint arXiv:2004.03877}, 2020.

\bibitem{zeng2020federated}
T.~Zeng, O.~Semiari, M.~Mozaffari, M.~Chen, W.~Saad, and M.~Bennis, ``Federated
  learning in the sky: Joint power allocation and scheduling with {UAV}
  swarms,'' \emph{arXiv preprint arXiv:2002.08196}, 2020.

\bibitem{zhang2020maritime}
J.~Zhang, M.~M. Wang, T.~Xia, and L.~Wang, ``Maritime {IoT}: An architectural
  and radio spectrum perspective,'' \emph{IEEE Access}, vol.~8, pp.
  93\,109--93\,122, 2020.

\bibitem{xia2020maritime}
T.~Xia, M.~M. Wang, J.~Zhang, and L.~Wang, ``Maritime {Internet of Things}:
  Challenges and solutions,'' \emph{IEEE Wireless Communications}, vol.~27,
  no.~2, pp. 188--196, 2020.

\bibitem{gupta2015survey}
L.~Gupta, R.~Jain, and G.~Vaszkun, ``Survey of important issues in {UAV}
  communication networks,'' \emph{IEEE Communications Surveys \& Tutorials},
  vol.~18, no.~2, pp. 1123--1152, 2015.

\bibitem{yuan2019potential}
Y.~Yuan, Y.~Zhao, B.~Zong, and S.~Parolari, ``Potential key technologies for
  {6G} mobile communications,'' \emph{Science China Information Sciences},
  vol.~63, pp. 1--19, 2020.

\bibitem{han2019terahertz}
C.~Han, Y.~Wu, Z.~Chen, and X.~Wang, ``Terahertz communications ({TeraCom}):
  Challenges and impact on 6{G} wireless systems,'' \emph{arXiv preprint
  arXiv:1912.06040}, 2019.

\bibitem{chevalier2019widely}
P.~Chevalier, A.~Armizhan, F.~Wang, M.~Piccardo, S.~G. Johnson, F.~Capasso, and
  H.~O. Everitt, ``Widely tunable compact terahertz gas lasers,''
  \emph{Science}, vol. 366, no. 6467, pp. 856--860, 2019.

\bibitem{HanTHz}
C.~Han and Y.~Chen, ``Propagation modeling for wireless communications in the
  terahertz band,'' \emph{IEEE Communications Magazine}, vol.~56, no.~6, pp.
  96--101, 2018.

\bibitem{KhalidTHz}
N.~Khalid and O.~B. Akan, ``Wideband {THz} communication channel measurements
  for 5{G} indoor wireless networks,'' in \emph{2016 IEEE International
  Conference on Communications (ICC)}.\hskip 1em plus 0.5em minus 0.4em\relax
  IEEE, 2016, pp. 1--6.

\bibitem{yang2014macro}
H.~Yang and T.~L. Marzetta, ``A macro cellular wireless network with uniformly
  high user throughputs,'' in \emph{2014 IEEE 80th Vehicular Technology
  Conference (VTC2014-Fall)}.\hskip 1em plus 0.5em minus 0.4em\relax IEEE,
  2014, pp. 1--5.

\bibitem{vu2020cell}
T.~T. Vu, D.~T. Ngo, N.~H. Tran, H.~Q. Ngo, M.~N. Dao, and R.~H. Middleton,
  ``Cell-free massive {MIMO} for wireless federated learning,'' \emph{IEEE
  Transactions on Wireless Communications}, vol.~19, no.~10, pp. 6377--6392,
  2020.

\bibitem{bashar2020deep}
M.~Bashar, A.~Akbari, K.~Cumanan, H.~Quoc~Ngo, A.~G. Burr, P.~Xiao, and
  M.~Debbah, ``Deep learning-aided finite-capacity fronthaul cell-free massive
  {MIMO} with zero forcing,'' \emph{IEEE ICC 2020}, 2020.

\bibitem{nawaz2020non}
S.~J. Nawaz, S.~K. Sharma, B.~Mansoor, M.~N. Patwary, and N.~M. Khan,
  ``Non-coherent and backscatter communications: Enabling ultra-massive
  connectivity in 6g wireless networks,'' \emph{arXiv preprint
  arXiv:2005.10937}, 2020.

\bibitem{bariah2020prospective}
L.~Bariah, L.~Mohjazi, S.~Muhaidat, P.~C. Sofotasios, G.~K. Kurt,
  H.~Yanikomeroglu, and O.~A. Dobre, ``A prospective look: Key enabling
  technologies, applications and open research topics in 6g networks,''
  \emph{arXiv preprint arXiv:2004.06049}, 2020.

\bibitem{mitola1999cognitive}
J.~Mitola and G.~Q. Maguire, ``Cognitive radio: making software radios more
  personal,'' \emph{IEEE personal communications}, vol.~6, no.~4, pp. 13--18,
  1999.

\bibitem{liu2013ambient}
V.~Liu, A.~Parks, V.~Talla, S.~Gollakota, D.~Wetherall, and J.~R. Smith,
  ``Ambient backscatter: Wireless communication out of thin air,'' \emph{ACM
  SIGCOMM Computer Communication Review}, vol.~43, no.~4, pp. 39--50, 2013.

\bibitem{van2018ambient}
N.~Van~Huynh, D.~T. Hoang, X.~Lu, D.~Niyato, P.~Wang, and D.~I. Kim, ``Ambient
  backscatter communications: A contemporary survey,'' \emph{IEEE
  Communications surveys \& tutorials}, vol.~20, no.~4, pp. 2889--2922, 2018.

\bibitem{8907447}
R.~{Long}, Y.~{Liang}, H.~{Guo}, G.~{Yang}, and R.~{Zhang}, ``Symbiotic radio:
  A new communication paradigm for passive internet of things,'' \emph{IEEE
  Internet of Things Journal}, vol.~7, no.~2, pp. 1350--1363, 2020.

\bibitem{amjad2017full}
M.~Amjad, F.~Akhtar, M.~H. Rehmani, M.~Reisslein, and T.~Umer, ``Full-duplex
  communication in cognitive radio networks: A survey,'' \emph{IEEE
  Communications Surveys \& Tutorials}, vol.~19, no.~4, pp. 2158--2191, 2017.

\bibitem{yuan2020high}
Z.~Yuan, Y.~Ma, Y.~Hu, and W.~Li, ``High-efficiency full-duplex {V2V}
  communication,'' in \emph{2020 2nd 6{G} Wireless Summit (6{G} SUMMIT)}.\hskip
  1em plus 0.5em minus 0.4em\relax IEEE, 2020, pp. 1--5.

\bibitem{xu2020resource}
D.~Xu, X.~Yu, Y.~Sun, D.~W.~K. Ng, and R.~Schober, ``Resource allocation for
  {IRS}-assisted full-duplex cognitive radio systems,'' \emph{arXiv preprint
  arXiv:2003.07467}, 2020.

\bibitem{shen2020beamforming}
H.~Shen, T.~Ding, W.~Xu, and C.~Zhao, ``Beamforming design with fast
  convergence for {IRS-Aided} full-duplex communication,'' \emph{IEEE
  Communications Letters}, 2020.

\bibitem{pan2020full}
G.~Pan, J.~Ye, J.~An, and M.-S. Alouini, ``When {Full-Duplex} transmission
  meets intelligent reflecting surface: Opportunities and challenges,''
  \emph{arXiv preprint arXiv:2005.12561}, 2020.

\bibitem{wood2014ethereum}
G.~Wood, ``Ethereum: {A} secure decentralised generalised transaction ledger,''
  \emph{Ethereum project yellow paper}, vol. 151, no. 2014, pp. 1--32, 2014.

\bibitem{mollah2020blockchain}
M.~B. Mollah, J.~Zhao, D.~Niyato, K.-Y. Lam, X.~Zhang, A.~M. Ghias, L.~H. Koh,
  and L.~Yang, ``Blockchain for future smart grid: A comprehensive survey,''
  \emph{IEEE Internet of Things Journal}, 2020.

\bibitem{porambagesec}
P.~Porambage, T.~Kumar, M.~Liyanage, J.~Partala, L.~Lov{\'e}n, M.~Ylianttila,
  and T.~Sepp{\"a}nen, ``Sec-{EdgeAI}: {AI} for edge security vs security for
  edge {AI}.''

\bibitem{ling2019blockchain}
X.~Ling, J.~Wang, T.~Bouchoucha, B.~C. Levy, and Z.~Ding, ``Blockchain radio
  access network ({B-RAN}): {T}owards decentralized secure radio access
  paradigm,'' \emph{IEEE Access}, vol.~7, pp. 9714--9723, 2019.

\bibitem{le2019prototype}
Y.~Le, X.~Ling, J.~Wang, and Z.~Ding, ``Prototype design and test of blockchain
  radio access network,'' in \emph{2019 IEEE International Conference on
  Communications Workshops (ICC Workshops)}.\hskip 1em plus 0.5em minus
  0.4em\relax IEEE, 2019, pp. 1--6.

\bibitem{ariyanti2020visible}
S.~Ariyanti and M.~Suryanegara, ``Visible light communication ({VLC}) for 6{G}
  technology: The potency and research challenges,'' in \emph{2020 Fourth World
  Conference on Smart Trends in Systems, Security and Sustainability
  (WorldS4)}.\hskip 1em plus 0.5em minus 0.4em\relax IEEE, 2020, pp. 490--493.

\bibitem{arfaoui2020physical}
M.~A. Arfaoui, M.~D. Soltani, I.~Tavakkolnia, A.~Ghrayeb, M.~Safari, C.~Assi,
  and H.~Haas, ``Physical layer security for visible light communication
  systems: A survey,'' \emph{IEEE Communications Surveys \& Tutorials}, 2020.

\bibitem{6gwhite}
``6{G} white paper.''
  \url{http://jultika.oulu.fi/Record/isbn978-952-62-2354-4}.

\bibitem{chowdhury2019role}
M.~Z. Chowdhury, M.~Shahjalal, M.~Hasan, and Y.~M. Jang, ``The role of optical
  wireless communication technologies in 5{G}/6{G} and {I}o{T} solutions:
  Prospects, directions, and challenges,'' \emph{Applied Sciences}, vol.~9,
  no.~20, p. 4367, 2019.

\bibitem{katz2020opportunities}
M.~Katz and I.~Ahmed, ``Opportunities and challenges for visible light
  communications in {6G},'' in \emph{2020 2nd {6G} Wireless Summit ({6G
  SUMMIT})}.\hskip 1em plus 0.5em minus 0.4em\relax IEEE, 2020, pp. 1--5.

\bibitem{zhang2020envisioning}
S.~Zhang, J.~Liu, H.~Guo, M.~Qi, and N.~Kato, ``Envisioning device-to-device
  communications in 6{G},'' \emph{IEEE Network}, vol.~34, no.~3, pp. 86--91,
  2020.

\bibitem{chen2020reconfigurable}
Y.~Chen, B.~Ai, H.~Zhang, Y.~Niu, L.~Song, Z.~Han, and H.~V. Poor,
  ``Reconfigurable intelligent surface assisted device-to-device
  communications,'' \emph{arXiv preprint arXiv:2007.00859}, 2020.

\bibitem{wyner1975wire}
A.~D. Wyner, ``The wire-tap channel,'' \emph{Bell system technical journal},
  vol.~54, no.~8, pp. 1355--1387, 1975.

\bibitem{csiszar1978broadcast}
I.~Csisz{\'a}r and J.~Korner, ``Broadcast channels with confidential
  messages,'' \emph{IEEE transactions on information theory}, vol.~24, no.~3,
  pp. 339--348, 1978.

\bibitem{shen2019secrecy}
H.~Shen, W.~Xu, S.~Gong, Z.~He, and C.~Zhao, ``Secrecy rate maximization for
  intelligent reflecting surface assisted multi-antenna communications,''
  \emph{IEEE Communications Letters}, vol.~23, no.~9, pp. 1488--1492, 2019.

\bibitem{hong2020artificial}
S.~Hong, C.~Pan, H.~Ren, K.~Wang, and A.~Nallanathan, ``Artificial-noise-aided
  secure {MIMO} wireless communications via intelligent reflecting surface,''
  \emph{arXiv preprint arXiv:2002.07063}, 2020.

\bibitem{yu2020robust}
X.~Yu, D.~Xu, Y.~Sun, D.~W.~K. Ng, and R.~Schober, ``Robust and secure wireless
  communications via intelligent reflecting surfaces,'' \emph{IEEE Journal on
  Selected Areas in Communications}, 2020.

\bibitem{yamamoto1991coding}
H.~Yamamoto, ``A coding theorem for secret sharing communication systems with
  two {THz} wiretap channels,'' \emph{IEEE Transactions on Information Theory},
  vol.~37, no.~3, pp. 634--638, 1991.

\bibitem{leung1978gaussian}
S.~Leung-Yan-Cheong and M.~Hellman, ``The {Gaussian} wire-tap channel,''
  \emph{IEEE transactions on information theory}, vol.~24, no.~4, pp. 451--456,
  1978.

\bibitem{klinc2011ldpc}
D.~Klinc, J.~Ha, S.~W. McLaughlin, J.~Barros, and B.-J. Kwak, ``{LDPC} codes
  for the {Gaussian} wiretap channel,'' \emph{IEEE Transactions on Information
  Forensics and Security}, vol.~6, no.~3, pp. 532--540, 2011.

\bibitem{sperandio2002wireless}
C.~Sperandio and P.~G. Flikkema, ``Wireless physical-layer security via
  transmit precoding over dispersive channels: optimum linear eavesdropping,''
  in \emph{MILCOM 2002. Proceedings}, vol.~2.\hskip 1em plus 0.5em minus
  0.4em\relax IEEE, 2002, pp. 1113--1117.

\bibitem{cobb2011intrinsic}
W.~E. Cobb, E.~D. Laspe, R.~O. Baldwin, M.~A. Temple, and Y.~C. Kim,
  ``Intrinsic physical-layer authentication of integrated circuits,''
  \emph{IEEE Transactions on Information Forensics and Security}, vol.~7,
  no.~1, pp. 14--24, 2011.

\bibitem{shan2013phy}
D.~Shan, K.~Zeng, W.~Xiang, P.~Richardson, and Y.~Dong, ``{PHY-CRAM}: Physical
  layer challenge-response authentication mechanism for wireless networks,''
  \emph{IEEE Journal on selected areas in communications}, vol.~31, no.~9, pp.
  1817--1827, 2013.

\bibitem{goel2008guaranteeing}
S.~Goel and R.~Negi, ``Guaranteeing secrecy using artificial noise,''
  \emph{IEEE transactions on wireless communications}, vol.~7, no.~6, pp.
  2180--2189, 2008.

\bibitem{Fortino14}
G.~Fortino and P.~Trunfio, \emph{Internet of Things Based on Smart Objects,
  Technology, Middleware and Applications}, 01 2014.

\bibitem{Poniszewska15}
A.~Poniszewska-Maranda and D.~Kaczmarek, ``Selected methods of artificial
  intelligence for {Internet of Things} conception,'' 10 2015, pp. 1343--1348.

\bibitem{Mucchi20}
L.~Mucchi, S.~Jayousi, S.~Caputo, E.~Paoletti, P.~Zoppi, S.~Geli, and
  P.~Dioniso, ``How {6G} technology can change the future wireless
  healthcare,'' 03 2020, pp. 1--6.

\bibitem{mollah2017secure}
M.~B. Mollah, M.~A.~K. Azad, and A.~Vasilakos, ``Secure data sharing and
  searching at the edge of cloud-assisted {I}nternet of {T}hings,'' \emph{IEEE
  Cloud Computing}, vol.~4, no.~1, pp. 34--42, 2017.

\bibitem{Hewa20}
T.~Hewa, G.~Gür, A.~Kalla, M.~Ylianttila, A.~Braeken, and M.~Liyanage, ``The
  role of blockchain in {6G}: Challenges, opportunities and research
  directions,'' 03 2020.

\bibitem{Berardinelli18}
G.~Berardinelli, N.~Mahmood, I.~Rodriguez~Larrad, and P.~Mogensen, ``Beyond
  {5G} wireless irt for industry 4.0: Design principles and spectrum aspects,''
  12 2018, pp. 1--6.

\bibitem{Sekaran20}
R.~Sekaran, R.~Patan, A.~Raveendran, F.~Al-Turjman, R.~MANIKANDAN, and
  L.~Mostarda, ``Survival study on blockchain based {6G}-enabled mobile edge
  computation for {IoT} automation,'' \emph{IEEE Access}, vol.~PP, 08 2020.

\bibitem{Peltonen20}
E.~Peltonen, M.~Bennis, M.~Capobianco, M.~Debbah, A.~Ding,
  F.~Gil-Casti{\~n}eira, M.~Jurmu, T.~Karvonen, M.~Kelanti, A.~Kliks,
  T.~Lepp{\"a}nen, L.~Lov{\`e}n, T.~Mikkonen, A.~Rao, S.~Samarakoon,
  K.~Sepp{\"a}nen, P.~Sroka, S.~Tarkoma, and T.~Yang, ``{6G} white paper on
  edge intelligence,'' \emph{arXiv preprint arXiv:2004.14850}, 2020.

\bibitem{Noury07}
N.~Noury, A.~Fleury, P.~Rumeau, A.~Bourke, G.~ÓLaighin, V.~Rialle, and J.-E.
  Lundy, ``Fall detection – principles and methods,'' \emph{Annual
  International Conference of the IEEE Engineering in Medicine and Biology
  Society}, vol. 2007, pp. 1663--6, 02 2007.

\bibitem{Mao20AI}
B.~Mao, Y.~Kawamoto, and N.~Kato, ``{AI}-based joint optimization of {QoS} and
  security for {6G} energy harvesting {Internet of Things},'' \emph{IEEE
  Internet of Things Journal}, vol.~PP, pp. 7032--7042, 03 2020.

\bibitem{Zheng19}
G.~Zheng, X.~Zang, N.~Xu, H.~Wei, Z.~Yu, V.~Gayah, K.~Xu, and Z.~Li,
  ``Diagnosing reinforcement learning for traffic signal control,'' \emph{arXiv
  preprint arXiv:1905.04716}, 2019.

\bibitem{Sun20}
Y.~Sun, J.~Liu, J.~Wang, Y.~Cao, and N.~Kato, ``When machine learning meets
  privacy in {6G}: A survey,'' \emph{IEEE Communications Surveys \& Tutorials},
  vol.~PP, 07 2020.

\bibitem{Zhang18}
T.~Zhang and Q.~Zhu, ``Distributed privacy-preserving collaborative intrusion
  detection systems for {VANETs},'' \emph{IEEE Transactions on Signal and
  Information Processing over Networks}, vol.~PP, 02 2018.

\bibitem{Dizdar20}
O.~Dizdar, Y.~Mao, W.~Han, and B.~Clerckx, ``{Rate-Splitting} multiple access:
  A new frontier for the {PHY} layer of {6G},'' \emph{arxiv}, 06 2020.

\bibitem{Mehta20}
P.~Mehta, R.~Gupta, and S.~Tanwar, ``Blockchain envisioned {UAV} networks:
  Challenges, solutions, and comparisons,'' \emph{Computer Communications},
  vol. 151, 02 2020.

\bibitem{Aygun19}
A.~Aygun, H.~Ghasemzadeh, and R.~Jafari, ``Robust interbeat interval and heart
  rate variability estimation method from various morphological features using
  wearable sensors,'' \emph{IEEE Journal of Biomedical and Health Informatics},
  vol.~PP, 12 2019.

\bibitem{Tian19}
S.~Tian, W.~Yang, J.~M. Le~Grange, P.~Wang, W.~Huang, and Z.~Ye, ``Smart
  healthcare: making medical care more intelligent,'' \emph{Global Health
  Journal}, vol.~3, 10 2019.

\bibitem{Torfs17}
G.~Torfs, H.~Li, S.~Agneessens, J.~Bauwelinck, L.~Breyne, O.~Caytan, W.~Joseph,
  S.~Lemey, H.~Rogier, A.~Thielens, D.~V. Ginste, X.~Yin, and P.~Demeester,
  ``{ATTO}: Wireless networking at fiber speed,'' \emph{Journal of Lightwave
  Technology}, vol.~36, no.~8, pp. 1468--1477, 2017.

\bibitem{Monserrat20}
J.~Monserrat, D.~Martin-Sacristan, F.~Bouchmal, O.~Carrasco, J.~Flores~de
  Valgas, and N.~Cardona, ``Key technologies for the advent of the {6G},'' 04
  2020, pp. 1--6.

\bibitem{Akyildiz20}
I.~Akyildiz, A.~Kak, and S.~Nie, ``{6G} and beyond: The future of wireless
  communications systems,'' \emph{IEEE Access}, vol.~PP, 07 2020.

\bibitem{Bizon14}
N.~Bizon, L.~Dascalescu, and N.~M.~Tabatabaei, \emph{Autonomous vehicles:
  Intelligent transport systems and smart technologies}, 01 2014.

\bibitem{Zhang19}
J.~Zhang, T.~Chen, S.~Zhong, J.~Wang, W.~Zhang, X.~Zuo, R.~Maunder, and
  L.~Hanzo, ``Aeronautical {Ad Hoc} networking for the
  {Internet-Above-The-Clouds},'' \emph{Proceedings of the IEEE}, 05 2019.

\bibitem{Lee19arx}
Y.~L. Lee, D.~Qin, L.-C. Wang, G.~Hong, and Sim, ``{6G} massive radio access
  networks: Key issues, technologies, and future challenges,'' 2019.

\bibitem{Solmaz20}
S.~Niknam, A.~Roy, H.~S. Dhillon, S.~Singh, R.~Banerji, J.~H. Reed, N.~Saxena,
  and S.~Yoon, ``Intelligent {O-RAN} for beyond {5G} and {6G} wireless
  networks,'' 2020.

\bibitem{Fadlullah20}
Z.~Fadlullah and N.~Kato, ``{HCP}: Heterogeneous computing platform for
  federated learning based collaborative content caching towards {6G}
  networks,'' \emph{IEEE Transactions on Emerging Topics in Computing},
  vol.~PP, 04 2020.

\bibitem{Karan20}
K.~Sheth, K.~Patel, H.~Shah, S.~Tanwar, R.~Gupta, and N.~Kumar, ``A taxonomy of
  ai techniques for 6{G} communication networks,'' \emph{Computer
  Communications}, vol. 161, pp. 279 -- 303, 2020.

\bibitem{Mudasar19}
M.~L. Memon, N.~Saxena, A.~Roy, and D.~R. Shin, ``Backscatter communications:
  Inception of the battery-free era—a comprehensive survey,''
  \emph{Electronics}, vol.~8, p. 129, 2019.

\bibitem{Deanweb}
``Predictions for 2030: telecoms, networks, spectrum \& 5{G}/6{G},''
  \url{https://disruptivewireless.blogspot.com/2020/01/predictions-for-next-decade-looking-out.html}.

\bibitem{williams2020communication}
R.~J. Williams, E.~De~Carvalho, and T.~L. Marzetta, ``A communication model for
  large intelligent surfaces,'' in \emph{2020 IEEE International Conference on
  Communications Workshops (ICC Workshops)}.\hskip 1em plus 0.5em minus
  0.4em\relax IEEE, 2020, pp. 1--6.

\end{thebibliography}

\end{document}